\documentclass[11pt]{article}
\usepackage{amsfonts,amssymb,amsmath,amsthm,amscd,latexsym}
\usepackage[latin1]{inputenc}

\def\Ad{\mathrm{Ad}}

\newcommand{\inv}[0]{{-1}}
\newcommand{\ntg}[0]{{n+2g}}
\newcommand{\cif}[0]{\mathcal{C}^\infty}
\newcommand{\surf}[0]{S_{g,n}}

\newcommand{\mindee}{\mbox{$\!\setminus D$}}

\newcommand{\grx}[1]{>_{\mbox{}_{{#1}}}}
\newcommand{\smx}[1]{<_{\mbox{}_{{#1}}}}

\newcommand{\gothg}{\mathfrak g }

\newcommand{\gothh}{\mathfrak h }

\newcommand{\RR}{\mathbb{R}}
\newcommand{\CC}{\mathbb{C}}

\newcommand{\mapcl}[0]{\text{Map}(S_{g,n})}
\newcommand{\pmapcl}[0]{\text{PMap}(S_{g,n})}
\newcommand{\mapcld}[0]{\text{Map}(S_{g,n}\mindee )}
\newcommand{\pmapcld}[0]{\text{PMap}(S_{g,n}\mindee )}

\newcommand{\dm}[0]{{\overline{M}}}
\newcommand{\da}[0]{{\overline{A}}}
\newcommand{\db}[0]{{\overline{B}}}

\newcommand{\dch}[0]{{\overline{h}}}
\newcommand{\dcm}[0]{{\overline{m}}}
\newcommand{\dca}[0]{{\overline{a}}}
\newcommand{\dcb}[0]{{\overline{b}}}
\newcommand{\dcx}[0]{{\overline{x}}}
\newcommand{\dcy}[0]{{\overline{y}}}

\newcommand{\mi}[0]{{M_i}}
\newcommand{\ai}[0]{{A_i}}
\newcommand{\bi}[0]{{B_i}}
\newcommand{\mj}[0]{{M_j}}
\newcommand{\aj}[0]{{A_j}}
\newcommand{\bjj}[0]{{B_j}}

\newtheorem{theorem}{Theorem}[section]
\newtheorem{lemma}[theorem]{Lemma}
\newtheorem{corollary}[theorem]{Corollary}

\begin{document}
\parskip 6pt
\parindent 0pt

\begin{center}
\baselineskip 24 pt {\Large \bf Dual generators of the fundamental
group and the moduli space of flat connections}

\baselineskip 18 pt

\vspace{1cm} {C.~Meusburger}\footnote{\tt  cmeusburger@perimeterinstitute.ca}\\
Perimeter Institute for Theoretical Physics\\
31 Caroline Street North,
Waterloo, Ontario N2L 2Y5, Canada\\

\vspace{0.5cm}

{ 23 July 2006}

\end{center}

\begin{abstract}
We define the dual of a set of generators of the fundamental group
 of an oriented two-surface $S_{g,n}$ of genus $g$ with $n$ punctures and the associated surface $S_{g,n}\!\setminus D$ with a
 disc $D$
 removed.  This dual is another set of generators related to
 the original generators via an involution and has the properties
 of a dual graph. In particular, it provides an algebraic
 prescription for determining the intersection points of a
 curve representing a general element of the fundamental group $\pi_1(S_{g,n}\!\setminus D)$
with the representatives of the generators and the order in which
these intersection points occur on the generators.We apply this
dual to the moduli space of flat connections on $S_{g,n}$ and show
that when expressed in terms both, the holonomies along a set of
generators and their duals, the Poisson structure on the moduli
space takes a particularly simple form. Using
 this description of the Poisson structure, we
 derive explicit expressions  for the Poisson brackets of general
 Wilson loop observables associated to closed, embedded curves on
 the surface and determine the associated flows on phase space. We
demonstrate that the observables constructed from the pairing in
the
 Chern-Simons action generate of infinitesimal Dehn
 twists and show that the mapping class group acts by Poisson
 isomorphisms.
\end{abstract}


\section{Introduction}

Moduli spaces of flat connections on orientable two-surfaces arise
in many contexts. Our main motivation is their role as the phase
space of Chern-Simons gauge theory,  in particular the application
to the Chern-Simons formulation of
 (2+1)-dimensional gravity with gauge groups $ISO(2,1)$, $SL(2,\CC)$, $SL(2,\RR)\times
SL(2,\RR)/\mathbb{Z}_2$ \cite{AT,Witten1}. Although obtained as
quotients of infinite dimensional spaces of flat connections on
the surface, moduli spaces are finite dimensional, which reflects
the topological nature of the underlying Chern-Simons theory. This
absence of local degrees of freedom allows one to parametrise them
 in terms of the holonomies of curves on the surface.
This parametrisation provides a complete set of gauge invariant
observables, in the following referred to as generalised Wilson
loop observables, given as conjugation invariant functions of the
holonomies. Furthermore, the parametrisation of the moduli space
in terms of holonomies gives rise to an efficient description of
its Poisson structure discovered by Fock and Rosly \cite{FR}. In
Fock and Rosly's formalism, the moduli space is parametrised by
the holonomies along a set of generators of the surface's
fundamental group and its Poisson structure is given by an
auxiliary Poisson structure on an extended phase space which is
obtained by associating one copy of the gauge group to each
generator.

Due to its simplicity, Fock and Rosly's description of the moduli
space has proven useful in the investigation of the phase space of
Chern-Simons theory, in particular in the Chern-Simons formulation
of (2+1)-dimensional gravity \cite{we1, ich}. Moreover, it serves
as the starting point for the quantisation of Chern-Simons theory.
Most quantisation approaches such as the combinatorial
quantisation formalism  for Chern-Simons theory with compact,
semisimple gauge groups \cite{AGSI,AGSII,AS} and the related
approaches in \cite{BNR,we2} for the case of, respectively, gauge
group $SL(2,\CC)$ and $G\ltimes\gothg^*$ are based on Fock and
Rosly's description of the moduli space and take the holonomies
along a set of generators of the fundamental group as their basic
variables.

The drawback of this description is that it obscures the
geometrical nature of the theory and thereby complicates its
physical interpretation. For instance, it is well known that the
Poisson bracket of Wilson loop observables associated to closed
curves on the surface is determined by the intersection behaviour
of these curves, i.~e.~the number of intersection points and the
associated oriented intersection numbers. It is shown in
\cite{ich} for the Chern-Simons formulation of (2+1)-dimensional
gravity with vanishing cosmological constant that this dependence
on intersection points is crucial for the geometrical
interpretation of the observables and the associated phase space
transformations they generate via the Poisson bracket.
 However, in Fock and Rosly's formalism \cite{FR} based on the
 holonomies along
a set of generators of the fundamental group, this dependence on
intersection points is not directly apparent. The main reason is
the lack of a direct link between the expressions of elements of
the fundamental group in terms of the generators and the
intersection points of their representatives on the surface. Given
 the expression of an element of the fundamental group in terms of
 the generators, it is in general difficult to determine how its
 representative intersects the representatives of the generators
 without explicitly drawing these curves on the surface. This
 difficulty manifests itself in Fock and Rosly's description of the Poisson structure, where one
 uses these expressions of in terms of the generators to calculate
 the Poisson bracket of the associated Wilson loop observables.

This problem and its relevance for the physical applications are
the main motivation of the present paper. We consider oriented
two-surfaces $S_{g,n}$ of general genus $g$ and with $n$ punctures
and the associated surfaces $\surf\mindee$ with a disc $D$ removed
and introduce the concept of a dual for a set of generators of the
fundamental groups $\pi_1(\surf)$, $\pi_1(\surf\mindee)$. This
dual is another set of generators related to the original
generators via an involution which has the properties of a dual
graph. It allows us to keep track of the intersection points of
general curves on the surface with representatives of the
generators. More precisely, we show that for any element of the
fundamental group $\pi_1(\surf\mindee)$, the intersection points
of a representing curve with the representatives of the generators
of $\pi_1(\surf\mindee)$ are labelled by the factors in its
expression as a product in the associated dual generators and
their inverses. Moreover, for elements with embedded
representatives, this expression allows one to determine
algebraically the order in which these intersection points occur
on the generators.

We then apply this duality for the fundamental group to Fock and
Rosly's description \cite{FR} of the moduli space of flat
connections. We find that, when expressed in terms of both, the
holonomies along the original set of generators of the fundamental
group and those along their duals, the Poisson structure takes a
particularly simple form in which its dependence on intersection
points is encoded algebraically and readily apparent. This allows
us to derive explicit expressions for the Poisson bracket of the
generalised Wilson loop observables associated to embedded curves
on the surface and to determine the flows these observables
generate via the Poisson bracket. In particular, we consider the
Wilson loop observables constructed from the pairing in the
Chern-Simons action and show that the flows generated by these
observables have the interpretation of infinitesimal Dehn twists.
We thus give an independent re-derivation of Goldman's classic
results in \cite{goldman} and generalise these results to surfaces
with punctures. However, while the results in \cite{goldman} are
presented in a more abstract and geometrical language, the
formulation in this paper is purely algebraic. As it gives
explicit expressions for the Poisson brackets of Wilson loop
observables and the associated phase space transformations in
terms of the holonomies along a set of generators of the
fundamental group, our formulation provides a direct link with the
description of the phase space and the quantisation formalisms in
\cite{AGSI,AGSII,AS,BNR,we1,we2}. This may prove useful in the
investigation of the associated
 observables and transformation in quantised Chern-Simons
 theory. Moreover, in the Chern-Simons formulation of
 (2+1)-dimensional gravity the explicit parametrisation in
 terms of holonomies  allows one to establish a link with the
 geometrical formalism and to
 relate these flows to the geometrical construction of
 (2+1)-spacetimes via grafting \cite{ich3}.

The paper is structured as follows.

In Sect.~\ref{dualint}, we motivate and define the concept of a
dual for a set of generators of the fundamental groups
$\pi_1(\surf)$, $\pi_1(\surf\mindee)$ and investigate the
involution which maps a set of generators to its dual. We show
that the intersection points of a general curve on the surface
$\surf\mindee$ with the representatives of the generators are
labelled by the factors in the expression of its homotopy
equivalence class in terms of their duals.

In Sect.~\ref{combsect} we investigate the combinatorial and
geometrical properties of the dual generators. For elements of
$\pi_1(\surf\mindee)$ with embedded representatives, we show that
their expression in terms of the dual generators allows one to
algebraically determine the order in which these intersection
points occur on the representatives of the generators.
Furthermore, we demonstrate that the involution defines an
(almost) unique assignment of these intersection points  between
the different factors in the expression of this element as a
product in the original generators. We show how this assignment of
intersection points corresponds to a graphical decomposition of
the associated curve on $\surf\mindee$ into representatives of the
generators.

In Sect.~\ref{modspacesect}, we apply the dual generators to the
description of the moduli space of flat connections on $\surf$ in
terms of the holonomies of a set of generators of the fundamental
group. We summarise the relevant facts about Chern-Simons theory
and Fock and Rosly's description \cite{FR} of the moduli space of
flat connections on $\surf$. We show that Fock and Rosly's
auxiliary Poisson structure can be cast in a particularly simple
form when expressed in both, the holonomies along the original set
of generators and their duals, and discuss how this reflects its
dependence on intersection points. Finally, we determine the
transformation of this Poisson structure under the involution that
maps the original generators to their duals.

In Sect.~\ref{wloopsect}, we use this description of the Poisson
structure to determine  the Poisson brackets of the generalised
Wilson loop observables associated to elements of $\lambda\in
\pi_1(\surf\mindee)$ and conjugation invariant functions on the
gauge group. For Wilson loop observables associated to elements
with embedded representatives, we derive the flows on phase space
these observables generate via the Poisson bracket. By using both,
the graphical assignment of intersection points and the ordering
algorithm given in Sect.~\ref{combsect}, we then obtain explicit
expressions for the transformation of the holonomies along our set
of generators  in terms of the expression of $\lambda$ as a
product in the generators of $\pi_1(\surf\mindee)$ and their
duals.

In Sect.~\ref{mapsect} we consider a  set of generic Wilson loop
observables  associated to the $\Ad$-invariant symmetric bilinear
form in the Chern-Simons action. We show that the flows generated
by these observables represent infinitesimal Dehn twists and that
the mapping class group $\mapcld$ acts by Poisson isomorphisms. We
then use this identity to determine the transformation of Fock and
Rosly's Poisson structure under a general automorphism of the
fundamental group $\pi_1(\surf\mindee)$ which acts on the
punctures by conjugation and maps a curve around the disc to
itself or its inverse.

Sect.~\ref{outlook} contains our outlook and conclusions, and in
the appendix we list a set of generators for the mapping class
groups $\mapcld$, $\mapcl$.


\section{The dual of a set of generators of the fundamental group}

\label{dualint}

\begin{figure}
\vskip .3in \protect\input epsf \protect\epsfxsize=12truecm
\protect\centerline{\epsfbox{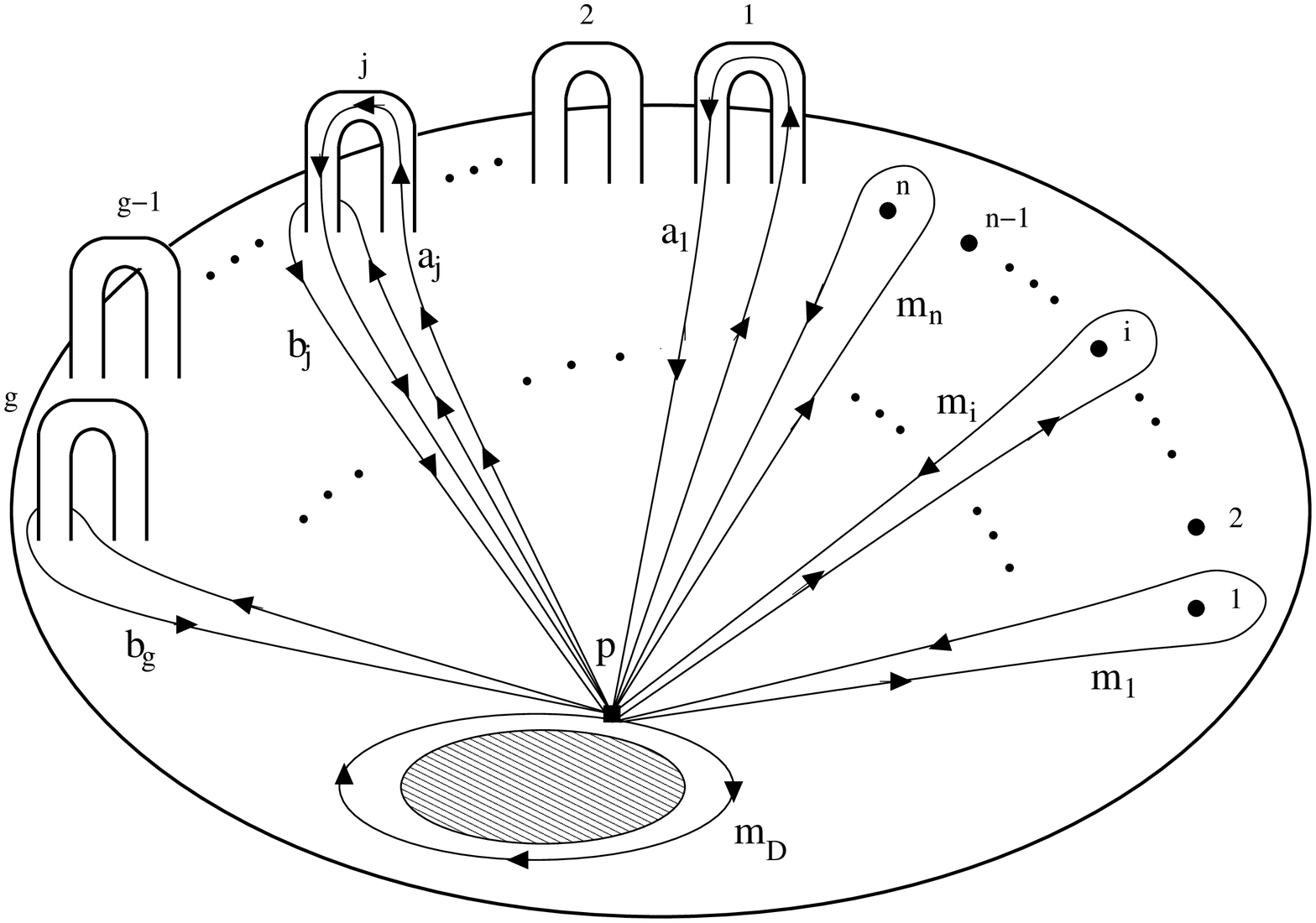}} \caption{The generators of
the fundamental group $\pi_1(\surf\mindee)$} \label{pi1fig}
\end{figure}
In this paper we consider orientable two-surfaces $\surf$ of genus
$g$ and with $n$ punctures and the associated surfaces
$\surf\mindee$ obtained from $\surf$ by removing a disc $D$. Both
the fundamental groups $\pi_1(\surf)$ and $\pi_1(\surf\mindee)$
are generated by the homotopy equivalence classes of a set of
loops $m_i$, $i=1,\ldots,n$, around each puncture and two curves
$a_j, b_j$, $j=1,\ldots,g$, for each handle as shown in
Fig.~\ref{pi1fig}.  While the fundamental group
$\pi_1(\surf\mindee)$ is the free group generated by the homotopy
equivalence classes of $m_i,a_j,b_j$
\begin{align}
\label{fundgroup2} &\pi_1(\surf\mindee)=\langle
m_1,\ldots,m_n,a_1,b_1,\ldots, a_g,b_g\rangle,
\end{align} the fundamental group $\pi_1(\surf)$ is obtained by imposing
a single defining relation
\begin{align}
\label{fundgroup} &\pi_1(\surf)=\langle
m_1,\ldots,m_n,a_1,b_1,\ldots, a_g,b_g\;;\;
[b_g,a_g^\inv]\circ\ldots\circ[b_1,a_1^\inv]\circ
m_n\circ\ldots\circ m_1=1\rangle\nonumber\\
&[b_i,a_i^\inv]=b_i\circ a_i^\inv\circ b_i^\inv\circ a_i,
\end{align}
which amounts to the requirement that the loop around the disc $D$
representing
\begin{align}
\label{relator}
m_D=[b_g,a_g^\inv]\circ\ldots\circ[b_1,a_1^\inv]\circ
m_n\circ\ldots\circ m_1
\end{align}
is contractible. Throughout the paper, we work with a fixed set of
generators as depicted in Fig.~\ref{pi1fig} and representatives
based at a fixed point $p\in\surf$, $p\in\surf\mindee$, which all
homotopies keep fixed. In the following we will often not mention
the dependence on the basepoint explicitly and do not distinguish
notationally  between curves on the surfaces $\surf$,
$\surf\mindee$ and their equivalence classes in the fundamental
groups $\pi_1(\surf)$, $\pi_1(\surf\mindee)$. We will also denote
by the same letter elements of the fundamental group
$\pi_1(\surf\mindee)$ and the corresponding elements of
$\pi_1(\surf)$ obtained via the canonical map
$\pi_1(\surf\mindee)\rightarrow\pi_1(\surf)$.

To define the dual of a set of generators of the fundamental
groups $\pi_1(\surf)$, $\pi_1(\surf\mindee)$, we first establish
the desired properties of this dual and then address its existence
and uniqueness. Heuristically, the dual of a set of generators
$\{m_1,\ldots,m_n,a_1,b_1,\ldots, a_g,b_g\}$ of the fundamental
groups $\pi_1(\surf)$, $\pi_1(\surf\mindee)$ should be another set
of generators $\{\dcm_1,\ldots,\dcm_n,\dca_1,\dcb_1,\ldots,
\dca_g,\dcb_g\}$ related to the original generators by an
automorphism $I\in\text{Aut}(\pi_1(\surf))$,
$I\in\text{Aut}(\pi_1(\surf\mindee))$ and with the following
properties.

\begin{figure}
\label{dualgen} \vskip .3in \protect\input epsf
\protect\epsfxsize=12truecm
\protect\centerline{\epsfbox{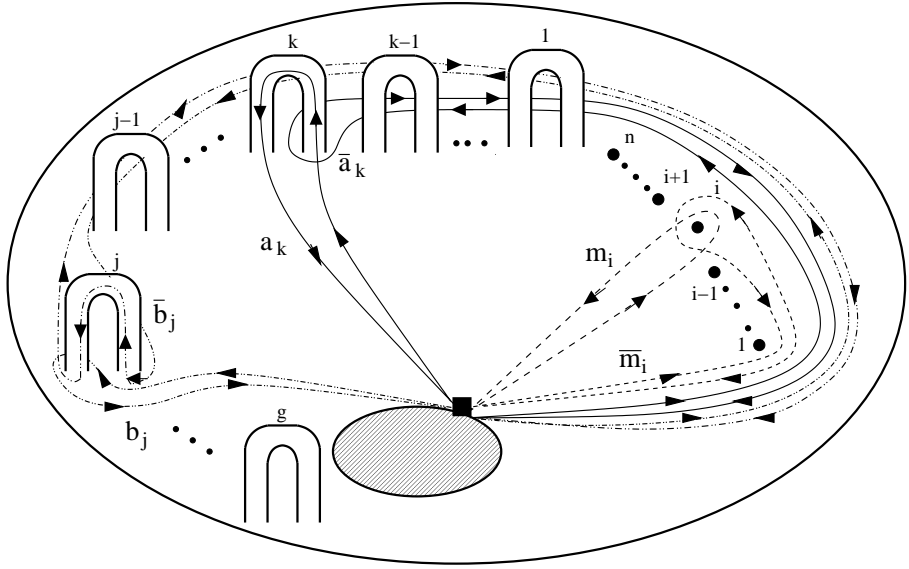}} \caption{The generators
and dual generators of the fundamental group
$\pi_1(\surf\mindee)$}
\end{figure}

\begin{enumerate}
\item The dual of the dual set of generators should be the
original set of generators, i.~e.~the automorphism $I$ should be
an involution $ I^2=1$.

\item The dual set of generators should be geometrically
equivalent to our original set of generators, i.~e.~the
representatives of the original and the dual generators should be
related by homeomorphisms of $\surf$, $\surf\mindee$. In other
words, we require an involution $I\in\text{Aut}(\surf\mindee)$
which is induced by a homeomorphism of $\surf\mindee$ which fixes
the punctures as a set and the boundary of the disc and which
induces an involution of $\pi_1(\surf)$ associated to a
homeomorphism of $\surf$ which fixes the punctures as a set.
Results from combinatorial group theory and geometric topology
(see for instance Theorems 3.4.5, 3.4.6, 3.4.7 in \cite{combgr})
imply that is the case if and only if the automorphism
$I\in\text{Aut}(\surf\mindee)$ satisfies
\begin{align}
\label{invprops2}
&I(m_i)=w_im_{\sigma(i)}^{\epsilon_i}w_i^\inv\quad I(m_D)=
wm_D^\epsilon w^\inv\quad
\omega(w_1)\epsilon_1=\ldots=\omega(w_n)\epsilon_n=\omega(w)\epsilon\in\{\pm
1\}
\end{align}
where  $w_i,w\in\pi_1(\surf\mindee)$, $\epsilon,\epsilon_i\in\{\pm
1\}$, $\sigma\in S_n$ is a permutation of the punctures and
$\omega(x)=1$ if $x\in\pi_1(\surf\mindee)$ corresponds to a
separating curve and $\omega(x)=-1$ otherwise. By applying an
inner automorphism of $\pi_1(\surf\mindee)$ which conjugates all
elements with a fixed element $\lambda\in\pi_1(\surf\mindee)$, we
can set $w=1$ in \eqref{invprops2}, which we will assume in the
following. The corresponding homeomorphism is  orientation
preserving and orientation reversing, respectively, if
$\epsilon=1$ and $\epsilon=-1$.

\item Finally, we require that the dual of our set of generators
of the fundamental group should have properties similar to those
of a dual graph and should allow one to keep track of the
intersection points of general curves on the surface with our set
of generators $m_i,a_j,b_j$. The intersection points of a general
curve on the surface with the representatives of the generators
$m_i,a_j,b_j$ should be labelled by the factors in the expression
of its homotopy equivalence class in terms of the dual generators.
As the expression of a general element in terms of the dual
generators is unique only for the fundamental group
$\pi_1(\surf\mindee)$, we  impose this condition on the involution
$I\in\text{Aut}(\pi_1(\surf\mindee))$. More precisely, for each
element $\lambda\in\pi_1(\surf\mindee)$ we consider the unique
expression of $\lambda$ as a {\em reduced word} in the dual
generators $\dcm_i,\dca_j,\dcb_j$
\begin{align}
\label{lambdadecomp3}
\lambda=\dcx_r^{\alpha_r}\cdots\dcx_1^{\alpha_1}=I(x_r^{\alpha_r}\cdots
x_1^{\alpha_1})\qquad x_k\in\{m_1,\ldots, b_g\}\;,\;
\dcx_k=I(x_k)\;,\;\alpha_k\in\{\pm 1\}
\end{align}
with $\dcx_k^{\alpha_k}\neq\dcx_{k+1}^{-\alpha_{k+1}}$ for
$k=1,\ldots,r-1$ and require that the intersection points of a
curve representing $\lambda$ with the representatives of the
generators $a_j,b_j$ are in one-to-one correspondence with factors
$\dcx_k=\dca_j$, $\dcx_k=\dcb_j$ in \eqref{lambdadecomp3}. For the
generators $m_i$ associated to the punctures, we require that each
factor $\dcx_k=\dcm_i$ in \eqref{lambdadecomp3} corresponds to a
pair of intersection points of this curve with a representative of
$m_i$.
\end{enumerate}
Together the first and the second requirement determine the
involution $I\in\text{Aut}(\pi_1(\surf\mindee))$ up to composition
$I\mapsto \rho\circ I$ with an automorphism
$\rho\in\text{Aut}(\pi_1(\surf\mindee))$ which satisfies
\eqref{invprops2} with $w=1$ and the additional condition
\begin{align}
\rho^\inv=I\rho I.
\end{align}
The third requirement defines $I$ up  to conjugation with an
automorphism $\rho\in\text{Aut}(\pi_1(\surf\mindee))$
\begin{align}
I\mapsto \rho\circ I\circ \rho^\inv,
\end{align}
which satisfies \eqref{invprops2} with $w=1$, since the
intersection points of $\rho(x)$, $x\in\{m_1,\ldots,b_g\}$, with
$\rho(\lambda)$ correspond one-to-one to intersection points of
$x$ with $\lambda$ and are labelled by the factors in the
expression of $\rho(\lambda)$ as a reduced word in $\rho\circ
I\circ \rho^\inv(m_i)$, $\rho\circ I\circ \rho^\inv(a_j)$,
$\rho\circ I\circ \rho^\inv(b_j)$. Furthermore, all involutions
conjugated to a given involution $I$ in that way are obtained from
automorphisms $\rho\in\text{Aut}(\pi_1(\surf\mindee))$ which
satisfy \eqref{invprops2} with $\epsilon=1$.
 We will discuss in Sect.~\ref{mapsect}
that such automorphisms of $\pi_1(\surf\mindee)$ represent
elements of the mapping class group $\mapcld$. Hence, two
involutions satisfying the requirements above and related by
conjugation with such an automorphism correspond to the choice of
an alternative set of generators $\{\rho(m_1),\ldots,\rho(b_g)\}$,
and we have the following theorem.
\begin{theorem}
The requirements above determine the involution
$I\in\text{Aut}(\pi_1(\surf\mindee))$ uniquely up to the initial
choice of the generators of $\pi_1(\surf\mindee)$.
\end{theorem}
After formulating our concept of the dual of a set of generators
of the fundamental group $\pi_1(\surf)$, $\pi_1(\surf\mindee)$ and
discussing its uniqueness, we will now demonstrate that such a set
of generators with the required properties exists. We define an
automorphism $I\in\text{Aut}(\pi_1(\surf\mindee))$ explicitly by
its action on the generators $m_i,a_j,b_j$ and then verify  that
it satisfies the requirements above.
\begin{lemma}
Let $I\in\text{Aut}(\surf\mindee)$ be defined by its action on our
set of generators
\begin{align}
\label{dualcurves} &I(m_i)=\dcm_i=m_1^\inv\circ\ldots\circ m_{i-1}^\inv \circ m_i^\inv \circ m_{i-1} \ldots \circ m_1\\
&I(a_j)=\dca_j=m_1^\inv\circ \ldots \circ m_n^\inv \circ
h_1^\inv\circ \ldots \circ h_{j-1}^\inv h_j^\inv \circ b_j\circ
h_{j-1}\circ\ldots\circ h_1 \circ m_n\circ \ldots\circ  m_1\nonumber\\
&I(b_j)=\dcb_j=m_1^\inv\circ \ldots \circ m_n^\inv \circ
h_1^\inv\circ \ldots \circ h_{j-1}^\inv\circ h_j^\inv \circ
a_j\circ h_{j-1}\circ\ldots\circ h_1 \circ m_n\circ \ldots\circ
m_1\nonumber\\
&\text{with}\quad h_j=[b_j,a_j^\inv]=b_j\circ a_j^\inv\circ
b_j\circ a_j\nonumber.
\end{align}
Then, $I$ is an involution and satisfies the requirements
\eqref{invprops2} with $w=1$, $\epsilon=-1$. It therefore arises
from an orientation reversing homeomorphism of $\surf\mindee$ and
induces an automorphism of $\pi_1(\surf)$ which arises from an
orientation reversing homeomorphism of $\surf$.
\end{lemma}
It remains to show that the dual generators $\dcm_i,\dca_j,\dcb_j$
defined by \eqref{dualcurves} are dual to our original generators
$m_i,a_j,b_j$ in a geometrical sense, i.~e.~ that they satisfy the
third requirement and allow us to determine the intersection
points of a general element $\lambda\in\pi_1(\surf\mindee)$ with
the generators $m_i,a_j,b_j$. For this, we  consider a set of
representing curves on $\surf\mindee$ as depicted in
Fig.~\ref{dualgen}, and note that the curves representing the
generators $\dca_j,\dcb_j$, respectively, intersect only $a_j$ and
$b_j$, in a single point. Similarly, the representatives of the
generators $\dcm_i$ intersect only $m_i$, but in two points and
with opposite oriented intersection numbers.
\begin{figure}
 \vskip .3in \protect\input epsf
\protect\epsfxsize=12truecm
\protect\centerline{\epsfbox{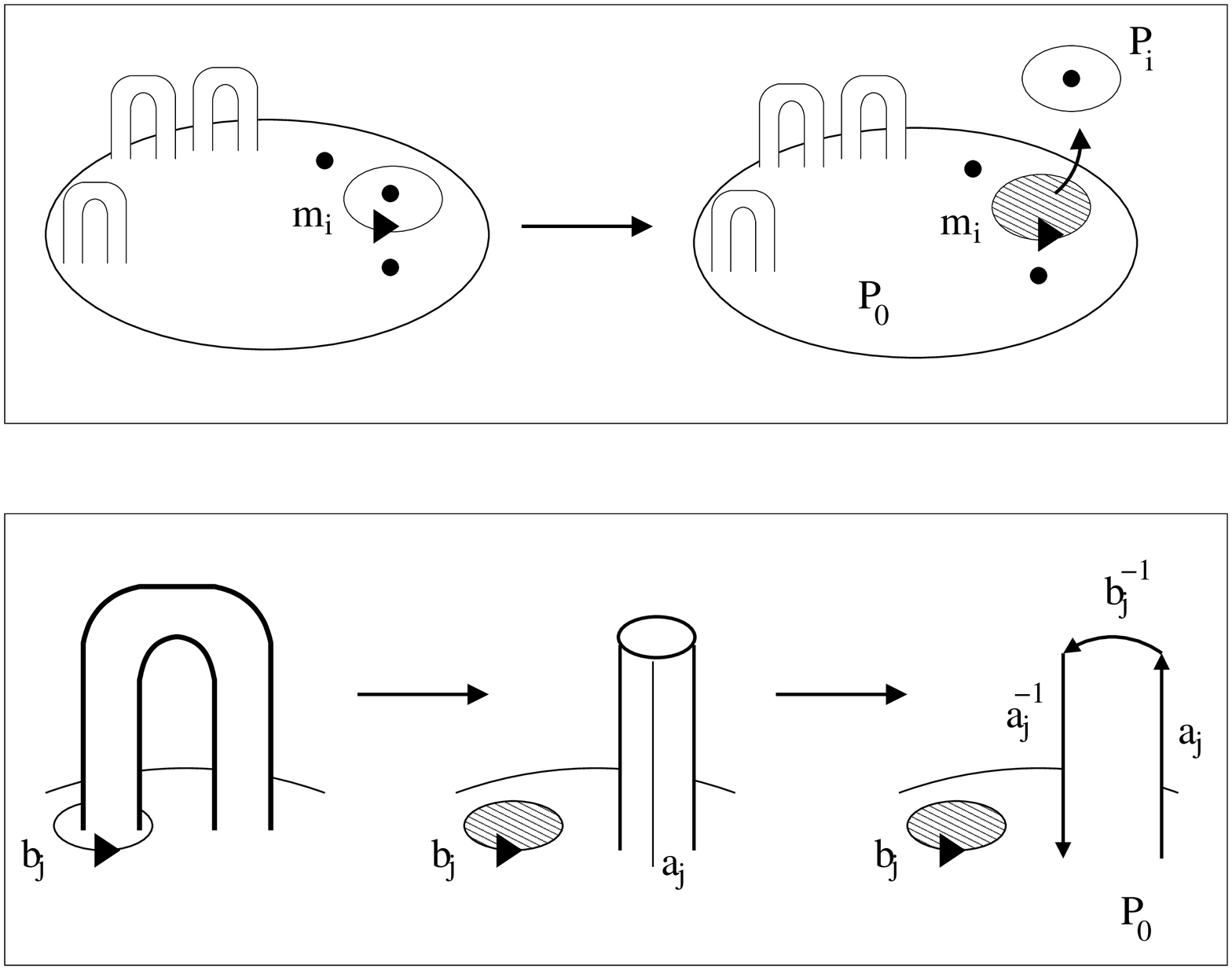}} \caption{Cutting the
surface  $\surf\mindee$ along the generators of the fundamental
group $\pi_1(\surf\mindee)$} \label{cut1}
\end{figure}
\begin{figure}
 \vskip .3in \protect\input epsf
\protect\epsfxsize=12truecm
\protect\centerline{\epsfbox{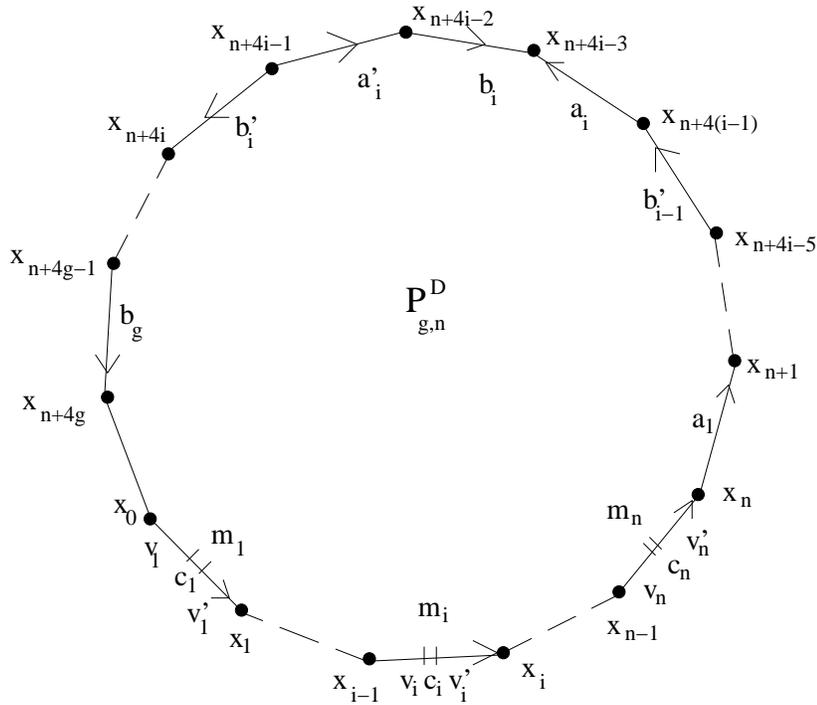}} \caption{The polygon
$P^D_{g,n}$} \label{cut2}
\end{figure}

\begin{figure}
 \vskip .3in \protect\input epsf
\protect\epsfxsize=12truecm
\protect\centerline{\epsfbox{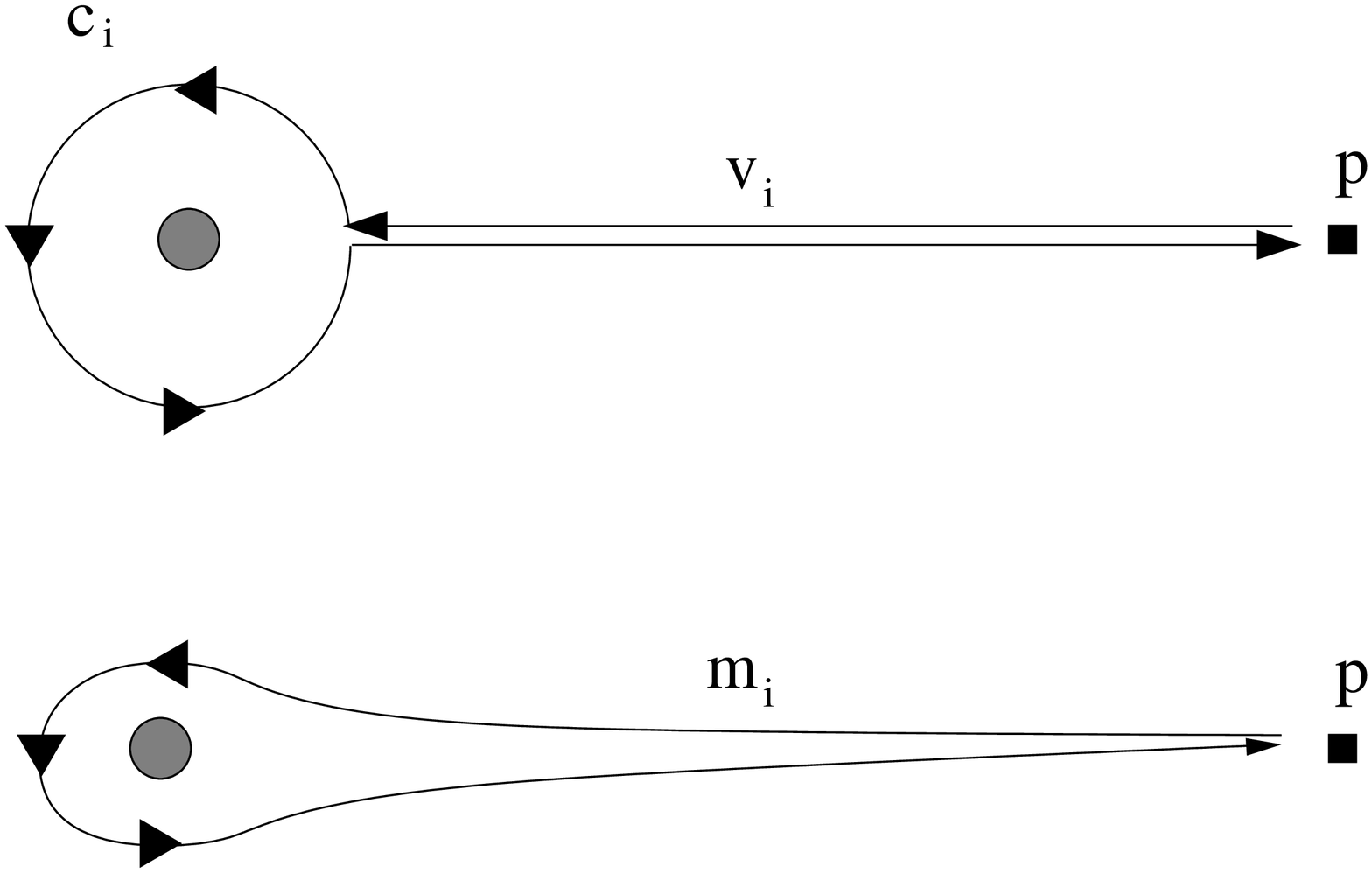}} \caption{The decomposition
of a generator $m_i$ into an arc $v_i$ and an infinitesimal circle
$c_i$} \label{spcs}
\end{figure}

To explore the geometric properties of the dual generators
further, we cut the surface $\surf\mindee$ along a set of curves
representing the generators $m_i,a_j,b_j\in\pi_1(\surf\mindee)$ as
shown in Fig.~\ref{cut1}. We then obtain $n$ punctured discs $D_i$
and a simply connected $(4g+n+1)$-gon $P^D_{g,n}$ depicted in
Fig.~\ref{cut2}. Each of the corners $x_i$, $i=0,\ldots,n+4g$, of
the polygon $P^D_{g,n}$ corresponds to the basepoint
$p\in\surf\mindee$. The side between $x_0$ and $x_{n+4g}$
represents the boundary of the disc $D$, the sides between
$x_{i-1}$ and $x_i$, $i=1,\ldots,n$, the generators $m_i$, and the
remaining $4g$ sides $a_j,a_j'$, $b_j,b_j'$ correspond pairwise to
the generators $a_j,b_j$, $j=1,\ldots,g$. In the following it will
be useful to represent the generators $m_i$ by a curve obtained by
composing an arc $v_i$ from the basepoint $p\in\surf\mindee$ to
the $ith$ puncture and an infinitesimal circle $c_i$ around the
puncture as shown in Fig.~\ref{spcs}. This yields a subdivision of
the side $m_i$ into three segments $v_i,v_i'$, $c_i$ as shown in
Fig.~\ref{cut2}.

We now consider the oriented segments $x_0x_k$ from the corner
$x_0$ to corners $x_k$, $k=1,\ldots,n+4g$ of the polygon
$P^D_{g,n}$. Each of these segments represents a certain element
of the fundamental group $\pi_1(\surf\mindee)$ which is given in
terms of the generators $m_i,a_j,b_j$ and their duals by
\begin{align}
\label{cornerpt}
&x_0x_i\cong m_i\circ\ldots\circ m_1=\dcm_1^\inv\circ\ldots\circ\dcm_i^\inv\qquad\text{for}\;1\leq i\leq n\\
&x_0x_{n+4j-3}\cong a_j\circ h_{j-1}\circ\ldots\circ h_1\circ m_n\circ\ldots \circ m_1=\dcm_1^\inv\circ\ldots\circ\dcm_n^\inv\circ\dch_1^\inv\circ\ldots\circ\dch_{j-1}\circ\dca_j^\inv\circ\dcb_j\circ\dca_j\nonumber\\
&x_0x_{n+4j-2}\cong b_j^\inv\circ a_j\circ h_{j-1}\circ\ldots\circ h_1\circ m_n\circ\ldots \circ m_1=\dcm_1^\inv\circ\ldots\circ\dcm_n^\inv\circ\dch_1^\inv\circ\ldots\circ\dch_{j-1}\circ\dca_j^\inv\circ\dcb_j\nonumber\\
&x_0x_{n+4j-1}\cong a_j^\inv\circ b_j^\inv\circ a_j\circ h_{j-1}\circ\ldots\circ h_1\circ m_n\circ\ldots \circ m_1=\dcm_1^\inv\circ\ldots\circ\dcm_n^\inv\circ\dch_1^\inv\circ\ldots\circ\dch_{j-1}\circ\dca_j^\inv\nonumber\\
&x_0x_{n+4j}\cong h_j\circ h_{j-1}\circ\ldots\circ h_1\circ
m_n\circ\ldots \circ
m_1=\dcm_1^\inv\circ\ldots\circ\dcm_n^\inv\circ\dch_1^\inv\circ\ldots\circ\dch_{j}\nonumber
\quad\text{for}\; 1\leq j\leq g\nonumber.
\end{align}
Both, the generators $m_i,a_j,b_j$ and their duals
$\dcm_i,\dca_j,\dcb_j$ are obtained by composing two segments
$x_0x_k$. For the generators associated to the punctures this
representation is unique
\begin{align}
\label{mseg} m_i= (x_0x_i)\circ (x_0x_{i-1})^\inv\qquad
\dcm_i=(x_0x_i)^\inv\circ(x_0x_{i-1}),
\end{align}
while there are two possibilities for each generator $a_j,b_j$,
$\dca_j,\dcb_j$
\begin{align}
\label{aseg}
a_j=&(x_0x_{n+4j-3})\circ(x_0x_{n+4j-4})^\inv=(x_0x_{n+4j-2})\circ(x_0x_{n+4j-1})^\inv\\
\dca_j
=&(x_0x_{n+4j-1})^\inv\circ(x_0x_{n+4j-4})=(x_0x_{n+4j-2})^\inv\circ(x_0x_{n+4j-3})\nonumber\\
\nonumber\\
\label{bseg}
b_j=&(x_0x_{n+4j-3})\circ(x_0x_{n+4j-2})^\inv=(x_0x_{n+4j})\circ(x_0x_{n+4j-3})^\inv\\
\dcb_j
=&(x_0x_{n+4j-1})^\inv\circ(x_0x_{n+4j-2})=(x_0x_{n+4j})^\inv\circ(x_0x_{n+4j-3}).\nonumber
\end{align}
To demonstrate that the dual generators $\dcm_i,\dca_j,\dcb_j$
allow us to determine the intersection points of general embedded
curves on $\surf\mindee$ with the generators $m_i,a_j,b_j$, we
consider an embedded curve $c:[0,1]\rightarrow\surf\mindee$,
$c(0)=c_\lambda(1)=q$ which does not contain the basepoint
$p\in\surf\mindee$. Furthermore, we require that $c$ has a minimum
number  of intersection points with the representatives of the
generators $m_i,a_j,b_j$, i.~e.~that the number of intersection
points cannot be reduced by applying a homotopy which fixes the
basepoint.

\begin{figure}
 \vskip .3in \protect\input epsf
\protect\epsfxsize=12truecm
\protect\centerline{\epsfbox{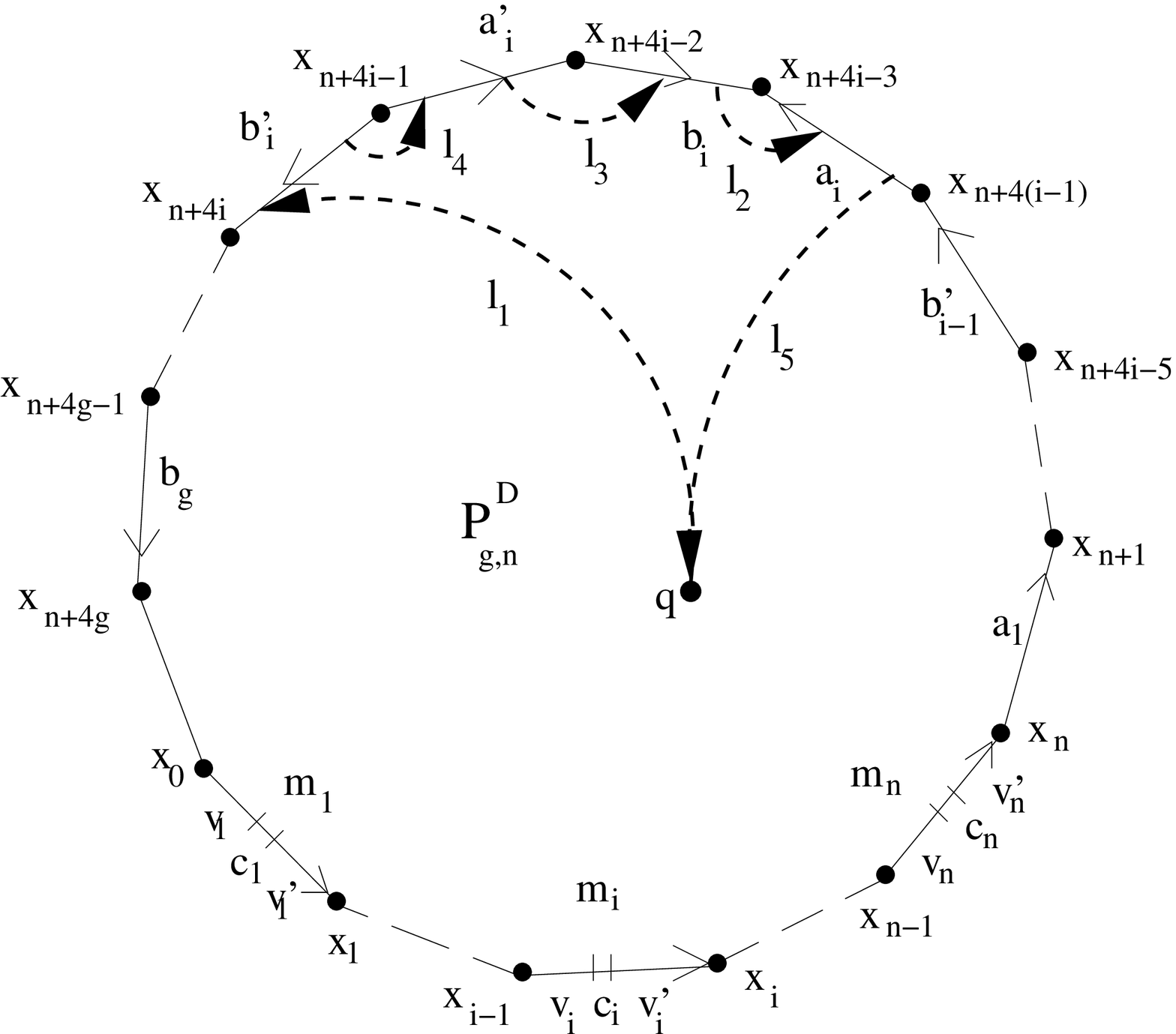}} \caption{A curve
representing
$\lambda=\dca_i^\inv\circ\dcb_i\circ\dca_i\circ\dcb_i^\inv$ on
$P_{g,n}^D$} \label{cut3}
\end{figure}
After the surface $\surf\mindee$ is cut along the representatives
of the generators $m_i,a_j,b_j$, the curve $c$ gives rise to a set
of oriented segments $l_i$, $i=1,\ldots,r+1$ on the polygon
$P_{g,n}^D$ as shown in Fig.~\ref{cut3}. We denote the starting
and endpoints of the segments $l_i$ by, respectively, $s_i$ and
$t_i$. With the exception of the starting point of the first and
the endpoint of the last segment $s_1=t_r=q$, all other starting
and endpoints lie on the sides $m_i$, $a_j,a'_j,b_j,b'_j$ of
$P_{g,n}^D$. Without changing the homotopy equivalence class of
$c$, we can ensure that all intersection points with the sides of
$P^D_{g,n}$ representing the generators $m_i$ occur on $v_i$ or
$v'_i$. For segments $l_k$ which end on $v_i$ or $v'_i$,  the next
segment $l_{k+1}$ then starts at the corresponding point on $v'_i$
and $v_i$, respectively. Similarly, for segments  that end in a
point on a side $a_j$ or $b_j$, the next segment then starts in
the corresponding point on $a'_j$ or $b'_j$ and vice versa.

We can now move the starting and endpoints of the segments $l_k$
towards the corners of the polygon $P_{g,n}^D$. For each segment
ending on a side $m_i$, we have $t_k=x_{i-1}$, $s_k=x_i$ if
$t_k\in v_i$ and $t_k=x_{i}$, $ s_{k+1}=x_{i-1}$ if $t_k\in v_i'$.
For a segment $l_k$ ending in a side $a_j$, we can either move its
endpoint $t_k$ to the starting point of $a_j$ and hence the
starting point $s_{k+1}$ to the starting point of $a'_j$ or move
$t_k$ and $s_{k+1}$ to the endpoints of respectively, $a_j$ and
$a_j'$. The first possibility yields $t_k=x_{n+4j-4}$,
$s_{k+1}=x_{n+4j-1}$, the second $t_k=x_{n+4j-3}$,
$s_{k+1}=x_{n+4j-2}$, and the corresponding expressions for a
segment ending on $a'_j$ are given by exchanging $t_k$ and
$s_{k+1}$. Similarly, for a segment $l_k$ ending on a side $b_j$,
we can move its endpoint $t_k$ and the starting point $s_{k+1}$ to
either the starting points of $b_j$ and $b'_j$, which implies
$t_k=x_{n+4j-2}$, $s_{k+1}=x_{n+4j-1}$ or to their endpoints which
yields $t_k=x_{n+4j-1}$, $s_{k+1}=x_{n+4j}$. Using the
decomposition \eqref{mseg}, \eqref{aseg}, \eqref{bseg} of the dual
generators in terms of the segments connecting the basepoint with
the corners of the polygon $P^D_{g,n}$, we then obtain
\begin{align}
\label{segid} (x_0s_{k+1})^\inv(x_0t_{k})=\begin{cases} \dcm_i
&\;\text{if}\quad t_k\in v_i, i=1,\ldots,n\\
\dcm_i^\inv
&\;\text{if}\quad t_k\in v_i', i=1,\ldots,n\\
\dca_j
&\;\text{if}\quad t_k\in a_j, j=1,\ldots,g\\
\dca_j^\inv &
\;\text{if}\quad t_k\in a'_j, j=1,\ldots,g\\
\dcb_j &
\;\text{if}\quad t_k\in b_j,j=1,\ldots,g\\
\dcb_j^\inv & \;\text{if}\quad t_k\in
b'_j,j=1,\ldots,g.\end{cases}
\end{align}
Note that the two possibilities of moving points on sides
$a_j,a'_j,b_j,b'_j$ to either the starting point or to the
endpoint of the sides $a_j,a'_j,b_j,b'_j$ correspond to the two
different expressions for $\dca_j,\dcb_j$ in \eqref{aseg},
\eqref{bseg}.

By expressing the segments $l_k$ on the polygon $P^D_{g,n}$ in
terms of the segments $x_0 s_k$, $x_0 t_k$, we find that the
homotopy equivalence class of $c$ is given by
\begin{align}
\label{segdecomp} c=l_{r+1}\circ l_{r-1}\circ\ldots\circ l_1=
(x_0q)\circ (x_0s_{r+1})^\inv(x_0
t_{r})\circ(x_0s_{r})^\inv(x_0t_{r-1})\circ\ldots\circ
(x_0s_2)^\inv(x_0t_1)\circ (qx_0),
\end{align}
where $x_0q$ stands for a segment connecting the point $q$ with
the basepoint $x_0$. By inserting identity \eqref{segid} into this
expression  we then obtain the unique expression of the homotopy
equivalence class of $c$ as a reduced word in the dual generators
$\dcm_i,\dca_j,\dcb_j$. As the starting and endpoints of segments
$l_k$ correspond to intersection points of $c$ with the generators
$m_i,a_j,b_j$, this implies that the intersection points of $c$
with the representatives of $a_j,b_j$ are in one-to-one
correspondence with factors $\dca_j,\dcb_j$ in the expression of
the homotopy equivalence class of $c$ as a product in the dual
generators $\dca_j,\dcb_j$. Similarly, each factor $\dcm_i$
corresponds to a pair of intersection points of $c$ with a
representative of $m_i$. Furthermore, if we define the oriented
intersection number $\epsilon(\lambda,\eta)$ of two curves
$\lambda,\eta\in\pi_1(\surf\mindee)$  to be positive if $\eta$
crosses $\lambda$ from the left to the right in the direction of
$\lambda$, we find that the exponents of $\dcm_i,\dca_j,\dcb_j$ in
\eqref{segid} determine the oriented intersection numbers
associated to these intersection points. We obtain the following
theorem.
\begin{theorem}
\label{intsecttheorem}

Consider an element $\lambda\in\pi_1(q,\surf\mindee)\cong\pi_1(p,
\surf\mindee)$ with an embedded representative  which intersects
the generators $m_i,a_j,b_j$ in a minimum number of points.
Express $\lambda$ as a reduced word in the generators
$\dcm_i,\dca_j,\dcb_j\in\pi_1(\surf\mindee)$
\begin{align}
\label{lambdadecomp0} \lambda=\dcx_r^{\alpha_r}\cdots
\dcx_1^{\alpha_1}\qquad
\dcx_k\in\{\dcm_1,\ldots,\dcm_n,\dca_1,\dcb_1,\ldots,\dca_g,\dcb_g\},\,\alpha_k\in\{\pm1\}.
\end{align} Then, the
intersection points of the curve representing $\lambda$ with the
generators $a_j,b_j$ are in one-to-one correspondence with factors
$\dcx_k=a_j$, $\dcx_k=b_j$ in the expression
\eqref{lambdadecomp0}. The associated exponents $\alpha_k$
determines the oriented intersection number
$\alpha_k=\epsilon(a_j, \lambda)$ for $\dcx_k=\dca_j$,
$\alpha_k=-\epsilon(b_j,\lambda)$ for $\dcx_k=\dcb_j$. Similarly,
each factor $\dcx_k=\dcm_i$ in \eqref{lambdadecomp0} corresponds
to a pair of intersection points of $\lambda$ with the generator
$m_i$ with opposite intersection numbers, and the exponent
$\alpha_k$ gives the oriented intersection number of the
intersection point which occurs first on $m_i$.
\end{theorem}

The expression of an  element $\lambda\in\pi_1(\surf\mindee)$ as a
reduced word in the
 dual generators $\dcm_i,\dca_j,\dcb_j$ therefore determines the intersection points
 of its representatives with the representatives of the generators $m_i,a_j,b_j$ and the
associated oriented intersection numbers. Note that since we
require that homotopy does not move the basepoint, each curve on
$\surf\mindee$ corresponds to an element
$\lambda\in\pi_1(\surf\mindee)$ and not a conjugacy class
$[\lambda]=\{\tau\lambda\tau^\inv\;|\;\tau\in\pi_1(\surf\mindee)\}$.
Different elements of a conjugacy class $[\lambda]$ therefore have
different numbers of intersection points with the generators
$m_i,a_j,b_j$. In particular, we find that this number is minimal
for those elements of $[\lambda]$ which are represented by a
cyclically reduced word in the generators $\dcm_i,\dca_j,\dcb_j$,
i.~e.~expressions of the form \eqref{lambdadecomp0} with
$\dcx_r^{\alpha_r}\neq \dcx_1^{-\alpha_1}$.

\section{The algebraic properties of the dual generators}
\label{combsect}

\subsection{The polygon picture: ordering the intersection points}
\label{intordersect} After showing how the expression of a general
element $\lambda\in\pi_1(\surf\mindee)$ as a reduced word in the
dual generators $\dcm_i,\dca_j,\dcb_j$ determines the number of
its intersection points  with the generators $m_i,a_j,b_j$ and the
oriented intersection numbers, we will now demonstrate that for
elements with embedded representatives it also determines the
order in which these intersection points occur on each generator
$m_i,a_j,b_j$. Moreover, we show that the dual generators allow
one to assign  these intersection points (almost) uniquely between
the different factors in the expression of $\lambda$ as a reduced
word in the generators $m_i,a_j,b_j$ and that this assignment of
the intersection points can be implemented via a graphical
procedure.

In the following we assume that $\lambda\in\pi_1(\surf\mindee)$
has an embedded representative and that its expression
\eqref{lambdadecomp0} as a reduced word in $\dcm_i,\dca_j,\dcb_j$
is also cyclically reduced $\dcx_r^{\alpha_r}\neq
\dcx_1^{-\alpha_1}$. To determine the order in which the
intersection points of $\lambda$ with the generators $m_i,a_j,b_j$
occur on each generator, we determine the order of the associated
points on the sides of the polygon $P^D_{g,n}$. For this, we
 recall the discussion  from the last section, where it was shown
 that after cutting the surface $\surf\mindee$ along the
 generators $m_i,a_j,b_j$, a curve representing $\lambda$ gives
 rise to a set of segments on the polygon $P^D_{g,n}$.
Each factor $\dcx_k^{\alpha_k}$ in the expression
\eqref{lambdadecomp0} of $\lambda$ as a cyclically reduced word in
the dual generators corresponds to two intersection points of
$\lambda$ with the sides of polygon $P^D_{g,n}$. One of these
intersection points is realised as the endpoint $t_k$ of a segment
$l_k$ and the other one as the starting point $s_{k+1}$ of the
next segment $l_{k+1}$, and the factor $x_k^{\alpha_k}$ identifies
these points. This implies that intersection points of a
representative of $\lambda$ with the sides of the polygon
$P^D_{g,n}$ are in one-to-one correspondence with cyclic
permutations of $\lambda$ and its inverse. This allows us to
identify the intersection points $\{t_1,s_2,t_2,\ldots, s_r, t_r,
s_{r+1}\}$ on polygon $P_{g,n}^D$ with elements of the set of
cyclic permutations
\begin{align}
&\text{CPerm}(\lambda)=\{\lambda_k\;|\; k=1,\ldots,r\}\cup
\{(\lambda_k)^\inv\;|\;
k=1,\ldots,r\}\label{cperm}\\
 \label{lambdak}
&\text{where}\quad\lambda_k=\dcx_k^{\alpha_k}\cdots\dcx_1^{\alpha_1}\dcx_r^{\alpha_r}\cdots\dcx_{k+1}^{\alpha_{k+1}}\,,\,
\qquad k=1,\ldots,r,
\end{align}
by setting
\begin{align}
\label{permident} &P: \{t_1,s_2,t_2,\ldots, s_r, t_r,
s_{r+1}\}\rightarrow
\text{CPerm}(\lambda)\\
 &P(t_k)=\lambda_{k-1}\quad P(s_{k+1})=\lambda_k^\inv\qquad
k=1,\ldots, r.\nonumber
\end{align}
We find that an intersection point $p \in \{t_1,s_2,t_2,\ldots,
s_r, t_r, s_{r+1}\}$ of a representative of $\lambda$ with the
boundary of the polygon $P^D_{g,n}$ lies on a side $a_j,b_j$,
respectively, if the last factor in $P(p)$ is $\dca_j$, $\dcb_j$
and on $a'_j,b'_j$ if it is $\dca_j^\inv$, $\dcb_j^\inv$.
Similarly, an intersection point with a side $m_i$ lies on $v_i$
and $v_i'$, respectively,  if it is $\dcm_i$ and $\dcm_i^\inv$
\begin{align}
&p\in a_j\Leftrightarrow LF(P(p))=\dca_j & &p\in
b_j\Leftrightarrow LF(P(p))=\dcb_j & &p\in v_i\Leftrightarrow
LF(P(p))=\dcm_i\\
&p\in a'_j\Leftrightarrow LF(P(p))=\dca_j^\inv  & &p\in
b'_j\Leftrightarrow LF(P(p))=\dcb_j^\inv  & &p\in
v'_i\Leftrightarrow LF(P(p))=\dcm_i^\inv\nonumber.
\end{align}
Furthermore, we note that if $p=t_k$ is the endpoint of the
segment $l_k$, the starting point of the segment $l_k$ lies on
$v_i,a_j,b_j$, respectively, if
$\dcx_{k-1}^{-\alpha_{k-1}}=\text{LF}(P(p)^\inv)$ is
$\dcm_i,\dca_j,\dcb_j$ and on $v_i',a_j',b_j'$ if  it is
$\dcm_i^\inv,\dca_j^\inv,\dcb_j^\inv$. Similarly, if $p=s_{k+1}$
is the starting point of the segment $l_{k+1}$, its endpoint
$t_{k+1}$ lies on $v_i,a_j,b_j$ if
$\dcx_{k+1}^{\alpha_{k+1}}=\text{LF}(P(p)^\inv)=\dcm_i,\dca_j,\dcb_j$
and on $v_i',a_j',b_j'$ for
$\dcx_{k+1}^{\alpha_{k+1}}=\text{LF}(P(p)^\inv)=\dcm_i^\inv,\dca_j^\inv,\dcb_j^\inv$.
Hence, the last factor $\text{LF}(P(p)^\inv)$ of $P(p)^\inv$
determines the location of the other end of the segment which
contains $p$.

To pursue this reasoning further, we  consider the elements
$\lambda_p^{(s)}\in\text{CPerm}(\lambda)$ obtained as cyclical
permutations of  $\lambda_p^{(1)}=P(p)$
\begin{align}
\label{sper} \lambda_p^{(s)}=\begin{cases} \lambda_{k-s} & p=t_k\\
\lambda_{k+s}^\inv & p=s_{k+1}
\end{cases}=\begin{cases}
\dcx_{k-s}^{\alpha_{k-s}}\cdots\dcx_1^{\alpha_1}
\dcx_r^{\alpha_r}\cdots\dcx_{k-s+1}^{\alpha_{k-s+1}} & p=t_k\\
\dcx_{k+s}^{-\alpha_{k+s}}\cdots\dcx_r^{-\alpha_r}\dcx_1^{-\alpha_1}\cdots\dcx_{k+s-1}^{-\alpha_{k+s-1}}
& p=s_{k+1}\end{cases} \qquad k,s=1,\ldots,r
\end{align}
where we identify $r+s=s$, $-s=r-s$. By inspecting the associated
segments on the polygon $P_{g,n}^D$, we find that for if $p=t_k$
is the endpoint of the segment $l_k$,
 the last factors in $\lambda_p^{(s)}$ and  $(\lambda_p^{(s)})^\inv$, respectively,
 determine on which side of the polygon $P^D_{g,n}$ the endpoint
$t_{k-s+1}$ of the segment $l_{k-s+1}$ and the starting point
$s_{k+s-1}$ of the segment $l_{k+s-1}$ are located. Similarly, for
starting points $p=s_{k+1}$, the last factors in $\lambda_p^{(s)}$
and $(\lambda_p^{(s)})^\inv$ give the location of, respectively,
the starting point $s_{k+s}$ and the endpoint $t_{k+s}$.

We now consider the intersection points of a representative of
$\lambda$ with the sides of the polygon $P^D_{g,n}$ with the
ordering obtained by traversing the boundary $\partial P^D_{g,n}$
counterclockwise starting at $x_0$. Note that this ordering is
unique, since embedded curves are represented by non-intersecting
segments on $P^D_{g,n}$ and exchanging two intersection points
would give rise to an intersection of the associated segments. For
two intersection points $p,q \in\{t_1,s_2,t_2,\ldots, s_r,t_r,
s_{r+1}\}$ which are located on different sides of polygon or on
different parts $v_i,v'_i$ of a side $m_i$, it follows from
Fig.~\ref{cut2} that $p$ occurs before $q$ if and only if the side
containing $p$ occurs before side containing $q$. Since these
sides are given by last factors in the associated cyclic
permutations $P(p)=\lambda^{(1)}_p$, $P(q)=\lambda^{(1)}_q$, this
is the case if and only if
\begin{align}
 \text{LF}(\lambda_p^{(1)})<_D \text{LF}(\lambda_q^{(1)})
\end{align}
with respect to the ordering
\begin{align}
\label{discorder} \dcb_g^\inv\!\!\grx {D}\!\!\dca_g^\inv \!\!\grx
{D}\!\! \dcb_g \!\!\grx {D}\!\!\dca_g \!\!\grx {D}\!\!
\dcb_{g-1}^\inv \!\!\grx {D}\!\!\ldots\!\!\grx {D}\!\!\dcb_1^\inv
\!\!\grx {D}\!\!\dca_1^\inv\!\!\grx {D}\!\!\dcb_1\!\!\grx
{D}\!\!\dca_1\!\!\grx {D}\!\!\dcm_n^\inv \!\!\grx
{D}\!\!\dcm_n\!\!\grx {D}\!\!\ldots \!\!\grx
{D}\!\!\dcm_1^\inv\!\!\grx {D}\!\!\dcm_1.
\end{align}

We now consider the case where both $p$ and $q$ are either located
on the same side $x\in\{a_1,a_1',\ldots,b_g,b_g'\}$ of the polygon
$P^D_{g,n}$ or, in the case of the punctures, on a single segment
$x\in\{v_1,v_1',\ldots, v_n,v_n'\}$. This is the case if and only
if the last factor in the associated permutations agree
$\text{LF}(\lambda_p^{(1)})=LF(\lambda_q^{(1)})$. The fact that
the associated segments cannot intersect then implies that the
order of $p$ and $q$ depends on the location of the other end of
these segments, which we will denote by $p',q'$ and which are
given by $\text{LF}((\lambda_p^{(1)})^\inv)$,
$\text{LF}((\lambda_q^{(1)})^\inv)$. Note that the fact that
expression \eqref{lambdadecomp0} for $\lambda$ is cyclically
reduced implies that $p'$ and $q'$  cannot lie on the side $x$
\begin{align}
\text{LF}(\lambda_p^{(s)})\neq
\text{LF}((\lambda_p^{(s)})^\inv)\qquad\forall
p\in\{t_1,s_2,t_2,\ldots, s_{r}, t_r,s_{r+1}\}, \;s=1,\ldots, r.
\end{align}
If we have
$\text{LF}((\lambda_p^{(1)})^\inv)\neq\text{LF}((\lambda_q^{(1)})^\inv)$,
then $p'$ and $q'$ are located either on different sides of
$P^D_{g,n}$ or on different parts $v_i,v_i'$ associated to a side
$m_i$. By drawing the associated segments on the polygon
$P^D_{g,n}$, we then find that $p$ occurs before $q$ if  either
$p'$ is located on a side before side $x$ and $q'$ on side after
$x$ or if both the sides containing $p'$ and $q'$ occur before or
after $x$ with the one containing $q'$ before the one for $p'$.
This can be implemented by defining an ordering $\grx x$ of the
set of sides $\{v_1,v_1',\ldots,b_g,b_g'\}-\{x\}$ obtained by
removing the factor associated to $x$ from the ordering
\eqref{discorder} and performing a cyclic permutation which moves
the factors to the right of the factor associated to $x$ in
\eqref{discorder} to the left. Explicitly, we have for the factors
$\dcm_i^{\pm 1}, \dca_j^{\pm 1}, \dcb_j^{\pm 1}$
\begin{align}
\label{miorder} &\dcm_{i-1}^\inv \!\!\grx
{m_i}\!\!\dcm_{i-1}\!\!\grx {m_i}\!\!\ldots\!\!\grx
{m_i}\!\!\dcm_1\!\!\grx {m_i}\!\!\dcb_g^\inv\!\!\grx
{m_i}\!\!\dca_g^\inv\!\!\grx {m_i}\!\!\dcb_g\!\!\grx
{m_i}\!\!\dca_g\!\!\grx {m_i}\!\!\ldots\!\!\grx {m_i}\!\!
\dcm_{i+1}^\inv\!\!\grx
{m_i}\!\!\dcm_{i+1}\!\!\grx {m_i}\!\!\dcm_i^\inv\\
&\dcm_i \!\!\grx {m_i^\inv}\!\!\dcm_{i-1}^\inv \!\!\grx
{m_i^\inv}\!\!\dcm_{i-1}\!\!\grx {m_i^\inv}\!\!\ldots\!\!\grx
{m_i^\inv}\!\!\dcm_1\!\!\grx {m_i^\inv}\!\!\dcb_g^\inv\!\!\grx
{m_i^\inv}\!\!\dca_g^\inv\!\!\grx {m_i^\inv}\!\!\dcb_g\!\!\grx
{m_i^\inv}\!\!\dca_g\!\!\grx {m_i^\inv}\!\!\ldots\!\!\grx
{m_i^\inv}\!\! \dcm_{i+1}\nonumber\\
 &\dcb_{j-1}^\inv\!\!\grx{a_j}\!\!
\dca_{j-1}^\inv\!\!\grx{a_j}\!\!\dcb_{j-1}\!\!\grx{a_j}\!\!\dca_{j-1}\!\!\grx{a_j}\!\!\ldots\!\!\grx{a_j}\!\!\dcm_1\!\!\grx{a_j}\!\!\dcb_g^\inv\!\!\grx{a_j}\!\!\ldots\!\!\grx{a_j}\!\!\dca_{j+1}\!\!\grx{a_j}\!\!\dcb_j^\inv\!\!\grx{a_j}\!\!\dca_j^\inv\!\!\grx{a_j}\!\!\dcb_j\nonumber\\
&\dcb_j\!\!\grx{a_j^\inv}\!\!\dca_j\!\!\grx{a_j^\inv}\!\!
\dcb_{j-1}^\inv\!\!\grx{a_j^\inv}\!\!
\dca_{j-1}^\inv\!\!\grx{a_j^\inv}\!\!\dcb_{j-1}\!\!\grx{a_j^\inv}\!\!\dca_{j-1}\!\!\grx{a_j^\inv}\!\!\ldots\!\!\grx{a_j^\inv}\!\!\dcm_1\!\!\grx{a_j^\inv}\!\!\dcb_g^\inv\!\!\grx{a_j^\inv}\!\!\ldots\!\!\grx{a_j^\inv}\!\!\dca_{j+1}\!\!\grx{a_j^\inv}\!\!\dcb_j^\inv\nonumber\\
 &\dca_j
\!\!\grx{b_j}\!\!\dcb_{j-1}^\inv\!\!\grx{b_j}\!\!
\dca_{j-1}^\inv\!\!\grx{b_j}\!\!\dcb_{j-1}\!\!\grx{b_j}\!\!\dca_{j-1}\!\!\grx{b_j}\!\!\ldots\!\!\grx{b_j}\!\!\dcm_1\!\!\grx{b_j}\!\!\dcb_g^\inv\!\!\grx{b_j}\!\!\ldots\!\!\grx{b_j}\!\!\dca_{j+1}\!\!\grx{b_j}\!\!\dcb_j^\inv\!\!\grx{b_j}\!\!\dca_j^\inv\nonumber\\
&\dca_j^\inv \!\!\grx{b_j^\inv}\!\!\dcb_j
\!\!\grx{b_j^\inv}\!\!\dca_j
\!\!\grx{b_j^\inv}\!\!\dcb_{j-1}^\inv\!\!\grx{b_j^\inv}\!\!
\dca_{j-1}^\inv\!\!\grx{b_j^\inv}\!\!\dcb_{j-1}\!\!\grx{b_j^\inv}\!\!\dca_{j-1}\!\!\grx{b_j^\inv}\!\!\ldots\!\!\grx{b_j^\inv}\!\!\dcm_1\!\!\grx{b_j^\inv}\!\!\dcb_g^\inv\!\!\grx{b_j^\inv}\!\!\ldots\!\!\grx{b_j^\inv}\!\!\dca_{j+1}\nonumber.
\end{align}
The intersection point $p$ then occurs before $q$ on $x$ if and
only if and only if the last factor
$\text{LF}((\lambda_p^{(1)})^\inv) $ is greater than the last
factor $\text{LF}((\lambda_q^{(1)})^\inv)$ with respect to the
ordering associated to the side $x$
\begin{align}
\text{LF}((\lambda_p^{(1)})^\inv) \grx x
\text{LF}((\lambda_q^{(1)})^\inv).
\end{align}
Finally, we consider the case where both $p,q$ are located on a
side $x$ of $P^D_{g,n}$ and the other ends of the associated
segments both lie on another side
$y\in\{v_1,v_1',\ldots,b_g,b_g'\}-\{x\}$.  This is the case if and
only if
$\text{LF}((\lambda_p^{(1)})^\inv)=\text{LF}((\lambda_q^{(1)})^\inv)$
or, equivalently,
$\text{LF}(\lambda_p^{(2)})=\text{LF}(\lambda_q^{(2)})$. Then fact
that segments cannot intersect implies that $p$ occurs before $q$
on $x$ if and only if $q'$ occurs before $p'$ on $y$. We then
consider the corresponding points $p''$, $q''$ on the side $y'$ of
the polygon $P^D_{g,n}$
 identified with $y$. As  the orientation of
two sides $y,y'$ in Fig.~\ref{cut2} are the opposite, we find that
$q'$ occurs before $p'$ on $y$ if and only if $p''$ occurs before
$q''$ on $y'$. Hence, we can repeat the reasoning of the last
paragraphs for the points $p''$ and $q''$. In the case where the
original intersection point $p$ is an endpoint $p=t_k$, the
corresponding point on $y'$ is the endpoint $p''=t_{k-1}$ of the
previous segment, and for a starting point $p=s_{k+1}$ it is the
starting point of the next segment $p''=s_{k+2}$. Hence, the side
containing $p'',q''$ is given by the last factors in
$\lambda_p^{(2)}$ ,$\lambda_q^{(2)}$ and other ends  of the
associated segments  by last factors in $(\lambda_p^{(2)})^\inv$,
$(\lambda_q^{(2)})^\inv$. If these ends are located on different
sides, we have
$\text{LF}((\lambda_p^{(2)})^\inv)\neq\text{LF}((\lambda_q^{(2)})^\inv)$
and $p$ occurs before $q$ if and only if
\begin{align}
\text{LF}((\lambda_p^{(2)})^\inv)\grx{\text{LF}(\lambda_p^{(2)})}\text{LF}((\lambda_q^{(2)})^\inv).
\end{align}
Otherwise we apply the same reasoning to the intersection points
associated with $\lambda_p^{(3)}$, $\lambda_q^{(3)}$. By
repeatedly applying this argument to the permutations
$\lambda^{(s)}_p$, $\lambda^{(s)}_q$ with increasing $s$, we
either obtain permutations $\lambda^{(k)}_p$, $\lambda^{(k)}_q$
whose last factors differ or the expression \eqref{lambdadecomp0}
of $\lambda$ as a cyclically reduced word in the dual generators
$\dcm_i,\dca_j,\dcb_j$ is periodic $\lambda=q^k$, $1<k<r$, with a
cyclically reduced word $q=\dcy_s^{\beta_s}\cdots\dcy_1^{\beta_1}$
in $\dcm_i,\dca_j,\dcb_j$. However, it is easy to see that
periodic words cannot have embedded representatives. The
representation of $\lambda=q^k$ by segments on the polygon
$P^D_{g,n}$ is obtained from the representation of $q$ by drawing
each of the segments associated to pairs of factors in $q$ $k$
times and omitting the segment associated to $\dcy_1^{\beta_1}$
and $\dcy_s^{\beta_s}$. One then has to add $k$ segments each
connecting one of the $k$ points associated to factor
$\dcy_1^{\beta_1}$ with one of the points associated to
$\dcy_s^{\beta_s}$. However, it is clear  that it is impossible to
do this without creating intersections of the segments.

Hence, the algorithm described above terminates and there exists a
$k\in\{1,\ldots,r-1\}$ such that
$\text{LF}(\lambda_p^{(s)})=\text{LF}(\lambda_q^{(s)})$ for $s\leq
k$ and
$\text{LF}((\lambda_p^{(k)})^\inv)\neq\text{LF}((\lambda_q^{(k)})^\inv)$.
The intersection point  $p$ then occurs before $q$ if and only if
the last factor $\text{LF}((\lambda_p^{(k)})^\inv)$ is greater
than the last factor $\text{LF}((\lambda_q^{(k)})^\inv)$ with
respect to the ordering $\grx {\text{LF}(\lambda_p^{(k)})}$, and
we obtain the following theorem.
\begin{theorem}
\label{intorderlem} Consider an element
$\lambda\in\pi_1(\surf\mindee)$ with an embedded representative
and two intersection points $p,q\in\{t_1,s_2,t_2,\ldots, s_r,
t_r,s_{r+1}\}$ of this representative with the polygon
$P^D_{g,n}$. Denote by $\lambda_p^{(s)}$, $\lambda_q^{(s)}$ the
 cyclic permutations of $\lambda$ and its inverse assigned to the
 points $p$ and $q$
  as defined in \eqref{sper}.
Then, the intersection point $p$ occurs before $q$ with respect to
the ordering obtained by traversing the polygon $P^D_{g,n}$
counterclockwise starting at $x_0$ if and only if either
\begin{align}
\label{cond1} \text{LF}(\lambda_p^{(1)})\smx D
\text{LF}(\lambda_q^{(1)}),\end{align} in which case the points
$p$, $q$ are located on different sides of the polygon $P^D_{g,n}$
or on different parts $v_i,v'_i$ of a side $m_i$,
 or
 \begin{align}
\label{cond2} \exists k\in\{1,\ldots,r-1\}:\quad
\text{LF}(\lambda_p^{(s)})=\text{LF}(\lambda_q^{(s)})\;\forall
s\leq k,\quad \text{LF}((\lambda_p^{(k)})^\inv)\grx
{\text{LF}(\lambda_p^{(k)})} \text{LF}((\lambda_q^{(k)})^\inv).
\end{align}
\end{theorem}
Hence, for any element $\lambda\in\pi_1(\surf\mindee)$ with an
embedded representative, the unique order in which the starting
and endpoints of the associated segments occur on the polygon
$P^D_{g,n}$ gives rise to an ordering  of the set of cyclic
permutations $\text{CPerm}(\lambda)$ defined by the conditions
\eqref{cond1}, \eqref{cond2}. In particular, this induces an
ordering of the intersection points of $\lambda$ with each of the
generators $m_i,a_j,b_j\in\pi_1(\surf\mindee)$. Intersection
points of $\lambda$ with the generators $a_j, b_j$ are in
one-to-one correspondence with intersection points of $\lambda$
with the sides $a_j,b_j$ of $P^D_{g,n}$ and therefore with
elements $\tau\in\text{CPerm}(\lambda)$ satisfying, respectively,
$\text{LF}(\tau)=\dca_j$ and $\text{LF}(\tau)=\dcb_j$. Taking into
account the orientation of the sides $a_j,b_j$ on $P^D_{g,n}$ we
find that the order in which these intersection points occur on
$a_j$ agrees with the one on the polygon, while for $b_j$ it is
the opposite. Similarly, intersection points of $\lambda$ with the
generator $m_i$ correspond one-to-one to intersection points of
$\lambda$ with the segments $v_i$ and $v_i'$ and hence to elements
$\tau\in\text{CPerm}(\lambda)$, $\text{LF}(\tau)=\dcm_i^{\pm 1}$,
and the order in which these intersection points occur on the
generator $m_i$ is the order of the corresponding points on
$P^D_{g,n}$. We therefore obtain a purely algebraic procedure,
which allows one to determine the number and order of intersection
points of general embedded curves on $\surf\mindee$ with the
representatives of the generators $m_i,a_j,b_j$ and the associated
oriented intersection numbers from the expression of its homotopy
equivalence class as a reduced word in the dual generators
$\dcm_i,\dca_j,\dcb_j$.


\subsection{The surface picture: assigning the intersection points between different factors}
\label{factassign}

  The involution $I\in\text{Aut}(\pi_1(\surf\mindee))$
  not only allows us to determine the number and order of intersection points of
  elements $\lambda\in\pi_1(\surf\mindee)$ with embedded
  representatives with the generators $m_i,a_j,b_j$ but
also assigns this intersection points (almost) uniquely between
the different factors in the expression of $\lambda$ as a reduced
word in the original $m_i,a_j,b_j$. Furthermore, we will show that
this assignment of intersection points can be implemented by
graphically decomposing a representative of $\lambda$ into curves
representing $m_i,a_j,b_j$.

The idea is the following. We consider the expression
\eqref{lambdadecomp0} of $\lambda$ as a reduced word in the dual
generators $\dcm_i,\dca_j,\dcb_j$ and the intersection point or
pair of intersection points associated to a factor
$\dcx_k^{\alpha_k}$ in \eqref{lambdadecomp0}. We then split the
expression of this factor as a reduced word in $m_i,a_j,b_j$ into
two reduced words  $\dcx_k^{\alpha_k}=w_2 w_1$, which correspond
to the two segments on the polygon $P^D_{g,n}$ in the
decompositions \eqref{mseg}, \eqref{aseg}, \eqref{bseg}. If $y_2$
and $y_1$, respectively, denote the expressions of the elements
$w_1\dcx_{k-1}^{\alpha_{k-1}}\cdots\dcx_1^{\alpha_1}\in\pi_1(\surf\mindee)$
and
$\dcx_r^{\alpha_r}\cdots\dcx_{k+1}^{\alpha_{k+1}}w_2\in\pi_1(\surf\mindee)$
in \eqref{lambdadecomp0} as reduced words in $m_i,a_j,b_j$, the
product $\lambda=y_2y_1$ then gives an expression of $\lambda$ as
a reduced word in the generators $m_i,a_j,b_j$ and we assign the
intersection point or pair of intersection points corresponding to
$\dcx_k^{\alpha_k}$ between the reduced words $y_2$ and $y_1$ in
this expression.

However, it is a priori not guaranteed that the product of the
reduced words $y_2,y_1$ in $m_i,a_j,b_j$ agrees with the
expression of $\lambda$ as a reduced word in $m_i,a_j,b_j$, since
it is possible that this product is not reduced. This is the case
if and only if the reduced words $y_2,y_1$ are of the form
$y_2=z_2x^\epsilon$, $y_1=x^{-\epsilon} z_1$, where
$x\in\{m_1,\ldots,b_g\}$, $\epsilon\in\{\pm 1\}$ and $z_1,z_2$ are
reduced words in $m_i,a_j,b_j$. In order to obtain a well-defined
assignment of the intersection points between the different
factors in the expression of $\lambda$ as a reduced word in
$m_i,a_j,b_j$, we therefore have to show that this situation can
be avoided. Furthermore, since there are two ways of splitting the
factors $\dca_j,\dcb_j$ in \eqref{aseg}, \eqref{bseg}, the
question arises if the requirement that the resulting expression
for $\lambda$ is a reduced word in $m_i,a_j,b_j$ removes this
ambiguity in the splitting. It turns out that up to a small
residual ambiguity, this is the case. We obtain an almost unique
assignment of intersection points between the different factors
$m_i,_j,b_j$ in $\lambda$ which is summarised in the following
theorem.

\begin{theorem}
\label{intasth}

Consider an embedded element $\lambda\in\pi_1(\surf\mindee)$ and a
factor $\dcx_k^{\alpha_k}$ in its expression \eqref{lambdadecomp0}
as a reduced word in the dual generators $\dcm_i,\dca_j,\dcb_j$.
\begin{enumerate}
\item If $\dcx_k=\dcm_i$, split the factor according to
\begin{align}
\dcm_i=(m_1^\inv\cdots m_{i}^\inv)(m_{i-1}\cdots m_1)
\end{align}
and let $y_s^{\beta_s}\cdots y_{l+1}^{\beta_{l+1}}$ and
$y_l^{\beta_l}\cdots y_1^{\beta-1}$, $y_k\in\{m_1,\ldots, b_g\}$,
$\beta_k\in\{\pm 1\}$ be the reduced words in $m_i,a_j,b_j$
obtained by setting
\begin{align}
\label{msplit1} &y_l^{\beta_l}\cdots y_1^{\beta_1}=\begin{cases}
m_{i-1}\cdots m_1 \dcx_{k-1}^{\alpha_{k-1}}\cdots
\dcx_1^{\alpha_1} &
\text{if}\quad\alpha_k=1\\
m_{i}\cdots m_1 \dcx_{k-1}^{\alpha_{k-1}}\cdots \dcx_1^{\alpha_1}&
\text{if}\quad\alpha_k=-1
\end{cases}\\
\label{msplit2} &y_s^{\beta_s}\cdots
y_{l+1}^{\beta_{l+1}}=\begin{cases}
\dcx_r^{\alpha_r}\cdots\dcx_{k+1}^{\alpha_{k+1}}m_1^\inv\cdots
m_{i}^\inv & \text{if}\quad\alpha_k=1\\
\dcx_r^{\alpha_r}\cdots\dcx_{k+1}^{\alpha_{k+1}}m_1^\inv\cdots
m_{i-1}^\inv & \text{if}\quad\alpha_k=-1.\end{cases}
\end{align}
Then, the expression for $\lambda$ as a reduced word in the
generators $m_i,a_j,b_j$ is given by the product
$\lambda=y_s^{\beta_s}\cdots y_{l+1}^{\beta_{l+1}}
y_l^{\beta_l}\cdots y_1^{\beta_1}$ and we assign the two
corresponding intersection points of $\lambda$ with $m_i$
 between the factors $y^{\beta_{l+1}}_{l+1}$ and $y_l^{\beta_l}$
and to the starting and endpoint of $m_i$.

\item If $\dcx_k\in\{\dca_j,\dcb_j\}$, let $y_l^{\beta_l}\cdots
y_1^{\beta_1}$ and $y_s^{\beta_s}\cdots y_{l+1}^{\beta_{l+1}}$
denote the reduced words in $m_i,a_j,b_j$ obtained by splitting
$\dcx_k^{\alpha_k}$ according to
\begin{align}
\label{absplit1} &\dca_j=(m_1^\inv\!\!\!\cdots h_{j-1}^\inv
a_j^\inv b_j a_j)(h_{j-1}\cdots
m_1)\quad\dcb_j=(m_1^\inv\!\!\!\cdots h_{j-1}^\inv a_j^\inv b_j
a_j)(b_j^\inv a_j h_{j-1}\cdots m_1)
\end{align}
\begin{align}
\label{splitabword1} y_l^{\beta_l}\cdots
y_1^{\beta_1}=&\begin{cases} h_{j-1}\cdots m_1
\dcx_{k-1}^{\alpha_{k-1}}\cdots \dcx_1^{\alpha_1} &
\text{if}\quad \dcx_k^{\alpha_k}=\dca_j\\
a_j^\inv b_j^\inv a_j h_{j-1}\cdots m_1
\dcx_{k-1}^{\alpha_{k-1}}\cdots \dcx_1^{\alpha_1}&
\text{if}\quad\dcx_k^{\alpha_k}=\dca_j^\inv\\
b_j^\inv a_jh_{j-1}\cdots m_1 \dcx_{k-1}^{\alpha_{k-1}}\cdots
\dcx_1^{\alpha_1} &
\text{if}\quad \dcx_k^{\alpha_k}=\dcb_j\\
a_j^\inv b_j^\inv a_jh_{j-1}\cdots m_1
\dcx_{k-1}^{\alpha_{k-1}}\cdots \dcx_1^{\alpha_1} & \text{if}\quad
\dcx_k^{\alpha_k}=\dcb_j^\inv
\end{cases}\\
\label{splitabword2} y_s^{\beta_s}\cdots
y_{l+1}^{\beta_{l+1}}=&\begin{cases}
\dcx_r^{\alpha_r}\cdots\dcx_{k+1}^{\alpha_{k+1}}m_1^\inv\cdots
h_{j-1}^\inv a_j^\inv b_j a_j & \text{if}\quad\dcx_k^{\alpha_k}=\dca_j\\
\dcx_r^{\alpha_r}\cdots\dcx_{k+1}^{\alpha_{k+1}}m_1^\inv\cdots
h_{j-1}^\inv & \text{if}\quad\dcx_k^{\alpha_k}=\dca_j^\inv\\
\dcx_r^{\alpha_r}\cdots\dcx_{k+1}^{\alpha_{k+1}}m_1^\inv\cdots
h_{j-1}^\inv a_j^\inv b_j a_j & \text{if}\quad\dcx_k^{\alpha_k}=\dcb_j\\
\dcx_r^{\alpha_r}\cdots\dcx_{k+1}^{\alpha_{k+1}}m_1^\inv\cdots
h_{j-1}^\inv a_j^\inv b_j &
\text{if}\quad\dcx_k^{\alpha_k}=\dcb_j^\inv,
\end{cases}
\end{align}
and let $z_m^{\delta_m}\cdots z_1^{\delta_1}$ and
$z_t^{\delta_t}\cdots z_{m+1}^{\beta_{m+1}}$
 be the reduced words obtained by splitting $\dcx_k^{\alpha_k}$
as
\begin{align}
\label{absplit2} \dca_j=(m_1^\inv\!\!\!\cdots h_{j-1}^\inv
a_j^\inv b_j )&( a_jh_{j-1}\cdots
m_1)\quad\dcb_j=(m_1^\inv\!\!\!\cdots h_{j-1}^\inv a_j^\inv b_j
a_j b_j^\inv)(a_j h_{j-1}\cdots m_1)\\
z_m^{\delta_m}\cdots z_1^{\delta_1}=&\begin{cases}
a_jh_{j-1}\cdots m_1 \dcx_{k-1}^{\alpha_{k-1}}\cdots
\dcx_1^{\alpha_1} &
\text{if}\quad \dcx_k^{\alpha_k}=\dca_j\\
 b_j^\inv a_j h_{j-1}\cdots m_1 \dcx_{k-1}^{\alpha_{k-1}}\cdots
\dcx_1^{\alpha_1}&
\text{if}\quad\dcx_k^{\alpha_k}=\dca_j^\inv\\
 a_jh_{j-1}\cdots m_1 \dcx_{k-1}^{\alpha_{k-1}}\cdots
\dcx_1^{\alpha_1} &
\text{if}\quad \dcx_k^{\alpha_k}=\dcb_j\\
 h_jh_{j-1}\cdots m_1
\dcx_{k-1}^{\alpha_{k-1}}\cdots \dcx_1^{\alpha_1} & \text{if}\quad
\dcx_k^{\alpha_k}=\dcb_j^\inv
\end{cases}\\
z_t^{\delta_t}\cdots z_{m+1}^{\beta_{m+1}}=&\begin{cases}
\dcx_r^{\alpha_r}\cdots\dcx_{k+1}^{\alpha_{k+1}}m_1^\inv\cdots
h_{j-1}^\inv a_j^\inv b_j  & \text{if}\quad\dcx_k^{\alpha_k}=\dca_j\\
\dcx_r^{\alpha_r}\cdots\dcx_{k+1}^{\alpha_{k+1}}m_1^\inv\cdots
h_{j-1}^\inv a_j^\inv & \text{if}\quad\dcx_k^{\alpha_k}=\dca_j^\inv\\
\dcx_r^{\alpha_r}\cdots\dcx_{k+1}^{\alpha_{k+1}}m_1^\inv\cdots
h_{j-1}^\inv h_j^\inv & \text{if}\quad\dcx_k^{\alpha_k}=\dcb_j\\
\dcx_r^{\alpha_r}\cdots\dcx_{k+1}^{\alpha_{k+1}}m_1^\inv\cdots
h_{j-1}^\inv a_j^\inv &
\text{if}\quad\dcx_k^{\alpha_k}=\dcb_j^\inv.
\end{cases}
\end{align}
Then the expression of $\lambda$ as a reduced word in
$m_i,a_j,b_j$ is either given by $\lambda=y_s^{\beta_s}\cdots
y_{l+1}^{\beta_{l+1}}y_l^{\beta_l}\cdots y_1^{\beta_1}$ and we
assign the corresponding intersection point between the factors
$y^{\beta_{l+1}}_{l+1}$ and $y_l^{\beta_l}$ and to the starting
point of $a_j$ or $b_j$, or it is given by
$\lambda=z_t^{\delta_t}\cdots
z_{m+1}^{\beta_{m+1}}z_m^{\delta_m}\cdots z_1^{\delta_1}$ and we
assign the corresponding intersection point between the factors
$z_{m+1}^{\beta_{m+1}}$ and $z_m^{\delta_m}$ and to the endpoint
of $a_j$ or $b_j$. Ambiguity in the sense that both
$y_s^{\beta_s}\cdots y_1^{\beta_1}$ and $z_t^{\delta_t}\cdots
z_1^{\delta_1}$ are reduced words in $m_i,a_j,b_j$ arises if and
only if either $y_{l+1}^{\beta_{l+1}}=z_m^{\delta_m}=x_k$ or $
y_l^{\beta_l}=z_{m+1}^{\delta_{m+1}}=x_k^\inv$.
\end{enumerate}

\end{theorem}

 The proof of Theorem \ref{intasth} is rather lengthy and technical and
 makes use of the following lemma.
\begin{lemma}
\label{helplem}

Consider an element of $\pi_1(\surf\mindee)$ given as a reduced
word in the generators $m_i,a_j,b_j$ and their duals by
\begin{align}
\dcx_r^{\alpha_r}\cdots\dcx_1^{\alpha_1}=wzw'
\qquad\dcx_k\in\{\dcm_1,\ldots,\dcb_g\}, \alpha_k\in\{\pm 1\},
z\in\{m_1^{\pm 1},\ldots,b_g^{\pm 1}\},
\end{align}
where $w,w'$ are reduced words in $m_i,a_j,b_j$. Then, we have the
following implications
\begin{align}
\label{mid} &w'=m_i\cdots m_1,\, \dcx_1^{\alpha_1}\neq\dcm_i^\inv
&
&\Rightarrow & &z\in\{m_{i+1}^{\pm 1},\ldots,b_g^{\pm 1}\}\\
&w'=m_{i-1}\cdots m_1,\, \dcx_1^{\alpha_1}\neq\dcm_i &
&\Rightarrow & &z\in\{m_1^{\pm 1},\ldots, m_{i-1}^{\pm 1}, m_i\}\\
\nonumber\\
\label{aid} &w'= h_{j-1}\cdots m_1,\,
\dcx_1^{\alpha_1}\neq\dca_j& &\Rightarrow & &z\in\{m_1^{\pm 1},\ldots, b_{j-1}^{\pm 1}, a_j\}\\
&w'= a_j h_{j-1}\cdots m_1,\, \dcx_1^{\alpha_1}\neq\dca_j &
&\Rightarrow & &z\in\{b_j^\inv,  a_{j+1}^{\pm 1},\ldots, b_g^{\pm 1}\}\label{aid2}\\
&w'=a_j^\inv b_j^\inv a_j h_{j-1}\cdots m_1,\,
\dcx_1^{\alpha_1}\neq\dca_j^\inv &
&\Rightarrow & &z\in\{a_j^\inv, b_j^{\pm 1}, a_{j+1}^{\pm 1},\ldots, b_g^{\pm 1}\}\label{aid3}\\
&w'= b_j^\inv a_j h_{j-1}\cdots m_1,\,
\dcx_1^{\alpha_1}\neq\dca_j^\inv& &\Rightarrow & &z\in\{m_1^{\pm 1},\ldots, b_{j-1}^{\pm 1}, a_j^{\pm 1}\}\label{aid4}\\
\nonumber\\
\label{bid} &w'= b_j^\inv a_j h_{j-1}\cdots m_1,\,
\dcx_1^{\alpha_1}\neq\dcb_j &
&\Rightarrow & &z\in\{a_j^\inv, b_j^\inv,  a_{j+1}^{\pm 1},\ldots, b_g^{\pm 1}\}\\
&w'= a_j h_{j-1}\cdots m_1,\, \dcx_1^{\alpha_1}\neq\dcb_j&
&\Rightarrow & &z\in\{m_1^{\pm 1},\ldots, b_{j-1}^{\pm 1},
a_j,b_j^{\pm 1}\}\\
&w'= a_j^\inv b_j^\inv a_j h_{j-1}\cdots m_1,\,
\dcx_1^{\alpha_1}\neq\dcb_j^\inv& &\Rightarrow & &z\in\{m_1^{\pm 1},\ldots, b_{j-1}^{\pm 1}, b_j\}\\
&w'= h_{j}\cdots m_1,\, \dcx_1^{\alpha_1}\neq\dcb_j^\inv &
&\Rightarrow & &z\in\{ a_{j+1}^{\pm 1},\ldots, b_g^{\pm 1}\}.
\end{align}
\end{lemma}
{\bf Proof:} The proof is by induction over the length $r$ of
$\lambda$ as a reduced word in $\dcm_i,\dca_j,\dcb_j$ and similar
for all of the implications above. We prove the  statement
\eqref{aid}. For $r=1$, expressions \eqref{dualcurves} for
$\dcm_i,\dca_j,\dcb_j$ imply $z\in\{a_j,a_{j-1}^\inv\}$, since the
word $wzw'$ is reduced.  Now assume the statement is true for
$r\leq k$ and there exists an element of $\pi_1(\surf\mindee)$
whose expression as a reduced word in $m_i,a_j,b_j$ and their
duals is given by $\dcx_{k+1}^{\alpha_{k+1}}\cdots
\dcx_1^{\alpha_1}=wz h_{j-1}\cdots h_1$, where $w$ is a reduced
word in $m_i,a_j,b_j$, $\dcx_1^{\alpha_1}\neq\dca_j$ and
$z\in\{a_j^\inv, b_j^{\pm 1}, a_{j+1}^{\pm 1}\cdots b_g^{\pm
1}\}$. This implies that the elements $\dcx_{k}^{\alpha_{k}}\cdots
\dcx_1^{\alpha_1}$ and $\dcx_{k+1}^{\alpha_{k+1}}$, expressed as
reduced words in $m_i,a_j,b_j$ are of the form
$\dcx_{k}^{\alpha_{k}}\cdots \dcx_1^{\alpha_1}=y h_{j-1}\cdots
h_1$, $\dcx_{k+1}^{\alpha_{k+1}}=y'zy^\inv$, where $y,y'$ are
reduced words in $m_i,a_j,b_j$. If $z\in\{ a_{j+1}^{\pm 1}\cdots
b_g^{\pm 1}\}$, it follows from the expression \eqref{dualcurves}
  that the last letter in $y$ is an element of $\{b_j^\inv,
a_{j+1}^{\pm 1},\ldots,b_g^{\pm 1}\}$, and we obtain a
contradiction. Similarly,  for $z=a_j^\inv$ the last letter of $y$
is in $\{b_j^{\pm 1}\}$, for $z=b_j^\inv$ in $\{a_{j+1},
a_j^\inv\}$ which again contradicts the induction hypothesis.
Finally, for $z=b_j$, the last letter in $y$ is either again in
$\{a_j^\inv, a_{j+1}\}$ or we have
$\dcx_{k+1}^{\alpha_{k+1}}=\dcb_j^\inv$ $y=a_j^\inv b_ja_j
h_{j-1}\cdots m_1$, which implies
$\dcx_k^{\alpha_k}\cdots\dcx_1^{\alpha_1}=\dca_j$ and contradicts
the induction hypothesis. Hence, the statement is true for
$r=k+1$, which proves the claim.\hfill $\Box$

 {\bf Proof of Theorem \ref{intasth}}: We prove the
statement for the case $\dcx_k^{\alpha_k}=\dca_j$. The reasoning
for the other cases is analogous. Suppose the word
$y_s^{\beta_s}\cdots y_{l+1}^{\beta_{l+1}} y_l^{\beta_l}\cdots
y_1^{\beta_1}$ in $m_i,a_j,b_j$ defined by \eqref{splitabword1},
\eqref{splitabword2} is not reduced,
i.~e.~$y_l^{\beta_l}=y_{l+1}^{-\beta_{l+1}}$. Then, either the
expression for the product
$\dcx_r^{\alpha_r}\cdots\dcx_{k+1}^{\alpha_{k+1}}$ as a reduced
word in $m_i,a_j,b_j$ must be of the form
\begin{align}
\label{hform1} \dcx_r^{\alpha_r}\cdots\dcx_{k+1}^{\alpha_{k+1}}=ux
a_j^\inv b_j^\inv a_j h_{j-1}\cdots m_1,
\end{align}
 or the expression for
$\dcx_{1}^{-\alpha_1}\cdots\dcx_{k-1}^{-\alpha_{k-1}}$ as a
reduced word in $m_i,a_j,b_j$ must be of the form
\begin{align}
\label{hform2}
\dcx_{1}^{-\alpha_1}\cdots\dcx_{k-1}^{-\alpha_{k-1}}=u'
x'h_{j-1}\cdots m_1
\end{align}
 where $x,x'\in\{m_1,\ldots,b_g\}$ and $u,u'$ are reduced words in $m_i,a_j,b_j$.
Now note that the expression \eqref{lambdadecomp0} for $\lambda$
is reduced and therefore
$\dcx_{k-1}^{\alpha_{k-1}},\dcx_{k+1}^{\alpha_{k+1}}\neq\dca_j^\inv$,
which allows us to apply the identities \eqref{aid}, \eqref{aid3}
in Lemma \ref{helplem} to expressions \eqref{hform2},
\eqref{hform2}. Suppose  the expression for $
\dcx_r^{\alpha_r}\cdots\dcx_{k+1}^{\alpha_{k+1}}$
 as a reduced word in $m_i,a_j,b_j$ is of the form
 \eqref{hform1}. Then, the identity
\eqref{aid3} implies $x\in\{a_j^\inv, b_{j+1}^{\pm 1}, \ldots,
b_g^{\pm 1}\}$ and therefore $y_s^{\beta_s}\cdots y_1^{\beta_1}$
is reduced unless the expression for
$\dcx_{1}^{-\alpha_1}\cdots\dcx_{k-1}^{-\alpha_{k-1}}$ as a
reduced word in $m_i,a_j,b_j$ is of the form \eqref{hform2}. But
then identity \eqref{aid} implies $x'\in\{m_1^{\pm 1}, \ldots,
b_{j-1}^{\pm 1}, a_j\}$ and therefore $y_s^{\beta_s}\cdots
y_1^{\beta_1}$ is reduced.

Hence, the expression of
$\dcx_{1}^{-\alpha_1}\cdots\dcx_{k-1}^{-\alpha_{k-1}}$ as a
reduced word in the generators $m_i,a_j,b_j$ must be of the form
\eqref{hform2} with $x'=a_j$ and the corresponding expression for
$\dcx_r^{\alpha_r}\cdots\dcx_{k+1}^{\alpha_{k+1}}$ must be given
by
\begin{align}
\label{hform3}
\dcx_r^{\alpha_r}\cdots\dcx_{k+1}^{\alpha_{k+1}}=u''x'' b_j^\inv
a_j h_{j-1}\cdots m_1,
\end{align}
where $u''$ is a reduced word in $m_i,a_j,b_j$ and
$x''\in\{m_1^{\pm 1}, \ldots, b_g^{\pm 1}\}\setminus\{a_j^\inv\}$.
This implies $y_l^{-\beta_l}=y_{l+1}^{\beta_{l+1}}=a_j$.
Furthermore, by applying identities \eqref{aid2} and \eqref{aid4}
in Lemma \ref{helplem} to the expressions of , respectively,
$\dcx_{1}^{-\alpha_1}\cdots\dcx_{k-1}^{-\alpha_{k-1}}$ and
$\dcx_r^{\alpha_r}\cdots\dcx_{k+1}^{\alpha_{k+1}}$ as reduced
words in $m_i,a_j,b_j$, we find that they are of the form
\begin{align}
\label{hform4}
&\dcx_{1}^{-\alpha_1}\cdots\dcx_{k-1}^{-\alpha_{k-1}}=u''' x'''
a_jh_{j-1}\cdots m_1\qquad x'''\in\{ b_j^\inv, a_{j+1}^{\pm 1},
\ldots, b_g^{\pm 1}\}\\
&\dcx_r^{\alpha_r}\cdots\dcx_{k+1}^{\alpha_{k+1}}=u''x'' b_j^\inv
a_j h_{j-1}\cdots m_1\qquad x''\in\{m_1^{\pm 1}, \ldots,
b_{j-1}^{\pm 1}, a_j\}\nonumber.
\end{align}
This implies that the product of the reduced words
\begin{align}
&y_{l-1}^{\beta_{l-1}}\cdots y_1^{\beta_1}=a_j h_{j-1}\cdots m_1
\dcx_{k-1}^{\alpha_{k-1}}\cdots\dcx_1^{\alpha_1}\\
&y_r^{\beta_r}\cdots
y_{l+2}^{\beta_{l+2}}=\dcx_r^{\alpha_r}\cdots\dcx_{k+1}^{\alpha_{k+1}}b_j^\inv
a_j h_{j-1}\cdots m_1\nonumber
\end{align}
is  reduced and gives the expression of $\lambda$ as a reduced
word in $m_i,a_j,b_j$, which proves the claim.\hfill $\Box$

Hence, by splitting the dual generators as in \eqref{mseg},
\eqref{aseg}, \eqref{bseg}, we obtain an almost unique assignment
of the intersection points of a general embedded curve $\lambda$
with the generators $m_i,a_j,b_j$ between the different factors in
the expression of $\lambda$ as a reduced word in $m_i,a_j,b_j$ and
to the starting and endpoints of $m_i,a_j,b_j$. We will now show
that this assignment of intersection points corresponds to a
graphical decomposition of a curve on $\surf\mindee$ representing
 $\lambda$ into representatives of $m_i,a_j,b_j$.

\begin{figure}
 \vskip .3in \protect\input epsf
\protect\epsfxsize=12truecm
\protect\centerline{\epsfbox{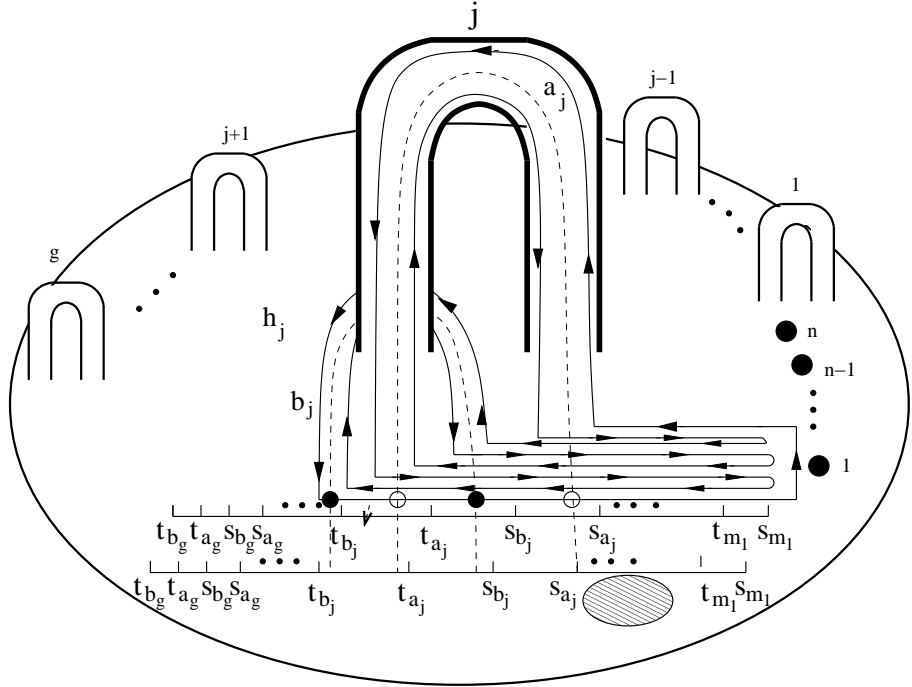}} \caption{The
representation of the element $\lambda=h_j=b_j \circ a_j^\inv\circ
b_j^\inv\circ a_j$ and its intersection points with the generators
$a_j$, $b_j$ .} \label{exhi2}
\end{figure}
\begin{figure}
\vskip .3in \protect\input epsf \protect\epsfxsize=12truecm
\protect\centerline{\epsfbox{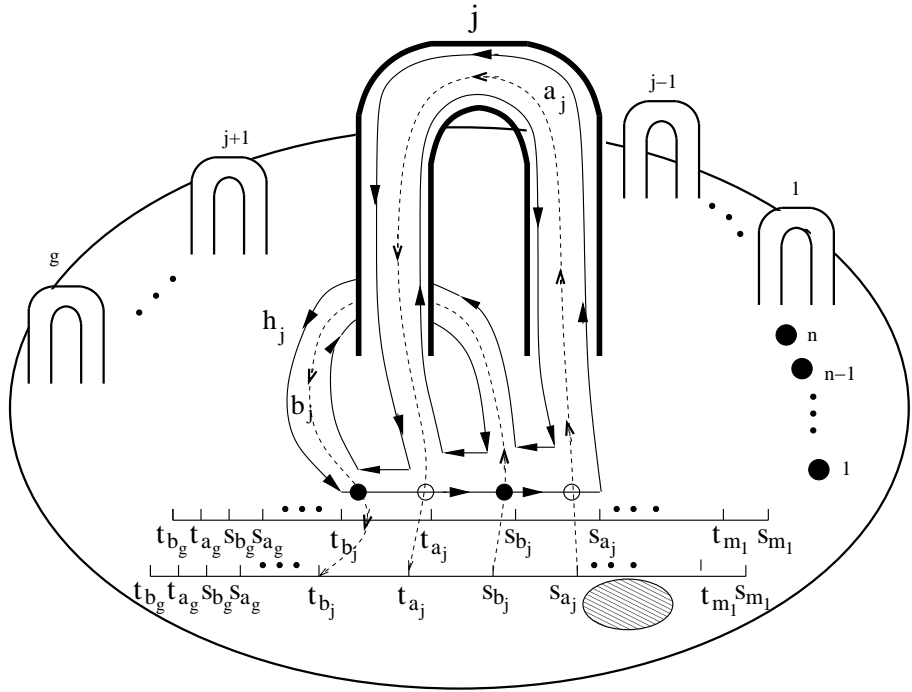}} \caption{ The
representation of the element $\lambda=h_j=b_j \circ a_j^\inv\circ
b_j^\inv\circ a_j$ and its nontrivial intersection points with the
generators $a_j$, $b_j$ .} \label{exhi}
\end{figure}

For this, we represent the generators  $m_i,a_j,b_j$,
$\dcm_i,\dca_j,\dcb_j$ by curves as in Fig.~\ref{pi1fig}, but
instead of a basepoint, we draw a line on which the starting
points $s_{m_i}, s_{a_j}, s_{b_j}$ and endpoints $t_{m_i},t_{a_j},
t_{b_j}$ are ordered from
 right to left according to
\begin{align}
\label{pi1genord}
s_{m_1}<t_{m_1}<\ldots<s_{m_n}<t_{m_n}<s_{a_1}<s_{b_1}<t_{a_1}<t_{b_1}<\ldots<s_{a_g}<s_{b_g}<t_{a_g}<t_{b_g}.
\end{align}
and the basepoint $p\in\surf\mindee$ is located to the right of
$s_{m_1}$. The curves representing the generators
$x\in\{m_i,a_j,b_j\}$ are decomposed into an oriented horizontal
segment from $p$ to the starting point $s_x$, a curve which starts
in $s_x$ and ends in $t_x$ and another horizontal segment from
$t_x$ back to $p$ as shown in Fig.~\ref{exhi2}. For their
inverses, we set $s_{x^\inv}=t_x$, $t_{x^\inv}=s_x$. To obtain an
embedded  representative  of an   element
$\lambda\in\pi_1(\surf\mindee)$ given uniquely as a reduced word
in the generators $m_i,a_j,b_j$ by
\begin{align}
\label{lambdadecomp2} \lambda=y_s^{\beta_s}\cdots
y_1^{\beta_1}\qquad y_k\in\{m_1,\ldots,b_g\}, \beta_k\in\{\pm 1\},
\end{align}
we draw consecutively the representatives of the factors
$y_k^{\beta_k}$  and contract the overlapping horizontal segments
such that the resulting curve has no self-intersections, see
Fig.\ref{exhi2}, \ref{exhi}. Thus, we obtain a curve representing
$\lambda$ which is composed of curves representing the factors
$y_k^{\beta_k}$ which start and end above the corresponding
starting points $s_k$ and endpoints $t_k$ on the horizontal line
and horizontal segments $t_{k}s_{k+1}$ connecting the starting and
endpoints of these factors.

To locate the intersection points of
$\lambda\in\pi_1(\surf\mindee)$ with the generators $m_i,a_j,b_j$
between the different factors of $\lambda$ and at the starting and
endpoints of $m_i,a_j,b_j$, we draw two such lines, one for the
generators $m_i,a_j,b_j$ and one for $\lambda$ such that the first
one is tangent to the disc, while the one for $\lambda$ is
displaced slightly. We represent the generators $m_i,a_j,b_j$ and
$\lambda$ graphically as described above such that all
intersection points of $\lambda$ with $m_i,a_j,b_j$ lie on the
horizontal segments $t_ks_{k+1}$ in the decomposition of $\lambda$
and above the starting and endpoints $s_{m_i}$, $s_{a_j}$,
$s_{b_j}$, $t_{m_i}$, $t_{a_j}$, $t_{b_j}$. We also require that
for each factor $m_i^{\pm 1}$ which gives rise to a pair of
intersection points with $m_i$, one of these points lies above the
starting point $s_{m_i}$ and one above the endpoint $t_{m_i}$.
 An intersection point $q_i$ is then said to occur between the factors
$y_i^{\beta_i}$ and $y_{i+1}^{\beta_{i+1}}$ on $\lambda$ if it
lies on the straight line connecting $t_i=t_{y_i^{\beta_i}}$ and
$s_{i+1}=s_{y_{i+1}^{\beta_{i+1}}}$, where we set $t_0=t_n$.
Furthermore, we say it occurs at the starting point of a
generators $m_i,a_j,b_j$ if it is located above, respectively,
$s_{m_i}$, $s_{a_j}$, $s_{b_j}$ and at its endpoint if it is
located above $t_{m_i}$, $t_{a_j}$, $t_{b_j}$. By comparing this
assignment of intersection points via the graphical procedure with
the assignment in Theorem \ref{intasth}, we find that they agree
for all $\lambda\in\pi_1(\surf\mindee)$ with embedded
representatives.
\begin{theorem} \label{grequith}
Consider an element $\lambda\in\pi_1(\surf\mindee)$  with an
embedded representative and given as a reduced word in the
generators $m_i,a_j,b_j$. Then, the assignment of intersection
points of $\lambda$ with $m_i,a_j,b_j$ between the different
factors in this expression and to the starting points and
endpoints of the generators $m_i,a_j,b_j$ via the graphical
procedure agrees with the one in Theorem \ref{intasth}. In
particular, the ambiguity for the assignment of an intersection
point of $\lambda$ with $x\in\{a_1,b_1,\ldots,a_g,b_g\}$ which
arises at a factor $x^{\pm 1}$ in \eqref{lambdadecomp2},
corresponds to sliding the intersection point along $x$. In the
following we assign such  intersection points to the right of
factors $a_j,b_j$ and to the left of factors $a_j^\inv,b_j^\inv$.

\end{theorem}

{\bf Proof:}

We first consider  a single factor $y\in\{m_1^{\pm
1},\ldots,b_g^{\pm 1}\}$ in the expression for $\lambda$ as a
reduced word in $m_i,a_j,b_j$ and the intersection points of the
associated curve with the representative of $m_i,a_j,b_j$.

\begin{enumerate}

\item  For the generator $a_j$, the graphical procedure implies
that $y$ has an intersection point with the starting point of
$a_j$ with positive  intersection number if and only if $s_y\geq
s_{a_j}$, which implies $y\in\{a_j^{\pm 1},b_j^{\pm
1},a_{j+1}^{\pm 1},\ldots,b_g^{\pm 1}\}$. Similarly, it has an
intersection point with the end point of $a_j$ with negative
intersection number if and only if $s_y\geq t_{a_j}$,
$y\in\{a_j^\inv, b_j^\inv, a_{j+1}^{\pm 1},\ldots,b_g^{\pm 1}\}$,
and in both cases the intersection points occur on the segment
$t_yp$. Intersection points at the starting point of $a_j$ with
negative intersection number and at the endpoint of $a_j$ with
positive intersection number lie on the segment $ps_y$ and occur,
respectively, for $t_y\geq s_{a_j}$, $y\in\{a_j^{\pm 1},b_j^{\pm
1},a_{j+1}^{\pm 1},\ldots,b_g^{\pm 1}\}$, and $t_y\geq t_{a_j}$,
$y\in\{a_j,b_j,a_{j+1}^{\pm 1},\ldots,b_g^{\pm 1}\}$.

By comparing with expression \eqref{dualcurves} for the generators
$m_i,a_j,b_j$ in terms of their duals, we find that $y$ gives rise
to an intersection point at the starting point of $a_j$ with
positive and negative intersection number if and only if its
expression as a reduced word in $\dcm_i,\dca_j,\dcb_j$,
respectively, ends in a sequence $\dca_j\dch_{j-1}\cdots\dcm_1$
and starts in a sequence $(\dca_j\dch_{j-1}\cdots\dcm_1)^\inv$.
Similarly, an intersection of $y$ with the endpoint of $\dca_j$
with negative and positive intersection number occurs if and only
if the expression for $y$ ends in a sequence
$\dca_j^\inv\dcb_j^\inv\dca_j\dch_{j-1}\cdots\dcm_1$ and starts in
a sequence
$(\dca_j^\inv\dcb_j^\inv\dca_j\dch_{j-1}\cdots\dcm_1)^\inv$,
respectively. Hence, each of the generators $\{a_{j+1}^{\pm
1},\ldots,b_g^{\pm 1}\}$ has two intersections with each the
starting and endpoint of $a_j$ and with opposite intersection
numbers, $b_j^{\pm 1}$ has two intersections with the starting
point $s_{a_j}$ with opposite intersection numbers and one with
the endpoint $t_{a_j}$, where the oriented intersection number is
positive for $b_j$ and negative for $b_j^\inv$. For the
intersections of $a_j$ with itself there is some ambiguity in
drawing the associated intersection diagram. We can either assign
two intersections with opposite intersection numbers to the
starting point of $a_i$ or one with positive intersection number
to the endpoint and one with negative intersection number to the
starting point. This corresponds to the fact that the two factors
$\dca_j,\dca_j^\inv$ in expression \eqref{dualcurves} for $a_j$
can be considered either part of the sequence $\dcm_1^\inv\cdots
\dch_{j-1}^\inv \dca_j^\inv \dcb_j\dca_j$ at the start of $a_j$ or
can be assigned to two sequences $\dcm_1^\inv\cdots
\dch_{j-1}^\inv \dca_j^\inv$ at the start of $a_j$ and
$\dca_j\dch_{j-1}\cdots\dcm_1$ at the end of $a_j$.

\item Similarly,  intersection points of $y$ with the starting
point of the representative $b_j$ with positive intersection
number occur for $y\in\{b_j^{\pm 1}, a_j^\inv, a_{j+1}^{\pm
1},\ldots, b_g^{\pm 1}\}$ and those with the endpoint of $b_j$ and
negative intersection number for $y\in\{b_j, a_{j+1}^{\pm
1},\ldots, b_g^{\pm 1}\}$ and lie on the segment $ps_y$. Those at
the starting point $s_{b_j}$ with negative intersection number and
at the endpoint $t_{b_j}$ with positive intersection number are
located on the segment $t_yp$ and occur for, respectively,
$y\in\{b_j^{\pm 1}, a_j, a_{j+1}^{\pm 1},\ldots, b_g^{\pm 1}\}$
and $y\in\{b_j^\inv, a_{j+1}^{\pm 1},\ldots, b_g^{\pm 1}\}$.
Hence, intersection points with the starting point of $b_j$ and,
respectively, positive and negative intersection numbers can be
identified with a sequence
$\dcb_j^\inv\dca_j\dch_{j-1}\cdots\dch_1$ at the end of the
expression for $y$ as a reduced word in $\dcm_i,\dca_j,\dcb_j$ and
with a sequence $(\dcb_j^\inv\dca_j\dch_{j-1}\cdots\dch_1)^\inv$
at its beginning. Intersection points at the end of $b_j$ with,
respectively, negative and positive intersection number correspond
sequences $\dch_{j}\cdots\dch_1$, $(\dch_{j}\cdots\dch_1)^\inv$.
Again, there is an ambiguity in the graphical  assignment of the
intersection points of $b_j$ with itself, which can be either
drawn as two intersection points at the starting point of $b_j$ or
one at the starting point and one at the endpoint. This ambiguity
corresponds to two different ways of assigning the two factors
$\dcb_j$, $\dcb_j^\inv$ in expression \eqref{dualcurves} for
$b_j$.

\item Finally, we consider intersection points of  $y$ with the
generators $m_i$. We find that $y\in\{m_i^\inv,m_{i+1}^{\pm
1},\ldots,b_g^{\pm 1}\}$ has both, an intersection point with the
starting point of $m_i$ and one with its end point, which lie on
the segment $ps_y$ and have, respectively, positive and negative
intersection number. Similarly, $y\in\{m_i,m_{i+1}^{\pm
1},\ldots,b_g^{\pm 1}\}$ has an intersection point with the
starting point of $m_i$ with negative intersection number and one
with its endpoint with positive intersection number, both located
on the segment $ps_y$. Hence, a pair of intersection points with
the starting and endpoint of $m_i$ and with opposite intersection
numbers  corresponds to a sequence $\dcm_i\cdots\dcm_1$ at the end
of  expression \eqref{dualcurves} for $y$ if the one at the
starting point has positive intersection number and to a sequence
$(\dcm_i\cdots\dcm_1)^\inv$ at the end of the expression if the
one at the starting point has negative intersection number. Note
that in contrast to the situation for the generators $a_j,b_j$
there is no ambiguity in assigning the intersection points of
$m_i$ with itself since we required that one of the two
intersection points lies at the starting point and one at the
endpoint of $m_i$.
\end{enumerate}

We now consider an element $\lambda\in\pi_1(\surf\mindee)$ with an
embedded representative and given as a reduced word in the
generators $m_i,a_j,b_j$
\begin{align}
\label{lambdproof} \lambda=y_s^{\beta_s}\cdots y_1^{\beta_1}\qquad
y_k\in\{m_1,\ldots,b_g\}, \beta_k\in\{\pm 1\}
\end{align}
 In the graphical procedure the intersection points of $\lambda$ with $m_i,a_j,b_j$ are obtained by decomposing it into
its factors and removing those intersection points of the starting
or endpoint of a generator $m_i,a_j,b_j$ which occur both on a
segment $t_{y_k}p$ and $ps_{y_{k+1}}$ with opposite intersection
numbers. To minimize the number of intersection points, one makes
use of the ambiguity in assigning the intersection points  of
$a_j$ and $b_j$ with themselves and removes the remaining
ambiguity by assigning ambiguous intersection points to the right
of $a_j$ and $b_j$.

From the discussion above it follows that a pair of intersection
points on the segments $t_{y_k^{\beta_k}}p$ and
$ps_{y_{k+1}^{\beta_{k+1}}}$ can be removed if and only if the
associated sequences $(\dcm_i\cdots\dcm_1)^{\pm 1}$,
$(\dca_j\dch_{j_1}\cdots\dch_1)^{\pm 1}$
$(\dca_j^\inv\dcb_j^\inv\dca_j\dch_{j_1}\cdots\dch_1)^{\pm 1}$,
$(\dcb_j^\inv\dca_j\dch_{j_1}\cdots\dch_1)^{\pm 1}$,
$\dch_{j}\cdots\dch_1)^{\pm 1}$ assigned to these intersection
points as described above cancel. Factors $\dcm_i^{\pm
1},\dca_j^{\pm 1},\dcb_j^{\pm 1}$ associated to intersection
points of $m_i,a_j,b_j$ with factors in \eqref{lambdproof}
therefore give rise to factors $\dcm_i^{\pm 1},\dca_j^{\pm
1},\dcb_j^{\pm 1}$ in the expression of $\lambda$ as a reduced
word in $\dcm_i,\dca_j,\dcb_j$ if and only if the corresponding
intersection points cannot be removed and remains in the
intersection diagram.  We then consider the precise form of these
sequences in the generators $\dcm_i,\dca_j,\dcb_j$ and compare
with the prescription in Theorem \ref{intasth}. A short
calculation using expressions \eqref{dualcurves} for the
generators $m_i,a_j,b_j$ and their duals then shows that the
assignment of intersection points in Theorem \ref{intasth} agrees
with the one from the graphical procedure for all intersection
points that do not lie on the segment $t_ss_1$.

For intersection points on the segment $t_ss_1$ we note that the
removal of intersection points by contracting the segments $ps_1$
and $pt_s$ in the graphical procedure amounts to moving the
basepoint of the curve representing $\lambda$. The number of
resulting intersection points is minimal and cannot be reduced
further by conjugating $\lambda$ with elements of
$\pi_1(\surf\mindee)$. In the discussion after Theorem
\ref{intsecttheorem} we found that this is the case if and only if
the expression for $\lambda$ as a reduced word in the generators
$\dcm_i,\dca_j,\dcb_j$ is cyclically reduced. Hence, the
assignment of intersection points obtained by graphically
representing $\lambda$ without contracting this segment agrees
with the assignment in Theorem \ref{intasth}. Contracting the
segment $t_s S_1$ amounts to applying the procedure in Theorem
\ref{intasth} to the cyclically reduced element associated to
$\lambda$. \hfill $\Box$

The graphical procedure described above therefore reproduces the
assignment of intersection points of
$\lambda\in\pi_1(\surf\mindee)$ with the generators $m_i,a_j,b_j$
between the different factors in the expression of $\lambda$ as a
reduced word in $m_i,a_j,b_j$ and to the starting and endpoint of
the representatives of $m_i,a_j,b_j$ in Theorem \ref{intasth}.
Furthermore, it allows us to consider general elements
$\eta,\lambda\in\pi_1(\surf\mindee)$ given as reduced words in the
generators $m_i,a_j,b_j$ and assign their intersection points
between the different factors in these expressions. For this, we
represent both $\eta$ and $\lambda$ graphically as a product of
curves representing $m_i,a_j,b_j$ as described above. We draw two
lines with starting and endpoints of associated to the generators
$m_i,a_j,b_j$, ordered according to \eqref{pi1genord} and
basepoints $p$ to the right of $s_{m_i}$ such that the one for
$\eta$ is tangent to the boundary and the one for $\lambda$
slightly displaced. After decomposing $\lambda$ into a set of
curves starting and ending above the corresponding starting and
endpoints and into horizontal segments parallel to the line for
$\lambda$, we can graphically determine the intersection points of
$\lambda$ with the generators $m_i,a_j,b_j$ as described above. To
obtain the intersection points of $\lambda$ and $\eta$, we
decompose $\eta$ by consecutively drawing its factors in the
expression as a reduced word in $m_i,a_j,b_j$ and obtain a
representative made up of curves starting and ending above the
starting and endpoints on the line for $\eta$ and of horizontal
segments. All intersection points of the representatives of
$\lambda$ and $\eta$ are then located above the starting and
endpoints for $\eta$ and on the horizontal segments in the
decomposition of $\lambda$.
\begin{figure}
 \vskip .3in \protect\input epsf
\protect\epsfxsize=12truecm
\protect\centerline{\epsfbox{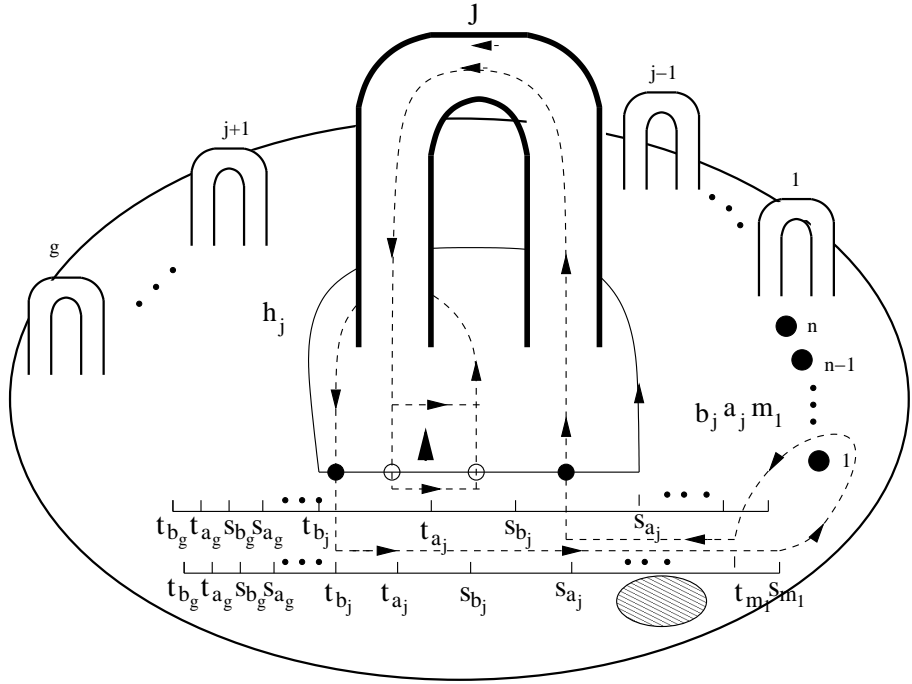}} \caption{ The
intersection points of curves representing $\lambda=h_j=b_j \circ
a_j^\inv\circ b_j^\inv\circ a_j$ (full line) and $\eta=b_j\circ
a_j\circ m_1$ (dashed line).} \label{exhi3}
\end{figure}
Finally, one removes any intersection point which occurs both, at
the endpoint of a factor and at the starting point of the next
factor in the decomposition of $\eta$ with opposite intersection
number by lifting the corresponding segment as shown in
Fig.~\ref{exhi3}.

After completing this procedure, one obtains two curves
representing $\eta$ and $\lambda$ with a minimum number of
intersection points, all of which are located above the starting
and endpoints on the line for $\eta$ and on the horizontal
segments for $\lambda$. By assigning an intersection that occurs
at the endpoint of a factor to the left of this factor and one
that occurs at its starting point to the right, we then obtain a
unique assignment of the intersection points of $\eta$ and
$\lambda$ between the factors in the expression of $\eta$ as a
reduced word in $m_i,a_j,b_j$.


\section{Dual generators and the moduli space of flat connections}
\label{modspacesect}

\subsection{Fock and Rosly's description of the moduli space of flat connections}

In this section, we apply the involution
$I\in\text{Aut}(\surf\mindee)$ to Fock and Rosly's  description of
the moduli space of flat connections on $\surf$ \cite{FR}. We show
that by expressing the Poisson structure on the moduli space in
terms of both  the generators $m_i,a_j,b_j\in\pi_1(\surf\mindee)$
and their duals $\dcm_i,\dca_j,\dcb_j$, one obtains a particularly
simple expression in which its dependence on intersection points
is apparent.

We start with a brief summary of moduli spaces of flat connections
and Fock and Rosly's formalism. In the following we consider a
finite dimensional Lie group $H$ with Lie algebra
$\gothh=\text{Lie}\;H$, viewed as a vector space over $\RR$. We
fix a basis $J_a$, $a=1,\ldots,\text{dim}\,\gothh$, of $\gothh$
and denote by  $L_a$, $R_a$, respectively, for the associated
right- and left-invariant vector fields on $H$
\begin{align}
\label{vecfields} L_af(u)=\frac{d}{dt}|_{t=0} f(e^{-tJ_a} u
)\qquad R_af(u)=\frac{d}{dt}|_{t=0} f(u e^{tJ_a} )\qquad\forall u
\in H, \forall f\in\cif(H).
\end{align}
Here and in the following, we denote by $ \exp:\;
 \gothh\ni x^a J_a \mapsto e^{x^aJ_a}\in H$ the
 exponential map, which we  require to be
 surjective.

The moduli space of flat $H$-connections on $\surf$ arises as the
phase space of a Chern-Simons theory with gauge group $H$ on the
three-manifold $\RR\times\surf$. To formulate this Chern-Simons
theory, one associates  to each puncture an orbit $\mathfrak{c}_i$
under the adjoint action of $H$ on $\gothh$
\begin{align}
u J_a u^\inv =\Ad(u)_a^{\;\;b} J_b
\end{align}
and  fixes a non-degenerate, $\Ad$-invariant, symmetric bilinear
form $\langle\,,\,\rangle$ on $\gothh$
\begin{align}
\label{pairing} \langle J_a,J_b\rangle=t_{ab}\qquad
t_{ab}t^{bc}=\delta_a^{\;\;c}.
\end{align}
A flat $H$-connection on $\surf$ is a one-form $A$ on $\surf$ with
values in the Lie algebra $\gothh$ whose curvature $F_A$  develops
a delta-function singularity at each puncture and vanishes
elsewhere
\begin{align}
\label{curv} F_A=dA+A\wedge A=\sum_{i=1}^n T_i\delta(z-z(i))\qquad
T_i\in\mathfrak{c}_i,
\end{align}
where $z(i)$, $i=1,\ldots,n$ denotes the coordinate of the
$i^{th}$ puncture. Gauge transformations are given by functions
$\gamma:\surf\rightarrow H$ and act on the connection according to
\begin{align}
\label{gtrafo} A\mapsto \gamma A\gamma^\inv+\gamma d\gamma^\inv.
\end{align}
The moduli space $\mathcal{M}^H_{g,n}$ is the quotient of the
space of flat $H$-connections on $\surf$ modulo gauge
transformations \eqref{gtrafo}. Although defined as a quotient of
an infinite dimensional space, the moduli space
$\mathcal{M}^H_{g,n}$ is finite dimensional and can be
parametrised by the holonomies along a set of generators of the
fundamental group $\pi_1(\surf)$. While the holonomies
$A_j=H_{a_j}$, $B_j=H_{b_j}$ of the generators associated to the
handles are general elements of the gauge group $H$, the
holonomies $M_i=H_{m_i}$ of the loops around the punctures are
restricted to conjugacy classes $\mathcal{C}_i\subset H$
associated to the corresponding orbits $\mathfrak{c}_i$.
Furthermore, the holonomies are subject to a constraint arising
from the defining relation of the fundamental group $\pi_1(\surf)$
\begin{align}
\label{holrel} [B_g,A_g^\inv]\cdots[B_1,A_1^\inv]\cdot M_n\cdots
M_1=1\qquad [B_j,A_j^\inv]=B_jA_j^\inv B_j^\inv A_j.
\end{align}
Gauge transformations \eqref{gtrafo} act on  the holonomies by
simultaneously conjugating them with the elements of the gauge
group $H$, and the moduli space $\mathcal{M}_{g,n}^H$ of flat
$H$-connections on $\surf$ is given as the quotient of the
holonomies modulo this simultaneous conjugation
\begin{align}
 \mathcal{M}_{g,n}^H\!=\!\{(M_1,...,M_n,A_1,B_1,..., A_g,B_g)\in H^\ntg\,|\,
M_i\in\mathcal{C}_i,[B_g,A_g^\inv]\cdots[B_1,A_1^\inv]M_n\cdots
M_1=1 \}/H\nonumber.
 \end{align}
On the level of connections,  the finite dimensionality of the
moduli space $\mathcal{M}_{g,n}^H$  manifests itself in the fact
that a flat $H$-connection can be trivialised, i.~e.~written as
pure gauge, on any simply connected region $R\subset\surf$
\begin{align}
A|_{R}=\gamma d\gamma^\inv\qquad \text{with}\quad\gamma:
R\rightarrow H.
\end{align} Maximal simply
connected regions are obtained by cutting the surface $\surf$
along a set of generators of the fundamental group. As in the case
of a surface $\surf\mindee$ discussed in Sect.~\ref{dualint}, this
yields a set of $n$ punctured discs and a polygon $P_{g,n}$, only
that now the points $x_0$ and $x_{n+4g}$ in Fig.~\ref{cut2} are
identified since the boundary of the disc $D$ is not present.

As shown by Alekseev and Malkin \cite{AMII}, a function $\gamma:
P_{g,n}\rightarrow H$ defines a flat gauge field on $\surf$  if
and only if it is such that the resulting holonomies around the
punctures are elements of the conjugacy classes $\mathcal{C}_i$
and it satisfies an overlap condition for each pair of sides
corresponding to a generator $y\in\{a_1,b_1,\ldots, a_g,b_g\}$
\begin{align}
\label{csident} & A|_{y'}=\gamma d\gamma^\inv|_{y'}=\gamma
d\gamma^\inv|_{y}=A|_y.\end{align} This requirement
\eqref{csident} is equivalent to the existence of constant
elements $N_{y}\in H$ such that
\begin{align}
\label{csident2} &\gamma^\inv|_{y'}=N_{y} \gamma^\inv|_{y} ,
\end{align} and the restriction of the holonomies around the punctures to
conjugacy classes $\mathcal{C}_i$ can be cast into the form
\begin{align}
\label{csident3}
\gamma^\inv(x_i)=N_{m_i}\gamma^\inv(x_{i-1})\qquad
N_{m_i}^\inv\in\mathcal{C}_i.
\end{align}
The elements $N_y$,
$y\in\{m_1,\ldots,m_n,a_1,b_1,\ldots,a_g,b_g\}$ in the overlap
condition \eqref{csident2}, \eqref{csident3} contain all
information about the physical state and are closely related to
the holonomies $M_i,A_j,B_j$ along our set of generators of the
fundamental group $\pi_1(\surf)$. It follows from Fig.~\ref{cut2}
that these holonomies are given by the values of the trivialising
function $\gamma$ at the corners of the polygon
\begin{align}
\label{polyhols} &M_i=\gamma(x_i)\gamma^\inv(x_{i-1})\\
&A_j=\gamma(x_{n+4j-3})\gamma(x_{n+4j-4})^\inv=\gamma(x_{n+4j-2})\gamma(x_{n+4j-1})^\inv\nonumber\\
&B_j=\gamma(x_{n+4j-3})\gamma(x_{n+4j-2})^\inv=\gamma(x_{n+4j})\gamma(x_{n+4j-1})^\inv.\nonumber
\end{align}
Using the overlap conditions \eqref{csident2}, \eqref{csident3},
we can express these holonomies in terms of the variables
$N_{m_i},N_{a_j}, N_{b_j}\in H$ and vice versa and obtain
\begin{align}
 \label{holexp4}
&N_{m_i}\!\!=\!\!\gamma^\inv(x_0) M_1^\inv\cdots M_{i}^\inv
M_{i-1}\!\cdots\! M_1\gamma(x_0) \\
&N_{a_j}\!\!=\!\!\gamma^\inv(x_0) M_1^\inv\cdots M_n^\inv
H_1^\inv\!\!\!\!\!\cdots\! H_j^\inv B_j H_{j-1}\!\cdots\!
H_1M_n\cdots M_1\gamma(x_0)\nonumber\\
&N_{b_j}\!\!=\!\!\gamma^\inv(x_0)M_1^\inv\cdots M_n^\inv
H_1^\inv\!\!\!\!\!\cdots\! H_j^\inv A_j H_{j-1}\!\cdots\! H_1
M_n\cdots M_1\gamma(x_0).\nonumber
\end{align}
Hence, up to conjugation with value of $\gamma^\inv$ at the
basepoint $x_0$, the variables $N_{m_i}$, $ N_{a_j}$, $N_{b_j}$
are the holonomies along the dual generators
$\dcm_i,\dca_j,\dcb_j$.

The moduli space $\mathcal{M}_{g,n}^H$ carries a canonical
symplectic structure induced by the canonical symplectic form
associated to the Chern-Simons action. An explicit and efficient
description of the symplectic structure on the moduli space is
provided by Fock and Rosly's formalism \cite{FR}. Fock and Rosly
parametrise the symplectic structure on the moduli space in terms
of an auxiliary Poisson structure on a finite-dimensional extended
phase space, namely the space of graph connections associated to
certain graphs on the surface $\surf$. After implementing a set of
residual constraints which amount to a flatness condition on the
graph connection and dividing by the associated graph gauge
transformations, this auxiliary poisson structure on the space of
graph connections then induces the canonical Poisson structure on
the moduli space $\mathcal{M}_{g,n}^H$.

In the following we will work with the formulation of Alekseev,
Grosse
 and Schomerus \cite{AGSI,AGSII,AS} who specialised Fock and Rosly's description of the moduli space
 to the simplest graph describing the spatial surface $\surf$, a set of generators of its fundamental
 group $\pi_1(\surf)$. In this case, the extended phase space is
 the manifold $H^\ntg$, and the different copies of $H$ correspond
 to the holonomies along the generators of the fundamental group. Fock and Rosly's description of the Poisson
 structure on the moduli space can then be summarised as follows.
\begin{theorem} (Fock, Rosly \cite{FR})
\label{extphsp}

Consider the manifold $H^\ntg$ with points parametrised according
to
\begin{align} \label{holgrpar}
(M_1,\ldots,M_n,A_1,B_1,\ldots,A_g,B_g)\in H^\ntg,
\end{align}
and denote by  $L^X_a, R^X_a$, $X\in\{M_1,\ldots,B_g\}$, the
right- and left invariant vector fields \eqref{vecfields}
associated to the different components of $H^\ntg$. Let
$r=r^{ab}J_a\otimes J_b\in \gothh\otimes\gothh$ be a classical
$r$-matrix for the Lie algebra $\gothh$, i.~e.~a solution of the
classical Yang-Baxter equation (CYBE)
\begin{align}
\label{CYBE}
&[[r,r]]=[r_{12},r_{13}]+[r_{12},r_{23}]+[r_{13},r_{23}]=0\\
&r_{12}:=r^{ab} J_a\otimes J_b\otimes 1,\, r_{13}:=r^{ab}
J_a\otimes 1\otimes  J_b,\,r_{23}:=r^{ab} 1\otimes J_a\otimes
J_b\nonumber,
\end{align}
whose symmetric component is the dual of the bilinear form
$\langle\,,\,\rangle$ in the Chern-Simons action
\begin{align}
\label{symmr} r^{ab}=r^{ab}_{(s)}+r^{ab}_{(a)}\quad
r^{ab}_{(a)}=\tfrac{1}{2}(r^{ab}-r^{ba})\quad
r_{(s)}^{ab}=\tfrac{1}{2}(r^{ab}+r^{ba})=\tfrac{1}{2}t^{ab}.\end{align}
Then, the
 Poisson bivector
\begin{align}
\label{frbivect2} B=&r^{ab}_{(a)}\!\!\left(\sum_{i=1}^n
R^{\mi}_a\!\!+\!L^{\mi}_a\!\!+\!\!\sum_{j=1}^g
R^{\aj}_a\!\!+\!L^{\aj}_a\!\!+\!R^{\bjj}_a\!\!+\!L^{\bjj}_a\right)\!\!\otimes\!\!\left(\sum_{i=1}^n
R^\mi_b\!\!+\!L^\mi_b\!\!+\!\sum_{j=1}^gR^{\aj}_b\!\!+\!L^{\aj}_b\!\!+\!R^{\bjj}_b\!\!+\!L^{\bjj}_b\right)\nonumber\\
+\tfrac{1}{2}&t^{ab}\left(\sum_{i=1}^n
R^{\mi}_a\!+\!L^{\mi}_a\!\right)\wedge\left(\sum_{j=1}^g
R^{\aj}_b\!+\!L^{\aj}_b\!+\!R^{\bjj}_b\!+\!L^{\bjj}_b\right)
+\tfrac{1}{2}t^{ab}\!\!\!\!\!\sum_{i,j=1,\;i<j}^n (R^{\mi}_a\!\!+\!L^{\mi}_a)\wedge(R^{M_j}_b\!\!+\!L^{M_j}_b)\nonumber\\
+\tfrac{1}{2}&t^{ab}\!\!\!\!\sum_{i,j=1,\;i<j}^g
(R^{\ai}_a\!+\!L^{\ai}_a\!+\!R^{\bi}_a\!+\!L^{\bi}_a)\wedge(R^{\aj}_b\!+\!L^{\aj}_b\!+\!R^{\bjj}_b\!+\!L^{\bjj}_b)\nonumber\\
+\tfrac{1}{2}&t^{ab} \sum_{i=1}^n R^\mi_a\wedge L_b^\mi+
t^{ab}\sum_{i=1}^g R^\ai_a\wedge(R^\bi_b+L^\ai_b
+L^\bi_b)+R^\bi_a\wedge (L^\ai_b+L^\bi_b)+L^\ai_a\wedge L^\bi_b
\end{align}
defines a Poisson structure on $H^\ntg$.
The symplectic structure on the moduli space $\mathcal{M}_{g,n}^H$
is obtained by restricting the components $M_i$ to the conjugacy
classes $\mathcal{C}_i$, by imposing the constraint \eqref{holrel}
and by dividing by the associated gauge transformation which act
by simultaneous conjugation of all components with $H$.
\end{theorem}

By realising the moduli space of flat  $H$-connections on $\surf$
as a quotient of the finite dimensional Poisson manifold $H^\ntg$,
Fock and Rosly's description of the moduli space \cite{FR}
provides a rather efficient description of its Poisson structure.
The Poisson bracket of functions $f\in\cif(\mathcal{M}_{g,n}^H)$
on the moduli space $\mathcal{M}_{g,n}^H$ is given by the
Fock-Rosly bracket \eqref{frbivect2} of the associated conjugation
invariant functions $f'\in\cif(H^\ntg)$
\begin{align} (\sum_{i=1}^n
R^\mi_a+L^\mi_a+\sum_{j=1}^g
R^{\aj}_a+L^\aj_a+R^\bjj_a+L^\bjj_a)f'=0\quad
a=1,\ldots,\text{dim}\;\gothh.
\end{align}
Note that although Fock and Rosly's formalism requires the choice
of a classical $r$-matrix for the gauge group, the bracket of such
conjugation invariant functions with general functions
$g\in\cif(H^\ntg)$ does not depend on the choice of the classical
$r$-matrix. As the term involving its antisymmetric component
$r_{(a)}$ in \eqref{frbivect2} vanishes if one of the functions is
invariant under simultaneous conjugation, the resulting bracket
depends only on the matrix $t^{ab}$ representing the
$Ad$-invariant bilinear form $\langle\,,\,\rangle$ in the
Chern-Simons action.

A particular set of functions on the moduli space
$\mathcal{M}^H_{g,n}$ is given by conjugation invariant functions
of the holonomies of closed curves of the surface $\surf$, which
in the following will be referred to as generalised Wilson loop
observables. As Fock and Rosly's Poisson structure is defined on
the extended phase space $H^\ntg$ where the constraint
\eqref{holrel} from the defining relation of the fundamental group
$\pi_1(\surf)$ is not imposed, such functions are obtained
 from elements of
the fundamental group $\pi_1(\surf\mindee)$ of the associated
surface with a disc removed. More precisely,  for each element
$\lambda\in\pi_1(\surf\mindee)$,
 given uniquely as a reduced word in the generators $m_i,a_j,b_j$
\begin{align}
\label{wlambda} \lambda=x_r^{\alpha_r}\cdots x_1^{\alpha_1}\qquad
x_k\in\{m_1,\ldots,b_g\},\alpha_k\in\{\pm 1\},
\end{align}
one defines a map $\rho_\lambda: H^\ntg\rightarrow H$, which
expresses the holonomy along $\lambda$ in terms of the holonomies
$M_i,A_j,B_j$ along the generators of $\pi_1(\surf\mindee)$
\begin{align}
\label{rhoxdef} \rho_\lambda: (M_1,\ldots,M_n,A_1,B_1,\ldots,
A_g,B_g) \mapsto H_\lambda=X_r^{\alpha_r}\cdots
X_1^{\alpha_1}\quad X_k\in\{M_1,\ldots,B_g\}.
\end{align}
The generalised Wilson loop observables associated to $\lambda$
are then obtained by composing  conjugation invariant functions
$f\in\cif(H)$ with the map $\rho_\lambda$
\begin{align}
\label{fxdef} f_\lambda=f\circ\rho_\lambda\;\;\in\cif(H^\ntg).
\end{align}
As this map satisfies the condition
\begin{align}
\rho_\lambda(u M_1 u^\inv,\ldots, u B_g u^\inv)=u
\rho_\lambda(M_1,\ldots, B_g) u^\inv,
\end{align}
it follows immediately that the  Wilson loop observables are
invariant under simultaneous conjugation of all arguments with
elements of the gauge group $H$ and hence define a function on the
moduli space $\mathcal{M}_{g,n}^H$
\begin{align}
(\sum_{i=1}^n R^\mi_a+L^\mi_a+\sum_{j=1}^g
R^{\aj}_a+L^\aj_a+R^\bjj_a+L^\bjj_a)f_\lambda=(R_a+L_a)f\circ\rho_\lambda=0\quad
a=1,\ldots,\text{dim}\;\gothh.
\end{align}


\subsection{The Poisson structure in terms of the dual generators}
\label{frdual}
 The drawback of Fock and Rosly's description
\cite{FR} of the Poisson structure on the moduli space is that it
obscures the geometrical nature of the theory. For instance, it is
known that the Poisson brackets of generalised Wilson loop
observables depend on the intersection behaviour of the associated
curves on the surface, i.~e.~ the number of intersection points
and the associated oriented intersection numbers. However, in Fock
and Rosly's description of the Poisson structure on the moduli
space in terms of the bivector \eqref{frbivect2} on the manifold
$H^\ntg$, this dependence on intersection points is not readily
apparent. In this subsection, we demonstrate that this problem can
be remedied by working with the dual generators of the fundamental
group. More precisely, we show that when expressed in terms of
both the holonomies along the generators $m_i,a_j,b_j$ and those
along their duals $\dcm_i,\dca_j,\dcb_j$, Fock and Rosly's Poisson
structure takes a particularly simple form in which the dependence
on intersection points is readily apparent.

For this, it is convenient to characterise Fock and Rosly's
Poisson structure by the brackets of functions of the holonomies
along our set of generators $m_i,a_j,b_j\in\pi_1(\surf\mindee)$.
 Using the notation introduced in the last subsection, we denote
 by $f_\lambda\in\cif(H^\ntg)$ the function obtained by composing
 a general (not necessarily conjugation invariant) function $f\in\cif(H)$
with the maps $\rho_\lambda: H^\ntg\rightarrow H$,
$\lambda\in\pi_1(\surf\mindee)$ as in \eqref{fxdef}. Using this
notation, the Poisson bracket given by \eqref{frbivect2} can be
expressed equivalently in terms of the functions $f_{m_i}$,
$f_{a_j}$, $f_{b_j}$ as
\begin{align}
\label{holfunctpb}
\{f_x,g_x\}&=r^{ab}(R_a+L_a)f_x(R_b+L_b)g_x-t^{ab}(R_a+L_a
)f_xR_bg_x\\
&=r^{ab}_{(a)}(R_a+L_a)f_x(R_b+L_b)g_x+\tfrac{1}{2}t^{ab}(R_af_xL_b
g_x-L_a
f_x R_b g_x )\quad \forall x\in\{m_1,\ldots,b_g\}\nonumber\\
\nonumber\\ \{f_x,
g_y\}&=r^{ab}(R_a+L_a)f_x(R_b+L_b)g_y\qquad\qquad\qquad\qquad\quad\forall
x,y\in\{m_1,\ldots,b_g\}, x<y\label{ord}\\
&=r^{ab}_{(a)}(R_a+L_a)f_x(R_b+L_b)g_y+\tfrac{1}{2}t^{ab}(R_a+L_a)f_x(R_b+L_b)g_y\nonumber\\
\nonumber\\
\{f_{a_j},g_{b_j}\}&=r^{ab}(R_a+L_a)f_{a_j}(R_b+L_b)g_{b_j}-t^{ab}L_af_{a_j}
R_bg_{b_j}\nonumber\\
&=r^{ab}_{(a)}(R_a+L_a)f_{a_j}(R_b+L_b)g_{b_j}+\tfrac{1}{2}t^{ab}(R_af_{a_j}
L_bg_{b_j}-L_a f_{a_j} R_b g_{b_j}),\label{finholf}
\end{align}
where $L_a,R_a$ denote the right- and left-invariant vector fields
\eqref{vecfields} on $H$ and  $<$ in \eqref{ord} stands for the
ordering
\begin{align}
\label{order} x<y\Leftrightarrow \begin{cases} x=m_i, y=m_j,
i,j\in\{1,\ldots,n\},\,
i<j\\x\in\{m_1,\ldots,m_n\},y\in\{a_1,b_1,\ldots, a_g,b_g\}\\
x\in\{a_i,b_i\}, y\in\{a_j,b_j\},\,i,j\in\{1,\ldots,g\},\,
i<j\end{cases}.
\end{align}
By using the expressions \eqref{dualcurves} for the dual
generators $\dcm_i,\dca_j,\dcb_j$ in terms of $m_i,a_j,b_j$, we
can  derive the Poisson brackets of functions $f_{m_i}, f_{a_j},
f_{b_j}$ of the holonomies along our generators with functions
$g_{\dcm_i}, g_{\dca_j}, g_{\dcb_j}$ of the holonomies along their
duals. A somewhat lengthy but straightforward calculation using
the identities \eqref{dualcurves}, \eqref{frbivect2} and the
$\Ad$-invariance of the bilinear form $\langle\,,\,\rangle$ then
yields the following theorem.
\begin{theorem}
\label{holdualtheor}

Consider functions $f,g\in\cif(H)$ and the associated functions
$f_{m_i}, f_{a_j}, f_{b_j}\in\cif(H^\ntg)$, $g_{\dcm_i},
g_{\dca_j}, g_{\dcb_j}\in\cif(H^\ntg)$ of the holonomies along the
generators $m_i,a_j,b_j\in\pi_1(\surf\mindee)$ and their duals
defined as in \eqref{fxdef}. Then, Fock and Rosly's Poisson
bracket on the manifold $H^\ntg$ is characterised by the following
brackets
\begin{align}
\label{holdualpb}
\{f_x,g_\dcy\}&=-r^{ba}(R_a+L_a)f_x(R_b+L_b)g_\dcy\quad\forall x\in\{m_1,\ldots,b_g\},\; \dcy\in\{\dcm_1,\ldots,\dcb_g\},\; x\neq y\\
\nonumber\\
\{f_{m_i},g_{\dcm_i}\}&=-r^{ba}(R_a+L_a)f_{m_i}(R_b+L_b)g_{\dcm_i}-t^{ab}(R_a+L_a)f_{m_i}\Ad(M_i
\cdots M_1)_b^{\;\;c}
L_cg_{\dcm_i}\\
&=-r^{ba}(R_a+L_a)f_{m_i}(R_b+L_b)g_{\dcm_i}+t^{ab}R_af_{m_i}\Ad(M_{i-1}\cdots
M_1)_b^{\;\;c}
(L_c+R_c)g_{\dcm_i}\nonumber\\
\nonumber\\
\{f_{a_j},g_{\dca_j}\}&=-r^{ba}(R_a+L_a)f_{a_j}(R_b+L_b)g_{\dca_j}-t^{ab}R_af_{a_j}\Ad(\bjj^\inv
H_j\cdots H_1 M_n\cdots M_1)_b^{\;\;c}
L_cg_{\dca_j}\nonumber\\
&=-r^{ba}(R_a+L_a)f_{a_j}(R_b+L_b)g_{\dca_j}+t^{ab}R_af_{a_j}\Ad(
H_{j-1}\cdots H_1 M_n\cdots M_1)_b^{\;\;c}
R_cg_{\dca_j}\\
\nonumber\\
\{f_{b_j},g_{\dcb_j}\}&=-r^{ba}(R_a+L_a)f_{b_j}(R_b+L_b)g_{\dcb_j}+t^{ab}R_af_{b_j}\Ad(\bjj^\inv
H_j\cdots H_1 M_n\cdots M_1)_b^{\;\;c} L_cg_{\dcb_j}\label{endholdualpb}\\
&=-r^{ba}(R_a+L_a)f_{b_j}(R_b+L_b)g_{\dcb_j}-t^{ab}R_af_{b_j}\Ad(\bjj^\inv
A_j H_{j-1}\cdots H_1 M_n\cdots M_1)_b^{\;\;c}
R_cg_{\dcb_j}\nonumber.
\end{align}
\end{theorem}
In expressions \eqref{holdualpb} to \eqref{endholdualpb},
 the Poisson brackets of the functions $f_x, g_{\dcy}\in\cif(H^\ntg)$ are given as a sum of a
global conjugation term involving the classical $r$-matrix and of
a term which depends only on the components $t^{ab}$ of the
bilinear form
 in the Chern-Simons action.  The former
vanishes if one of the two functions is conjugation invariant,
i.~e.~represents a function on the moduli space
$\mathcal{M}_{g,n}^H$. The latter is nontrivial only in the
brackets of functions of a generator $x\in\{m_1,\ldots,b_g\}$ with
functions of its dual $\dcx$. This reflects the fact that the
Wilson loop observables associated to different curves on the
spatial surface have non-vanishing Poisson brackets only if these
curves intersect. As shown in the previous section, the
intersection points of a general curve
$\lambda\in\pi_1(\surf\mindee)$ with the generators $m_i,a_j,b_j$
correspond to factors $\dcm_i^{\pm 1},\dca_j^{\pm 1},\dcb_j^{\pm
1}$ in the expression of $\lambda$ as a reduced word in the dual
generators. The formula \eqref{holdualpb} therefore implies that
each intersection point of a general embedded curve
$\lambda\in\pi_1(\surf\mindee)$ with the generators $a_j,b_j$ and
each pair of intersection points of $\lambda$ with a generator
$m_i$ gives rise to a summand in the Poisson brackets of a
generalised Wilson loop observable associated to $\lambda$ with
functions of the holonomies along the generators $m_i,a_j,b_j$. We
will investigate this dependence on intersection points in more
detail in Sect.~\ref{wloopsect}, where we derive a formula for the
Poisson brackets of generalised Wilson loop observables with
general functions $f\in\cif(H^\ntg)$.

To obtain a more general formulation which clarifies the role of
the involution $I\in\text{Aut}(\pi_1(\surf\mindee))$ in the
description of the moduli space, we consider the diffeomorphism
$\Phi_I: H^\ntg\rightarrow H^\ntg$ induced by $I$. This
diffeomorphism maps the components of $H$ which represent the
holonomies along the generators $m_i,a_j,b_j$ to the holonomies
along their duals
\begin{align}
\label{groupinvol}
&\Phi_I:\;(M_1,\ldots,M_n,A_1,B_1,\ldots,A_g,B_g)\rightarrow
(\dm_1,\ldots,\dm_n,\da_1,\db_1,\ldots,\da_g,\db_g )\\
&\dm_i=M_1^\inv\cdots M_i^\inv M_{i-1}\cdots
M_1\nonumber\\&\da_j=M_1^\inv\cdots H_j^\inv B_j H_{j-1}\cdots
M_1\nonumber\\&\db_j=M_1^\inv\cdots H_j^\inv A_j H_{j-1}\cdots
M_1\nonumber.
\end{align}
More generally, for any $\lambda\in\pi_1(\surf\mindee)$ with dual
$I(\lambda)$,  the holonomy along $I(\Lambda)$ is obtained by
composing the map $\rho_\lambda:H^\ntg\rightarrow H$ with $\Phi_I$
\begin{align}
\label{rhoinvol} \rho_{I(\lambda)}=\rho_\lambda\circ
\Phi_I\qquad\forall \lambda\in\pi_1(\surf\mindee).
\end{align}
Using this identity together with expressions \eqref{holdualpb} to
\eqref{endholdualpb} for the Poisson bracket, we find that the
bracket of any function $f\in\cif(H^\ntg)$ which is invariant
under simultaneous conjugation and general $g\in\cif(H^\ntg)$
takes the form
\begin{align}
\label{helppb} \{f, g\circ \Phi_I\}=& t^{ab}\sum_{i=1}^n R^\mi_a f
\; \Ad(M_{i-1}\cdots M_1)_b^{\;\;c}\left(
(R^\mi_c+L^\mi_c)g\right)\circ \Phi_I\\
+& t^{ab}\sum_{j=1}^g  R^{A_j}_a f\;  \Ad(H_{j-1}\cdots
M_1)_b^{\;\;\;c}\left( R^{A_j}_c g\right)\circ \Phi_I\nonumber\\
-& t^{ab}\sum_{j=1}^g  R^{B_j}_a f\;  \Ad(B_j^\inv
A_jH_{j-1}\cdots M_1)_b^{\;\;\;c} \left(R^{B_j}_c g\right)\circ
\Phi_I\nonumber.
\end{align}
To derive a general formula for the transformation of the Poisson
structure under the involution $I\in\text{Aut}(\surf\mindee)$ and
the associated diffeomorphism $\Phi_I:H^\ntg\rightarrow H^\ntg$,
we express Fock and Rosly's Poisson structure entirely in terms
 of functions $f_{\dcm_i}, f_{\dca_j},
f_{\dcb_j}$ associated to the holonomies along the dual generators
$\dcm_i,\dca_i,\dcb_i\in\pi_1(\surf\mindee)$. Using the
expressions \eqref{dualcurves} for $\dcm_i,\dca_j,\dcb_j$ in terms
of the original generators $m_i,a_j,b_j$ and formulas
\eqref{holdualpb} to \eqref{endholdualpb} for the Poisson bracket,
 we obtain

\begin{align}
\label{dualpb}
\{f_\dcx,g_\dcx\}&=-r^{ba}(R_a+L_a)f_\dcx(R_b+L_b)g_\dcx+t^{ab}(R_a+L_a
)f_\dcx R_bg_\dcx\\
&=r^{ab}_{(a)}(R_a+L_a)f_\dcx(R_b+L_b)g_\dcx-\tfrac{1}{2}t^{ab}(R_af_\dcx
L_b g_\dcx-L_a
f_\dcx R_b g_\dcx )\quad \forall \dcx\in\{\dcm_1,\ldots,\dcb_g\}\nonumber\\
\nonumber\\
\label{ordbrack}
\{f_\dcx,g_\dcy\}&=-r^{ba}(R_a+L_a)f_\dcx(R_b+L_b)g_\dcy\nonumber\qquad\qquad\qquad\qquad\qquad\forall \dcx,\dcy\in\{\dcm_1,\ldots,\dcb_g\}, \dcx<\dcy\\
&=r^{ab}_{(a)}(R_a+L_a)f_\dcx(R_b+L_b)g_\dcy-\tfrac{1}{2}t^{ab}(R_a+L_a)f_\dcx(R_b+L_b)g_\dcy\\
\nonumber\\
\{f_{\dca_j},g_{\dcb_j}\}&=-r^{ba}(R_a+L_a)f_{\dca_j}(R_b+L_b)g_{\dcb_j}+t^{ab}L_af_{\dca_j}
R_bg_{\dcb_j}\\
&=r^{ab}_{(a)}(R_a+L_a)f_{\dca_j}(R_b+L_b)g_{\dcb_j}-\tfrac{1}{2}t^{ab}(R_af_{\dca_j}
L_bg_{\dcb_j}-L_a f_{\dca_j} R_b g_{\dcb_j})\nonumber,
\end{align}
where the ordering in \eqref{ordbrack} is the one obtained by
replacing each generator in the ordering \eqref{order} with its
dual. By comparing these brackets with expressions \eqref{dualpb}
to \eqref{finholf} for the Fock-Rosly Poison brackets of the
functions $f_{m_i}, f_{a_j}, f_{b_j}$ associated to the original
generators, we find that they take the same form up to a flip and
a sign change in the classical $r$-matrix and obtain the following
theorem.
\begin{theorem}
\label{frdualth} The Fock-Rosly Poisson bivector \eqref{frbivect2}
is form-invariant under the simultaneous exchange of the
generators $m_i,a_j,b_j\in\pi_1(\surf\mindee)$ with their duals
$\dcm_i,\dca_j,\dcb_j\in\pi_1(\surf\mindee)$ and of the $r$-matrix
$r=r^{ab}J_a\otimes J_b$ with its flip $
-\sigma(r)=-r^{ba}J_a\otimes J_b$
\begin{align}
\{f\circ \Phi_I, g\circ \Phi_I\}_{r}=\{f,g\}_{-\sigma(r)}\circ
\Phi_I\qquad\forall f,g\in\cif(H^\ntg).
\end{align}
In particular, for any $f\in\cif(H^\ntg)$ invariant under
simultaneous conjugation and arbitrary $g\in\cif(H^\ntg)$ we have
\begin{align}
\label{invid} \{f\circ \Phi_I, g\circ \Phi_I\}=-\{f,g\}\circ
\Phi_I.
\end{align}
\end{theorem}
As the Poisson structure on the moduli space $\mathcal{M}_{g,n}^H$
is given by the Fock-Rosly Poisson brackets of conjugation
invariant functions on $H^\ntg$ which  do not depend on the
antisymmetric part of the $r$-matrix but only the components
$t^{ab}$ of the $\Ad$-invariant symmetric bilinear form, this
implies that the Poisson structure on the moduli space is
invariant under an exchange of the generators $m_i,a_j,b_j$ and
their duals up to a global minus sign. We will demonstrate in
Sect.~\ref{mapsect} that this is the case for any automorphism of
$\pi_1(\surf\mindee)$ which  satisfies the condition
\eqref{invprops2} with $w=1$, $\epsilon=-1$.


\section{Application: The phase space transformations generated by Wilson loop observables}

\label{wloopsect}

\subsection{The Poisson brackets of Wilson loop observables}

In this section, we use the dual generators of the fundamental
group to derive explicit expressions for the Poisson bracket of
generalised Wilson loop observables associated to embedded curves
on the surface $\surf\mindee$ and to determine the associated
flows on the extended phase space $H^\ntg$. For the case of a
surface without punctures, the Poisson brackets of generalised
Wilson loop observables and the associated flows on the moduli
space $\mathcal{M}_{g,0}^H$ were first determined by  Goldman
\cite{goldman} who uses cohomological methods and characterises
these quantities in terms of the intersection behaviour of the
associated curves on $\surf$. The dual generators of the
fundamental group allows us to generalise these results to
punctured surfaces. Moreover, we obtain a purely algebraic
formulation,
 in which these flows are characterised by the transformation of
 the holonomies along our set of generators of the fundamental
 group $\pi_1(\surf\mindee)$ and derived from the expression of
 the associated curves as reduced words in the dual generators.

The first step is to determine the Poisson bracket of a general
function $g\in\cif(H^\ntg)$ with the Wilson loop observable
$f_\lambda$  associated to a conjugation invariant function
$f\in\cif{H}$ and to an element $\lambda\in\pi_1(\surf\mindee)$
with an embedded representative. To calculate the bracket
$\{g,f_\lambda\}$, one inserts the expression
\eqref{lambdadecomp1} of $\lambda$ as a reduced word in the
generators $\dcm_i,\dca_j,\dcb_j$ into the formula \eqref{helppb}
for the Poisson bracket. As the maps $\rho_\lambda:
H^\ntg\rightarrow H$ satisfy the identity
\begin{align}
\label{rhoconj}
\rho_{\tau\circ\lambda\circ\tau^\inv}=\rho_\tau\cdot\rho_\lambda\cdot\rho_\tau^\inv,
\end{align}
we have for any conjugation invariant function $f\in\cif(H)$
\begin{align}
\label{invfid} f(\rho_\lambda(u) \cdot\rho_\tau^\inv(u)
g\rho_\tau(u))=f(\rho_{\tau\circ\lambda\circ\tau^\inv}(u)
g)\qquad\forall g\in H, u\in H^\ntg.
\end{align}
By applying this identity together with the expressions
\eqref{dualcurves} to the terms in \eqref{helppb} involving the
adjoint action, we then obtain the following theorem.

\begin{theorem}

\label{obspb}

Consider an element $\lambda\in\pi_1(\surf\mindee)$ with an
embedded representative given uniquely as a reduced word in the
generators $\dcm_i,\dca_j,\dcb_j\in\pi_1(\surf\mindee)$ by
\begin{align}
\label{lambdadecomp1}
\lambda=\dcx_r^{\alpha_r}\cdots\dcx_1^{\alpha_1}\qquad\dcx_k\in\{\dcm_1,\ldots,\dcb_g\},\;\alpha_k\in\{\pm
1\}
\end{align}
 and let $f\in\cif(H)$ be conjugation
invariant. Then, the Poisson bracket of a general function
$g\in\cif(H^\ntg)$ with the gauge invariant observable
$f_\lambda=f\circ\rho_\lambda$ is given by
\begin{align}
\label{caspb} \{g, f_\lambda\}&= \sum_{i=1}^n t^{ab}(R^\mi_ag
\!+\!L^\mi_a
g)\big(\!\!\!\!\!\!\!\!\!\!\sum_{\dcx_k=\dcm_i,\alpha_k=1}
\!\!\!\!\!\!\!\!\!\!\! R_bf_{\Ad(m_{i-1}...m_1
\dcx_{k-1}^{\alpha_{k-1}}\!...
\dcx_1^{\alpha_1})\lambda}\!-\!\!\!\!\!\!\!\!\!\!\!\!\!\!\sum_{\dcx_k=\dcm_i,\alpha_k=-1}\!\!\!\!\!\!\!\!\!\!\!
R_bf_{\Ad(m_{i}... m_{1} \dcx_{k-1}^{\alpha_{k-1}}\!...
\dcx_1^{\alpha_1})\lambda})\nonumber\\
&+\sum_{j=1}^gt^{ab}
R^\aj_ag\big(\!\!\!\!\!\!\!\!\!\sum_{\dcx_k=\dca_j,\alpha_k=1}
\!\!\!\!\!\!\!\!\!\! R_bf_{\Ad( h_{j-1}...m_1
\dcx_{k-1}^{\alpha_{k-1}}\!...
\dcx_1^{\alpha_1})\lambda}-\!\!\!\!\!\!\!\!\!\!\!\!\sum_{\dcx_k=\dca_j,\alpha_k=-1}\!\!\!\!\!\!\!\!\!\!
 R_bf_{\Ad(a_j^\inv b_j^\inv a_jh_{j-1} ...m_1 \dcx_{k-1}^{\alpha_{k-1}}\!...
\dcx_1^{\alpha_1})\lambda})\\
&-\sum_{j=1}^gt^{ab}
R^\bjj_ag\big(\!\!\!\!\!\!\!\!\!\!\sum_{\dcx_k=\dcb_j,\alpha_k=1}
\!\!\!\!\!\!\!\!\!\!\! R_bf_{\Ad(b_j^\inv a_jh_{j-1}...m_1
\dcx_{k-1}^{\alpha_{k-1}}\!... \dcx_1^{\alpha_1})\lambda}
\!-\!\!\!\!\!\!\!\!\!\!\!\!\!\!\sum_{\dcx_k=\dcb_j,\alpha_k=-1}\!\!\!\!\!\!\!\!\!\!\!
 R_bf_{\Ad(a_i^\inv b_j a_j h_{j-1}\cdots m_1 \dcx_{k-1}^{\alpha_{k-1}}\!...
\dcx_1^{\alpha_1})\lambda}),\nonumber
\end{align}
where we write
$f_{\Ad(\tau)\lambda}=f\circ\rho_{\tau\circ\lambda\circ\tau^\inv}$
\end{theorem}

Note that both, the observable $f_{\Ad(\tau)\lambda}=f_{\lambda}$
and the right-hand-side  of \eqref{caspb}, are invariant under
conjugation $\lambda\rightarrow \tau\circ\lambda\circ \tau^\inv$
with a general element $\tau\in\pi_1(\surf\mindee)$. Although the
factors $\dcm_i,\dca_j,\dcb_j$ in the decomposition of $\tau$ give
rise to additional summands in \eqref{caspb}, their contributions
cancel pairwise. Conversely, two summands in \eqref{caspb} which
cancel each other can arise only if $\lambda$ is of the form
$\lambda=\tau\circ\tilde\lambda\circ \tau^\inv$,
$\tilde\lambda\in\pi_1(\surf\mindee)$ and both of them result from
factors  in the decomposition of $\tau$. As the Poisson bracket of
$f_\lambda$ depends only on the conjugacy class
$[\lambda]=\{\tau\lambda\tau^\inv\;|\;\tau\in\pi_1(\surf\mindee)\}$,
we can simplify calculations by restricting attention to curves
$\lambda\in\pi_1(\surf\mindee)$ whose expression as a reduced word
in $\dcm_i,\dca_j,\dcb_j$ is also cyclically reduced. Hence, we
have obtained an explicit expression for the Poisson bracket of
the gauge invariant observable $f_\lambda$ with a general function
on $H^\ntg$ in terms of functions associated to certain elements
in the conjugacy class of $\lambda$. To achieve a more geometrical
interpretation of the formula \eqref{caspb}, we note that the
summands in \eqref{caspb} are in one-to-one correspondence with
factors $\dcx_k=\dcm_i,\dca_j,\dcb_j$ in the decomposition
\eqref{lambdadecomp1} of $\lambda$ and their signs - modulo an
overall sign for generators  $\dcb_j$ - are given by the
corresponding exponents $\alpha_k$. In Sect.~\ref{dualint}, we
showed that factors $\dcx_k=\dcm_i,\dca_j,\dcb_j$ in the
expression \eqref{lambdadecomp0} correspond to, respectively,
intersection points of $\lambda$ with the generators $m_i,a_j,b_j$
and their exponent $\alpha_k$ determines the oriented intersection
number. Furthermore, the factors $\tau$ in expressions of the form
$f_{\Ad(\tau)\lambda}$ in \eqref{caspb} are precisely the ones
 in expressions \eqref{msplit1}, \eqref{msplit2},
\eqref{splitabword1}, \eqref{splitabword2} in Theorem
\ref{intasth}, which give the splitting of the
 generators $\dcm_i,\dca_j,\dcb_j$ as reduced words in $m_i,a_j,b_j$  to assign these intersection points between the different
factors in the expression of $\lambda$. Note also that the
ambiguity in moving these intersection points to the either the
starting point or endpoint of the generators $a_j,b_j$ which we
encountered in Sect.~\ref{factassign} is reflected in formula
\eqref{caspb}. Using the $Ad$-invariance of the bilinear form
$\langle\,,\,\rangle$, the conjugation invariance of $f$ and the
identity \eqref{rhoconj}, we find
\begin{align}
\label{leftrightshift}
t^{ab}R^a_XgR^bf_{\Ad(\tau)\lambda}=-t^{ab}L^a_X g R^b
f_{\Ad(x\tau)\lambda}\qquad\forall x\in\{a_1,b_1,\ldots,a_g,b_g\},
g\in\cif(H^\ntg),
\end{align}
which corresponds to shifting an intersection point at the
starting point of $x\in\{a_1,b_1,\ldots,a_g,b_g\}$ to its
endpoint. By applying this identity to all factors
$\dcx_k^{\alpha_k}\in\{\dca_1,\dcb_1,\ldots,\dca_g,\dcb_g\}$ where
the corresponding intersection point is located at the endpoint of
a generator $a_j,b_j$, we reproduce the assignment in Theorem
\eqref{intasth} and obtain
\begin{corollary}
\label{holpbth} Consider an element
$\lambda\in\pi_1(\surf\mindee)$ with an embedded representative
and given uniquely as a product in the generators
$m_i,a_j,b_j\in\pi_1(\surf\mindee)$ and their duals by
\begin{align}
\label{lambdform}
\lambda=\dcx_r^{\alpha_r}\cdots\dcx_1^{\alpha_1}=
z_t^{\delta_t}\circ\ldots\circ z_1^{\delta_1}\qquad
x_i,z_j\in\{m_1,\ldots,b_g\}, \alpha_i,\delta_j\in\{\pm 1\}.
\end{align}
Assign the intersection points of $\lambda$  with $m_i,a_j,b_j$
 between the different factors $z_k^{\delta_k}$ in \eqref{lambdform}
and to the starting and endpoints of the generators $m_i,a_j,b_j$
as in Theorem \ref{intasth}.  Then, the Poisson bracket
\eqref{caspb} can be written as
\begin{align}
\label{intpb0} &\{g, f_\lambda\}=\\
&\sum_{i=1}^n
t^{ab}\left(\!\!\!\!\!\!\!\!\!\!\!\!\!\!\!\!\!\!\!\!\!\!\sum_{\qquad\qquad
k: t_{k}s_{k+1}\cap m_i=s_{m_i}}
\!\!\!\!\!\!\!\!\!\!\!\!\!\!\!\!\!\!\!\!\!\!
\epsilon(m_i,t_ks_{k+1})R^\mi_agR_bf_{\Ad(z_{k}^{\delta_k}\cdots
z_1^{\delta_1})\lambda}\!-\!\!\!\!\!\!\!\!\!\!\!\!\!\!\!\!\!\!\!\!\!\!\!\!\!\!\!\!\sum_{\qquad\qquad
k: t_{k+1}s_k\cap
m_i=t_{m_i}}\!\!\!\!\!\!\!\!\!\!\!\!\!\!\!\!\!\!\!\!\!\!
\epsilon(m_i,t_ks_{k+1}) L^\mi_agR_bf_{\Ad(z_{k}^{\delta_k}\cdots z_1^{\delta_1})\lambda}\right)\nonumber\\
&+\sum_{j=1}^g t^{ab}
\left(\!\!\!\!\!\!\!\!\!\!\!\!\!\!\!\!\!\!\!\!\!\!\!\sum_{\qquad\qquad
k: t_{k}s_{k+1}\cap a_j=s_{a_j}}\!\!\!\!\!\!\!\!\!\!\!
\!\!\!\!\!\!\!\!\!\!\!
\epsilon(a_j,t_ks_{k+1})R^\aj_agR_bf_{\Ad(z_{k}^{\delta_k}\cdots
z_1^{\delta_1})\lambda}\!-\!\!\!\!\!\!\!\!\!\!\!\!\!\!\!\!\!\!\!\!\!\!\!\!\!\!\!\sum_{\qquad\qquad
k: t_{k+1}s_k\cap
a_j=t_{a_j}}\!\!\!\!\!\!\!\!\!\!\!\!\!\!\!\!\!\!\!\!\!\!\epsilon(a_j,t_ks_{k+1})
L^\aj_agR_bf_{\Ad(z_{k}^{\delta_k}\cdots
z_1^{\delta_1})\lambda}\right)\nonumber\\
&+\sum_{j=1}^g
t^{ab}\left(\!\!\!\!\!\!\!\!\!\!\!\!\!\!\!\!\!\!\!\!\!\!\sum_{\qquad\qquad
k: t_{k}s_{k+1}\cap b_j=s_{b_j}}
\!\!\!\!\!\!\!\!\!\!\!\!\!\!\!\!\!\!\!\!\!\!
\epsilon(b_j,t_ks_{k+1})R_a^\bjj g
R_bf_{\Ad(z_{k}^{\delta_k}\cdots
z_1^{\delta_1})\lambda}\!-\!\!\!\!\!\!\!\!\!\!\!\!\!\!\!\!\!\!\!\!\!\!\!\!\!\!\!\sum_{\qquad\qquad
k: t_{k+1}s_k\cap b_j=t_{b_j}}\!\!\!\!\!\!\!\!\!\!\!
\!\!\!\!\!\!\!\!\!\!\!\epsilon(b_j,t_ks_{k+1})
L^\bjj_agR_bf_{\Ad(z_{k}^{\delta_k}\cdots
z_1^{\delta_1})\lambda}\right),\nonumber
\end{align}
where $\epsilon({x}, t_ks_{k+1})$ stands for the oriented
intersection number of $x\in\{m_1,\ldots,b_g\}$ with the oriented
segment $t_ks_{k+1}$, and we write $t_ks_{k+1}\cap x=s_x$ if the
intersection point of the oriented segment $t_ks_{k+1}$ with $x$
is located at the starting point of $x$ and $t_ks_{k+1}\cap x=t_x$
if it is located at the endpoint.
\end{corollary}

Corollary \eqref{holpbth}  allows us to derive an explicit
expression for the Poisson bracket of the Wilson loop observables
associated to elements $\lambda,\eta\in\pi_1(\surf\mindee)$ with
embedded representatives. By applying formula \eqref{intpb0} to
the Wilson loop observable $g_\eta$  associated to a conjugation
invariant function $g\in\cif(H)$ and $\eta\in\pi_1(\surf\mindee)$,
we find that if a given segment $t_ks_{k+1}$ intersects a factors
in the expression for $\eta$ as a reduced word in $m_i,a_j,b_j$ at
its endpoint and the next factor in its starting point with
opposite intersection number, the contributions of these factors
cancel. Recalling the discussion after Theorem \ref{grequith}, we
note that this corresponds to the graphical procedure for removing
unnecessary intersection points of the curves representing $\eta$
and $\lambda$.
 Using the identity
\eqref{leftrightshift}, we can express all vector fields acting on
$g_\eta$ in terms of the left-invariant vector fields for the
different arguments and transform them into expressions of the
form $g_{\Ad(\tau)\eta}$ via \eqref{invfid}. We obtain the
following corollary.

\begin{corollary}
\label{genholth} $\quad$

Consider conjugation invariant functions $f,g\in\cif(H)$ and
elements $\eta,\lambda\in\pi_1(\surf\mindee)$ with embedded
representatives and given uniquely as reduced words in the
generators $m_i,a_j,b_j$ by
\begin{align}
\label{etalambdadef} &\eta=y_s^{\beta_s}\circ \cdots\circ
y_1^{\beta_1}\qquad \lambda=z_t^{\delta_t}\circ\ldots\circ
z_1^{\delta_1}\qquad y_i,z_j\in\{m_1,\ldots,b_g\},
\beta_i,\delta_j\in\{\pm 1\}.
\end{align}
Let ${p_k}$, $k=1,\ldots,m$, denote the intersection points of
$\lambda$ and $\eta$ such that $p_k$ occurs between the factors
$y_{i_k}^{\beta_{i_k}}$ and $y_{i_k+1}^{\beta_{i_k+1}}$ in the
expression \eqref{etalambdadef} of $\eta$ and between the factors
$z_{j_k}^{\delta_{j_k}}$ and $z_{j_k+1}^{\delta_{j_k+1}}$ in the
expression of $\lambda$ and denote by
$\epsilon_k(\eta,\lambda)=\epsilon(\eta,\lambda,p_k)$ the oriented
intersection number of $\eta$ and $\lambda$ in $p_k$. Then, the
Poisson bracket of the observables $g_\eta$ and $f_\lambda$ is
given by
\begin{align}
\label{intform} \{g_\eta,f_\lambda\}=\sum_{k=1}^m
\epsilon_k(\eta,\lambda)\,
t^{ab}\,(R_ag)_{(y_{i_k}^{\beta_{i_k}}\cdots y_1^{\beta_1})\circ
\eta\circ(y_{i_k}^{\beta_{i_k}}\cdots y_1^{\beta_1})^\inv} (R_b
f)_{(z_{j_k}^{\delta_{j_k}}\cdots z_1^{\delta_1})\circ
\lambda\circ(z_{j_k}^{\delta_{j_k}}\cdots z_1^{\delta_1})^\inv}.
\end{align}
\end{corollary}
For surfaces $S_{g,0}$ without punctures, formula \eqref{intform}
gives an algebraic version of Goldman's product formula, see
Theorem 3.5. in \cite{goldman}. Moreover, it generalises this
formula to punctured surfaces not considered in \cite{goldman}.
The concept of duality for a set of generators of the fundamental
group therefore establishes a link between the purely algebraic
description of the Poisson brackets of generalised Wilson loop
observables and the geometrical formulation in terms of
intersection points derived in \cite{goldman}. We will find in the
next subsection, that this description can be used to obtain an
algebraic formula of the associated flows on phase space.

\subsection{The flows generated by the Wilson loop observables: the geometrical formulation}

After determining the Poisson bracket of generalised Wilson loop
observables with general functions $g\in\cif(H^\ntg)$, we will now
derive the associated one parameter groups of diffeomorphisms of
$H^\ntg$ these observables generate via the Poisson bracket.

In the following we will often parametrise elements of the group
$H$ in terms of the exponential map $\exp:\gothh\rightarrow H$.
While we require this map to be surjective throughout this paper,
we will not suppose that its injectivity. This implies that the
exponential map is locally but not globally bijective and that the
parametrisation of group elements in terms of Lie algebra elements
is in general not unique. Following Goldman \cite{goldman}, we
define for any conjugation
 invariant function $f\in\cif(H)$ a Lie-algebra valued map
$g_f:\,H\rightarrow \gothh$ and  an associated one-parameter group
of diffeomorphisms $G^t_f: H\rightarrow H$
\begin{align}
\label{gfdef} &\langle
g_f(u),J_a\rangle=R_{a}f(u)=\frac{d}{dt}|_{t=0}f(ue^{t J_a})\quad
G_f^t(u)=e^{tg_f(u)}\qquad\forall u\in H,\;
a=1,\ldots,\text{dim}\,\gothh,
\end{align}
where the parameter $t\in\RR$ is restricted appropriately to
ensure the bijectivity of the exponential map. As the bilinear
form  $\langle\,,\,\rangle$ is non-degenerate, equation
\eqref{gfdef} defines $g_f$ uniquely as
\begin{align}
g_f(u)=t^{ab}R_af(u)J_b,
\end{align} and from the $\Ad$-invariance of the bilinear
form $\langle\,,\,\rangle$ it follows that the Lie algebra valued
functions $g_f: H\rightarrow\gothh$ and the associated
diffeomorphisms $G^t_f:H\rightarrow H$ satisfy the covariance
conditions
\begin{align}
\label{covcond} &g_f(gug^\inv)=gg_f(u)g^\inv &
&G^t_f(gug^\inv)=gG^t_f(u)g^\inv\qquad\forall g,u\in H.
\end{align}
For the  functions
\begin{align}
\label{glamdef} G_{f,\lambda}^t=G_f^t\circ\rho_\lambda:
H^\ntg\rightarrow H,
\end{align}
 obtained by composing these diffeomorphisms
with the map $\rho_\lambda:H^\ntg\rightarrow H$ to the holonomy
along $\lambda$, this covariance condition and the identity
\eqref{rhoconj} imply
\begin{align}
\label{pi1gconj}
G_{f,\tau\circ\lambda\circ\tau^\inv}^t=\rho_\tau\cdot
G_{f,\lambda}^t\cdot \rho_\tau^\inv.
\end{align}
We now consider an element $\lambda\in\pi_1(\surf\mindee)$ with an
embedded representative and given as a reduced word in the
generators $m_i,a_j,b_j$ as in \eqref{etalambdadef}. Corollary
\ref{holpbth} then implies that the Poisson bracket of a Wilson
loop observable $f_\lambda$ associated to a conjugation invariant
function $f\in\cif(H)$ with functions $g_y$, $g\in\cif(H)$,
$y\in\{m_1,\ldots,b_g\}$ of the holonomies along the generators is
given by
\begin{align}
\label{firstpb} \{g_y,f_\lambda\}=& t^{ab}R_ag_y \left(\sum_{k:
t_ks_{k+1}\cap y=s_y}\!\!\!\!\!\!\!\!\epsilon(y, t_ks_{k+1}) R_b
f_{\lambda_k}\right)-t^{ab}L_ag_y \left(\sum_{k: t_ks_{k+1}\cap
y=t_y}\!\!\!\!\!\!\!\!\epsilon(y, t_ks_{k+1}) R_b
f_{\lambda_k}\right)\\
=&\!\!\!\!\!\!\!\!\sum_{k: t_ks_{k+1}\cap
y=s_y}\!\!\!\!\!\!\!\!\epsilon(y, t_ks_{k+1}) \frac{d}{dt}|_{t=0}
g\circ \left( \rho_y\cdot G^t_{f,\lambda_k}\right) +
\!\!\!\!\!\!\!\!\sum_{k: t_ks_{k+1}\cap
y=s_y}\!\!\!\!\!\!\!\!\epsilon(y, t_ks_{k+1}) \frac{d}{dt}|_{t=0}
g\circ \left( G^t_{f,\lambda_k}\cdot\rho_y \right),\nonumber
\end{align}
where $\lambda_k$ are the  cyclic permutations of the expression
\eqref{etalambdadef} of $\lambda$ as a reduced word in
$m_i,a_j,b_j$
\begin{align}
\label{lambdaidef} \lambda_k:=(z_k^{\delta_k}\cdots
z_1^{\delta_1})\lambda(z_k^{\delta_k}\cdots
z_1^{\delta_1})^\inv=z_k^{\delta_k}\cdots
z_1^{\delta_1}z_r^{\delta_r}\cdots
z_{k+1}^{\delta_{k+1}}\qquad\forall k=0,\ldots,t.
\end{align}
This suggests that the flow  generated by the Wilson loop
observable $f_\lambda$ acts on the holonomies along the generators
$m_i,a_j,b_j$ by right-multiplication with the functions
$G^t_{f,\lambda_k}$ associated to all segments $t_k s_{k+1}$ which
intersect the generators at its starting point and by
left-multiplication with the functions $G^t_{f,\lambda_k}$ for
segments which intersect its endpoint. It turns out that this is
the case and that the ordering of these factors
$G^t_{f,\lambda_k}$ is given by the order of the associated
intersection points on the generator.
\begin{theorem}
\label{flowth}

Let $\lambda\in\pi_1(\surf\mindee)$ be an element with an embedded
representative, given as a reduced word in the generators
$m_i,a_j,b_j$ as in \eqref{etalambdadef} and with associated
cyclic permutations \eqref{lambdaidef}. Represent $\lambda$
graphically as described in Sect.~\ref{factassign} and consider
its intersection points with a generator $y\in\{m_1,\ldots,b_g\}$.
Denote by $t_{i_1}s_{i_1+1},\ldots, t_{i_k}s_{i_k+1}$ the segments
of $\lambda$ which intersect $y$ above its starting point such
that $t_{i_n}s_{i_n+1}$ lies below $t_{i_m}s_{i_m+1}$ for $n<m$
and by $t_{j_1}s_{j_1+1},\ldots, t_{j_l}s_{j_l+1}$ the segments
which intersect $y$ above its endpoint such that
$t_{j_n}s_{j_n+1}$ lies below $t_{j_m}s_{j_m+1}$ for $n<m$. Let
$\epsilon_{i_n}=\epsilon(y, t_{i_n}s_{i_n+1})$,
$\epsilon_{j_n}=\epsilon(y, t_{j_n}s_{j_n+1})$ the associated
oriented intersection numbers. Then, the
 one-parameter group of
diffeomorphisms
\begin{align}
\label{poissid} &T_{f,\lambda}^t:H^\ntg\rightarrow H^\ntg\\
&\frac{d}{dt}\; g\circ
T_{f,\lambda}^t|_{t=0}=\{g,f_\lambda\}\qquad \forall
g\in\cif(H^\ntg).\nonumber
\end{align}
generated by the Wilson loop observable $f_\lambda$ acts on the
group element representing the holonomy along $y$ according to
\begin{align}
\label{transfdef} T_{f,\lambda}^t: Y\mapsto
G_{f,\lambda_{j_1}}^{t\epsilon_{j_1}}
G_{f,\lambda_{j_2}}^{t\epsilon_{j_{2}}}\cdots
G_{f,\lambda_{j_l}}^{t\epsilon_{j_l}}(M_1,\ldots,B_g)\cdot\;
Y\;\cdot G_{f,\lambda_{i_k}}^{t\epsilon_{i_k}}\cdots
G_{f,\lambda_{i_2}}^{t\epsilon_{i_2}}
G_{f,\lambda_{i_1}}^{t\epsilon_{i_1}}(M_1,\ldots,B_g),
\end{align}
where $G_{f,\lambda}^t: H^\ntg\rightarrow H$ is given by
\eqref{glamdef} and $\lambda_k$ by \eqref{lambdaidef}.
\end{theorem}

{\bf Proof:}  To prove the theorem, we need to show that the
derivatives at $t=0$ of the functions in \eqref{transfdef} agree
with the Poisson brackets \eqref{intpb0} and that the expression
\eqref{transfdef} defines a one-parameter group of diffeomorphism.
The first statement follows directly from formula \eqref{firstpb}.
To show
 that the maps $T_{f,\lambda}^t: H^\ntg\rightarrow
H^\ntg$  define a one-parameter group of diffeomorphisms, we have
to determine how the group elements associated to the curves
$\lambda_{i_n}$ and $\lambda_{j_n}$ transform under
$T_{f,\lambda}^t$. For this we note that these curves can be
obtained by conjugating the curve representing $\lambda$ with a
vertical segment from the line containing the starting and
endpoints to the segment $t_{z_{i_n}}s_{z_{i_{n}+1}}$,
$t_{z_{j_n}}s_{z_{j_{n}+1}}$ as shown in Fig.~\ref{csex}. This
vertical segment intersects  the horizontal segments
$t_{z_{i_m}}s_{z_{i_m}+1}$, $t_{z_{j_m}}s_{z_{j_m}+1}$ with $m<n$.
Hence, we find that the transformation of the holonomies along
$\lambda_{i_n}$, $\lambda_{j_n}$ is given by
\begin{align}
&\rho_{\lambda_{i_n}}\circ T^t_{f,\lambda}=
G_{f,\lambda_{i_1}}^{-t\epsilon_{i_1}}
G_{f,\lambda_{i_2}}^{-t\epsilon_{i_2}} \cdots
G_{f,\lambda_{i_{n-l}}}^{-t\epsilon_{i_{n-1}}}\cdot\;
\rho_{\lambda_{j_n}} \;\cdot
G_{f,\lambda_{i_{n-l}}}^{t\epsilon_{i_{n-1}}} \cdots
G_{f,\lambda_{i_2}}^{t\epsilon_{i_2}}
G_{f,\lambda_{i_1}}^{t\epsilon_{i_1}}\qquad n=0,\ldots,k\\
&\rho_{\lambda_{j_n}}\circ T^t_{f,\lambda}=
G_{f,\lambda_{j_1}}^{t\epsilon_{j_1}}
G_{f,\lambda_{j_2}}^{t\epsilon_{j_2}} \cdots
G_{f,\lambda_{j_{n-l}}}^{t\epsilon_{j_{n-1}}}\cdot\;
\rho_{\lambda_{j_n}} \;\cdot
G_{f,\lambda_{j_{n-l}}}^{-t\epsilon_{j_{n-1}}} \cdots
G_{f,\lambda_{j_2}}^{-t\epsilon_{j_2}}
G_{f,\lambda_{j_1}}^{-t\epsilon_{j_1}}\qquad
n=0,\ldots,l\nonumber,
\end{align}
which implies
\begin{align}
\label{gftrafo} G_{f,\lambda_{i_n}}^s\circ
T_{f,\lambda}^t=G_{f,\lambda_{i_1}}^{-t\epsilon_{i_1}}
G_{f,\lambda_{i_2}}^{-t\epsilon_{i_2}} \cdots
G_{f,\lambda_{i_{n-l}}}^{-t\epsilon_{i_{n-1}}}\cdot\;
G_{f,\lambda_{i_n}}^s \;\cdot
G_{f,\lambda_{i_{n-l}}}^{t\epsilon_{i_{n-1}}} \cdots
G_{f,\lambda_{i_2}}^{t\epsilon_{i_2}}
G_{f,\lambda_{i_1}}^{t\epsilon_{i_1}}\qquad n=0,\ldots,k\\
G_{f,\lambda_{j_n}}^s\circ
T_{f,\lambda}^t=G_{f,\lambda_{j_1}}^{t\epsilon_{j_1}}
G_{f,\lambda_{j_2}}^{t\epsilon_{j_2}} \cdots
G_{f,\lambda_{j_{n-l}}}^{t\epsilon_{j_{n-1}}}\cdot\;
G_{f,\lambda_{j_n}}^s \;\cdot
G_{f,\lambda_{j_{n-l}}}^{-t\epsilon_{j_{n-1}}} \cdots
G_{f,\lambda_{j_2}}^{-t\epsilon_{j_2}}
G_{f,\lambda_{j_1}}^{-t\epsilon_{j_1}}\qquad
n=0,\ldots,l\nonumber.
\end{align} By
inserting \eqref{gftrafo} into the definition \eqref{transfdef} of
$T^t_{f,\lambda}$ we obtain after some calculation
\begin{align}
T_{f,\lambda}^s\circ
T_{f,\lambda}^t=T_{f,\lambda}^{s+t}\qquad\forall
t,s\in\RR.\qquad\qquad\qquad\qquad\qquad\qquad\Box
\end{align}

Hence, we find that the one parameter group of diffeomorphisms of
$H^\ntg$ generated by a generalised Wilson loop observable
$f_\lambda$ acts on the holonomies along the generators of the
fundamental group by left- and right-multiplication with the
functions $G_{f,\lambda_k}^t$ associated to the segments
$t_ks_{k+1}$ which intersect the corresponding curve on the
surface. The ordering of the different factors $G_{f,\lambda_k}^t$
is given by the vertical ordering of these segments, which agrees
with the order in which the intersection points occur on the
generator for intersection points above its starting point and is
the opposite for intersection points above its endpoint. The
 unique vertical ordering of the segments $t_ks_{k+1}$ in the graphical
representation of $\lambda$ is therefore crucial to ensure that
expression \eqref{transfdef} defines a one-parameter group of
diffeomorphisms.  This unique ordering of the segments is a direct
consequence of the fact that $\lambda$ can be represented by an
embedded curve. While the formulas \eqref{caspb}, \eqref{intpb0},
\eqref{intform} for the Poisson brackets are valid for general
elements $\lambda\in\pi_1(\surf\mindee)$, the associated
one-parameter groups of diffeomorphisms on $H^\ntg$ generated by
the observables $f_\lambda$ via the Poisson bracket are not of the
form \eqref{transfdef} for elements without embedded
representatives.

The fact that the transformations $T^t_{f,\lambda}:
H^\ntg\rightarrow H^\ntg$ are generated via the Poisson bracket
allows one to immediately deduce some of their properties. We have
the following corollary, see also the discussion in \cite{goldman}
\begin{corollary} \label{properties}Consider
$\lambda,\eta\in\pi_1(\surf\mindee)$ with embedded representatives
and general conjugation invariant functions $f,h\in\cif(H)$. Then,
the associated one-parameter groups of diffeomorphisms
$T^t_{f,\lambda}, T^t_{h,\eta}: H^\ntg\rightarrow H^\ntg$ given by
\eqref{transfdef} have the following properties
\begin{enumerate}
\item $T^t_{f,\lambda}$ acts by Poisson isomorphisms
\begin{align}
\label{poissisom} \{g\circ T^t_{f,\lambda}, k\circ
T^t_{f,\lambda}\}=\{g,k\}\circ T^t_{f,\lambda}\qquad\forall
g,k\in\cif(H^\ntg).
\end{align}
\item $T^t_{f,\lambda}$ leaves the constraint \eqref{holrel}
invariant, commutes with simultaneous conjugation of all arguments
of $H^\ntg$ by $H$ and acts on the first $n$ components by
conjugation. \item The transformations $T^t_{f,\lambda}$,
$T^t_{h,\eta}$ commute if and only if
\begin{align}
\langle g_f(u), g_h(u)\rangle=0\qquad\forall u\in
H^\ntg\label{gcommute}
\end{align}
or  $\eta$ and $\lambda$ are conjugated to elements with
representatives that do not intersect.
\end{enumerate}
\end{corollary}

{\bf Proof: } Identity \eqref{poissisom} is a direct consequence
of the fact that the transformations $T^t_{f,\lambda}$ are
generated by  Hamiltonians.  That it leaves the constraint
invariant and commutes with the associated gauge transformations
follows from the fact that it is generated by a gauge invariant
function which Poisson commutes with functions of the constraint
\eqref{holrel}. That it acts on the first $n$ components of
$H^\ntg$ by conjugation follows directly from formula
\eqref{caspb}, \eqref{intpb0} for the Poisson bracket. Finally, we
note that
\begin{align}
\frac{d^2}{dsdt} h\circ T^t_{f,\lambda}\circ
T^s_{g,\eta}|_{t=s=0}-\frac{d^2}{dsdt} h\circ T^s_{g,\eta}\circ
T^t_{f,\lambda}|_{t=s=0}=\{\{h,f_\lambda\}, g_\eta\}-
\{\{h,g_\eta\}, f_\lambda\}=\{h,\{f_\lambda,g_\eta\}\}.\nonumber
\end{align}
From expression \eqref{intform} for the Poisson bracket and taking
into account the identity $t^{ab}R_af(u)R_bh(u)=\langle
g_f(u),g_h(u) \rangle$ $\forall u\in H$ one then obtains
\eqref{gcommute}. \hfill$\Box$

\subsection{The flows generated by the Wilson loop observables: the algebraic formulation}

Formula \eqref{transfdef} gives an explicit expression for the
transformation generated by the generalised Wilson loop
observables associated to  elements
$\lambda\in\pi_1(\surf\mindee)$ with embedded representatives.
However, it  still relies  on a graphical procedure to determine
the order in which the intersection points of $\lambda$ occur on
the representatives of each generator $m_i,a_j,b_j$. We will now
use the dual generators $\dcm_i,\dca_j,\dcb_j$ to derive a purely
algebraic formulation, in which the flows $T^t_{f,\lambda}$ are
characterised entirely in terms of the expression of $\lambda$ as
a reduced word in the dual generators $\dcm_i,\dca_j,\dcb_j$. For
this, we recall the discussion from Sect.~\ref{intordersect} where
we demonstrated how the order in which the intersection points
occur on each generator $m_i,a_j,b_j$ can be derived from the
expression of $\lambda$ as a cyclically reduced word in the dual
generators $\dcm_i,\dca_j,\dcb_j$
\begin{align}
\label{lambdadecomp4}
\lambda=\dcx_r^{\alpha_r}\cdots\dcx_1^{\alpha_1}\qquad\dcx_r^{\alpha_r}\neq\dcx_1^{-\alpha_1}.
\end{align}
It is shown there that the intersection points of $\lambda$ with
generators $x\in\{a_1,\ldots, b_g\}$ are in one to
one-correspondence with cyclic permutations
$\tau\in\text{CPerm}(\lambda)$ with $\text{LF}(\tau)=\dcx$ and
that each element  $\tau\in\text{CPerm}(\lambda)$ with
$\text{LF}(\tau)=\dcm_i$ corresponds to a pair of intersection
points of $\lambda$ with $m_i$, one at its starting point and one
at its endpoint and with opposite intersection numbers. For a
given factor $\dcx_k^{\alpha_k}$ in \eqref{lambdadecomp4}, the
associated element $\tau\in\text{CPerm}(\lambda)$ with
$\text{LF}(\tau)=\dcx_k$ is given by
\begin{align}
\label{tauassoc} \tau=\begin{cases}
\dcx_{k-1}^{\alpha_{k-1}}\cdots\dcx_1^{\alpha_1}\dcx_r^{\alpha_r}\cdots\dcx_k^{\alpha_k}
& \alpha_k=1\\
(\dcx_{k}^{\alpha_{k}}\cdots\dcx_1^{\alpha_1}\dcx_r^{\alpha_r}\cdots\dcx_{k+1}^{\alpha_{k+1}})^\inv
& \alpha_k=-1.
\end{cases}
\end{align}
The elements $\lambda_{i_n}$, $\lambda_{j_n}$  in
\eqref{transfdef} are obtained from the cyclic permutations $\tau$
via the assignment of intersection points between the different
factors in the expression of $\lambda$ as a reduced word in the
generators $m_i,a_j,b_j$ given in Theorem \ref{intasth}. Using the
 formulas \eqref{msplit1}, \eqref{msplit2}, \eqref{splitabword1},
\eqref{splitabword2} for the splitting of the dual generators and
setting
 $\text{sgn}(\tau)=1$ if $\tau$ is a cyclic
permutation of $\lambda$ as a cyclically reduced word in
$\dcm_i,\dca_j,\dcb_j$ and $\text{sgn}(\tau)=-1$ if it is a cyclic
permutation of $\lambda^\inv$, we find that the cyclic
permutations associated to the segments $\lambda_{i_n}$ which
intersect the generators at their starting points are given by
\begin{align}
\label{ltaurel} \lambda_{i_n}=\begin{cases} (m_{i-1}\cdots
m_1)\tau_{i_n}^{\text{sgn}(\tau_{i_n})}(m_{i-1}\cdots m_1)^\inv &
\text{for LF}(\tau_{i_n})=\dcm_i\\
(h_{j-1}\cdots
m_1)\tau_{i_n}^{\text{sgn}(\tau_{i_n})}(h_{j-1}\cdots m_1)^\inv &
\text{for LF}(\tau_{i_n})=\dca_j\\
(b_j^\inv a_jh_{j-1}\cdots
m_1)\tau_{i_n}^{\text{sgn}({\tau_{i_n}})}(b_j^\inv
a_jh_{j-1}\cdots m_1)^\inv & \text{for
LF}(\tau_{i_n})=\dcb_j\end{cases}.
\end{align}
Similarly, the cyclic permutations for segments $\lambda_{j_n}$
which intersect the generators at their endpoints take the form
\begin{align}
\label{ltaurel2} \lambda_{j_n}=\begin{cases} (m_{i-1}\cdots
m_1)\tau_{i_n}^{\text{sgn}(\tau_{j_n})}(m_{i-1}\cdots m_1)^\inv &
\text{for LF}(\tau_{j_n})=\dcm_i \\(a_jh_{j-1}\cdots
m_1)\tau_{j_n}^{\text{sgn}(\tau_{j_n})}(a_jh_{j-1}\cdots m_1)^\inv
& \text{for LF}(\tau_{j_n})=\in\{\dca_j,\dcb_j\}\end{cases}.
\end{align}
Furthermore, we note that the oriented intersection numbers
$\varepsilon_{i_n}$, $\varepsilon_{j_n}$ in \eqref{transfdef} are
given by $\text{sgn}(\tau)$ for $\text{LF}(\tau)=\dca_j$, by
$-\text{sgn}(\tau)$ for $\text{LF}(\tau)=\dcb_j$. For
$\text{LF}(\tau)=\dcm_i$, $\text{sgn}(\tau)$ gives the oriented
intersection number of the intersection point at the starting
point of $m_i$, which is the opposite of the one for its endpoint.

Hence, if we denote by $\tau_{i_n},
\tau_{j_n}\in\text{CPerm}(\lambda)$ the cyclic permutations
associated via \eqref{tauassoc}, respectively, to elements
$\lambda_{i_n}\in\pi_1(\surf\mindee)$ at the right of $Y$ and
elements $\lambda_{j_n}\in\pi_1(\surf\mindee)$  in
\eqref{transfdef}, we find using \eqref{pi1gconj}
\begin{align}
&G^{t\varepsilon_{i_n}}_{f,\lambda_{i_n}}=\begin{cases}
(M_{i-1}\cdots M_1) G^{t\,\text{sgn}(\tau_{i_n})}_{f,|\tau_{i_n}|}
(M_{i-1}\cdots M_1)^\inv & \text{for
LF}(\tau_{i_n})=\dcm_i\\
(H_{j-1}\cdots M_1) G^{t\,\text{sgn}(\tau_{i_n})}_{f,|\tau_{i_n}|}
(H_{j-1}\cdots M_1)^\inv & \text{for
LF}(\tau_{i_n})=\dca_j\\
(B_j^\inv A_jH_{j-1}\cdots M_1)
G^{-t\,\text{sgn}(\tau_{i_n})}_{f,|\tau_{i_n}|} (B_j^\inv A_j
H_{j-1}\cdots M_1)^\inv & \text{for LF}(\tau_{i_n})=\dcb_j
\end{cases}\\
&G^{t\varepsilon_{j_n}}_{f,\lambda_{j_n}}=\begin{cases}
(M_{i-1}\cdots M_1)
G^{-t\,\text{sgn}(\tau_{i_n})}_{f,|\tau_{i_n}|} (M_{i-1}\cdots
M_1)^\inv & \text{for
LF}(\tau_{j_n})=\dcm_i\\
(A_jH_{j-1}\cdots M_1)
G^{t\,\text{sgn}(\tau_{i_n})}_{f,|\tau_{j_n}|} ( A_j H_{j-1}\cdots
M_1)^\inv & \text{for LF}(\tau_{j_n})=\dca_j\\
(A_jH_{j-1}\cdots M_1)
G^{-t\,\text{sgn}(\tau_{i_n})}_{f,|\tau_{j_n}|} ( A_j
H_{j-1}\cdots
M_1)^\inv & \text{for LF}(\tau_{j_n})=\dcb_j\\
\end{cases}\nonumber
\end{align}
where $|\tau|=\tau^{\text{sgn}(\tau)}$ for
$\tau\in\text{CPerm}(\lambda)$. We now  move all factors
$G^{t\varepsilon_{j_n}}_{f, \lambda_{j_n}}$ in \eqref{transfdef}
to the right of $Y=A_j$ by conjugating them with $A_j^\inv$ and
all factors $G^{t\varepsilon_{i_n}}_{f, \lambda_{i_n}}$ to the
left of $Y=B_j$ by conjugating them with $B_j^\inv$.
Fig.~\ref{cut2} implies that the order of the intersection points
on the generator $a_j$ is the order of the associated intersection
points on the side $a_j$ of the polygon $P^D_{g,n}$ and that the
order of the intersection points on $b_j$ is the opposite of the
order of intersection points on the side $b_j$ of $P^D_{g,n}$. In
the case of the generators $m_i$, the order of the intersection
points at the starting point of $m_i$ agrees with the order of the
corresponding points on $P^D_{g,n}$, while the order of the
intersection points at its endpoint is the opposite.

\begin{theorem}
\label{combinttheor}

For any embedded $\lambda\in\pi_1(\surf\mindee)$ given as a
cyclically reduced words in the dual generators
$\dcm_i,\dca_j,\dcb_j$ and any conjugation invariant
$f\in\cif(H)$, the transformation $T^t_{f,\lambda}$ generated by
the observable $f_\lambda$ is given by
\begin{align}
\label{algdef}
&T^t_{f,\lambda}:\\
&M_i\mapsto (M_{i-1}\cdots M_1)\!\!\! \left(\!\!\prod^{\rightarrow}_{\substack{\tau\in\text{CPerm}(\lambda)\\ \text{LF}(\tau)=\dcm_i}}\!\!\!\!\!\!\!\!\!\!\!G^{t\,\text{sgn}(\tau)}_{f,|\tau|}\!\!\right)^\inv\!\!\!\!\!\!\!\!\!\!\;(M_{i-1}\cdots M_1)^\inv \! M_i(M_{i-1}\cdots M_1)\!\!\! \left(\!\!\prod^{\rightarrow}_{\substack{\tau\in\text{CPerm}(\lambda)\\ \text{LF}(\tau)=\dcm_i}}\!\!\!\!\!\!\!\!\!\!\!G^{t\,\text{sgn}(\tau)}_{f,|\tau|}\!\!\right)\!\!\!\!\;(M_{i-1}\cdots M_1)^\inv\nonumber\\
&A_j\mapsto A_j\cdot (H_{j-1}\cdots M_1)\! \left(\!\!\prod^{\rightarrow}_{\substack{\tau\in\text{CPerm}(\lambda)\\ \text{LF}(\tau)=\dca_j}}\!\!\!\!\!\!\!\!\!\!\!G^{t\,\text{sgn}(\tau)}_{f,|\tau|}\!\!\right)\!\!\;(H_{j-1}\cdots M_1)^\inv\nonumber\\
&B_j\mapsto  (A_jH_{j-1}\cdots M_1)\!
\left(\!\!\prod^{\rightarrow}_{\substack{\tau\in\text{CPerm}(\lambda)\\
\text{LF}(\tau)=\dcb_j}}\!\!\!\!\!\!\!\!\!\!\!G^{t\,\text{sgn}(\tau)}_{f,|\tau|}\!\!\right)^\inv\!\!\!\!\!\!\!\!\;(A_jH_{j-1}\cdots
M_1)^\inv\cdot B_j,\nonumber
\end{align}
where the ordering of the factors  is the one defined by
\eqref{cond1}, \eqref{cond2}. More precisely, for
$\tau,\eta\in\text{CPerm}(\lambda)$,
$\text{LF}(\tau)=\text{LF}(\eta)=\dcx$, we have
\begin{align}
\label{ordwils} \tau<\eta\qquad\Leftrightarrow\qquad
\text{LF}(\tau_s)=\text{LF}(\eta_s)\;\forall 1\leq s\leq k,
\;\text{LF}(\tau_k^\inv)\grx
{\text{LF}(\tau_k)}\text{LF}(\eta_k^\inv)
\end{align}
with the ordering  $\grx x$ defined as in \eqref{miorder} and
$\tau_k,\eta_k$ denoting the cyclic permutations \eqref{lambdak}.
If $\{\tau\in\text{CPerm}(\lambda)\;|\;
\text{LF}(\tau)=\dcx\}=\{\tau_1,\ldots,\tau_s\}$ and
$\tau_1<\ldots<\tau_s$ with respect to the ordering
\eqref{ordwils}, then the associated product is given by
\begin{align}
\prod^{\rightarrow}_{\substack{\tau\in\text{CPerm}(\lambda)\\
\text{LF}(\tau)=\dcx}}\!\!\!\!\!\!\!\!\!G^{t}_{f,\tau}=
G^t_{f,\tau_s}\cdot G^t_{f,\tau_2}\cdots G^t_{f,\tau_1}.
\end{align}
\end{theorem}
 By applying the involution $I\in\text{Aut}(\pi_1(\surf\mindee))$
to Fock and Rosly's description \cite{FR} of the moduli space of
flat connections, we therefore obtain explicit expressions for the
Poisson brackets of generalised Wilson loop observables associated
to curves on $\surf\mindee$  and the associated flows on phase
space. These expressions generalise the results of Goldman, see in
particular Theorem 3.5. and formulas (4.4), (4.6) in
\cite{goldman}, to the case of surfaces with punctures. While the
results in \cite{goldman} are obtained via cohomological methods
and characterise flows and Poisson brackets geometrically in terms
of the intersection behaviour of curves on the surfaces, Theorems
\ref{obspb}, \ref{flowth} and \ref{combinttheor} give concrete
expressions in terms of the holonomies along the generators of the
fundamental group $\pi_1(\surf\mindee)$. Moreover, the resulting
formulas are purely algebraic and characterise flows and Poisson
brackets in terms of the expression of the associated curves as
reduced words in the dual generators.

This explicitness presents an advantage in physical applications
of the theory such as the Chern-Simons formulation of
(2+1)-dimensional gravity and, more generally, the quantisation of
Chern-Simons theory. Most approaches to quantisation such as
Alekseev, Grosse and Schomerus' combinatorial quantisation
formalism \cite{AGSI, AGSII,AS} for Chern-Simons theory with
compact semisimple gauge groups and the quantisation procedures in
\cite{BNR} and \cite{we2,we3}
 for, respectively, gauge group $SL(2,\CC)$ and semidirect product gauge groups
$G\ltimes\gothg^*$, are based on Fock and Rosly's description of
the moduli space and take the holonomies along a set of generators
of the fundamental group as their basic variables. The formulation
in this paper could therefore serve as a framework for the
investigation of the associated  observables and transformations
in quantised Chern-Simons theory. Moreover, it is used in
\cite{ich3} to investigate the role of these flows in the
Chern-Simons formulation of (2+1)-dimensional gravity and to
relate them to the geometrical construction of evolving
(2+1)-spacetimes.

\subsection{Example}

\begin{figure}
 \vskip .3in \protect\input epsf
\protect\epsfxsize=12truecm
\protect\centerline{\epsfbox{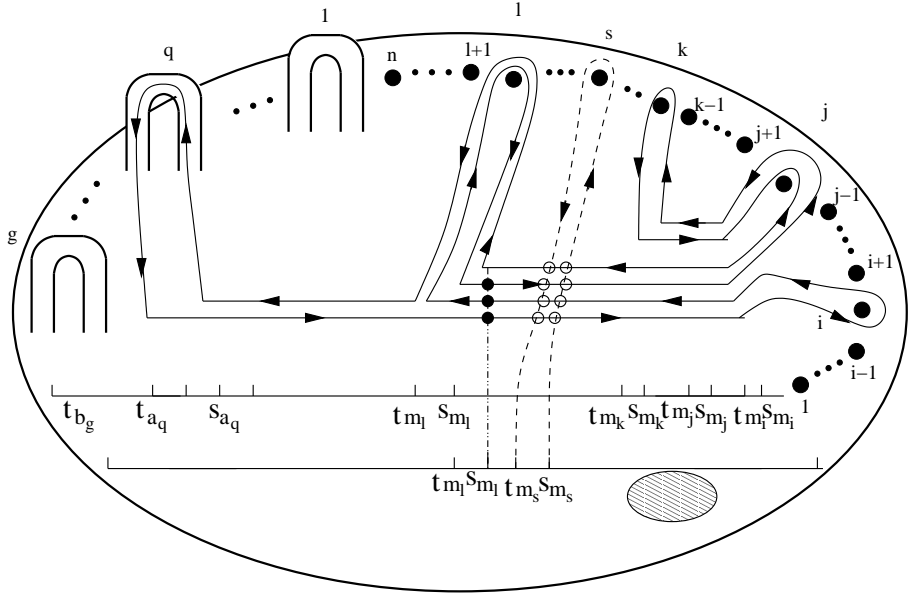}} \caption{ The graphical
representation of $\lambda=a_q\circ m_l\circ m_j^\inv\circ
m_k\circ m_j\circ m_l^\inv\circ m_i$ (full line), its intersection
points with the generators $m_s$ (dashed line, white circles) and
with a vertical line based at $s_{m_l}$ (black
circles).}\label{csex}
\end{figure}

We conclude this section by determining the transformations
$T^t_{f,\lambda}$ for a concrete example. We consider the element
\begin{align}
\label{lambdex} \lambda=&a_q\circ m_l\circ m_j^\inv\circ m_k\circ
m_j\circ m_l^\inv\circ m_i\qquad\qquad\qquad 1\leq i<j<k<l\leq n,
\;q\in\{1,\ldots,g\}\\
=&(\dcm_1^\inv\cdots\dcm_{i-1}^\inv)\circ(\dcm_{i}^\inv\cdots\dch_{q-1}^\inv \dca_q^\inv\dcb_q\dca_q)\circ(\dch_{q-1}\cdots\dcm_{l+1})\circ(\dcm_{l-1}\cdots\dcm_j)\circ(\dcm_{j+1}^\inv\cdots\dcm_k^\inv)\circ\nonumber\\
\circ
&(\dcm_{k-1}\cdots\dcm_{j+1})\circ(\dcm_{j}^\inv\cdots\dcm_{l-1}^\inv)\circ(\dcm_l\cdots\dcm_{i+1}
)\circ(\dcm_{i-1}\cdots\dcm_1),\nonumber
\end{align}
whose graphical representation is shown in Fig.~\ref{csex}. By
using the graphical representation in Fig.~\ref{csex} and the
definitions \eqref{transfdef}, \eqref{lambdaidef} we can determine
the action of the one-parameter group of diffeomorphisms
$T^t_{f,\lambda}$ associated to a conjugation invariant function
$f\in\cif(H^\ntg)$.  We find that their action on the holonomies
along the generators $m_i,a_j,b_j$ is given by
\begin{align}
\label{exholtr} &M_i\mapsto G_{f,\lambda}^{-t} \cdot M_i\cdot
G_{f,\lambda}^{t}\\
&M_s\mapsto G_{f,\lambda}^{-t}G_{f,\lambda_1}^{t}\cdot  M_s\cdot
G_{f,\lambda_1}^{-t}G_{f,\lambda}^{t}\qquad\forall i<s<j\nonumber\\
&M_j\mapsto G_{f,\lambda}^{-t}G_{f,\lambda_1}^{t}
G_{f,\lambda_2}^{-t}G_{f,\lambda_3}^{t}  \cdot M_j\cdot
 G_{f,\lambda_3}^{-t}
G_{f,\lambda_2}^{t} G_{f,\lambda_1}^{-t}G_{f,\lambda}^{t}\nonumber\\
&M_s\mapsto G_{f,\lambda}^{-t}G_{f,\lambda_1}^{t}
G_{f,\lambda_2}^{-t}G_{f,\lambda_3}^{t} G_{f,\lambda_4}^{-t}
G_{f,\lambda_5}^{t}\cdot M_s\cdot
G_{f,\lambda_5}^{-t}G_{f,\lambda_4}^{t} G_{f,\lambda_3}^{-t}
G_{f,\lambda_2}^{t} G_{f,\lambda_1}^{-t}G_{f,\lambda}^{t}\qquad j<s<k\nonumber\\
&M_k\mapsto G_{f,\lambda}^{-t}G_{f,\lambda_1}^{t}
G_{f,\lambda_2}^{-t}G_{f,\lambda_3}^{t} G_{f,\lambda_4}^{-t} \cdot
M_k\cdot G_{f,\lambda_4}^{t} G_{f,\lambda_3}^{-t}
G_{f,\lambda_2}^{t} G_{f,\lambda_1}^{-t}G_{f,\lambda}^{t}\nonumber\\
&M_s\mapsto G_{f,\lambda}^{-t}G_{f,\lambda_1}^{t}
G_{f,\lambda_2}^{-t}G_{f,\lambda_3}^{t}  \cdot M_s\cdot
G_{f,\lambda_3}^{-t}
G_{f,\lambda_2}^{t} G_{f,\lambda_1}^{-t}G_{f,\lambda}^{t}\qquad k<s<l\nonumber\\
&M_l\mapsto G_{f,\lambda}^{-t}G_{f,\lambda_1}^{t}\cdot  M_l\cdot
G_{f,\lambda_1}^{-t}G_{f,\lambda}^{t}\nonumber\\
 &Y\mapsto G_{f,\lambda}^{-t}G_{f,\lambda_6}^{t}
\cdot Y\cdot G_{f,\lambda_6}^{-t}G_{f,\lambda}^{t}\qquad
Y\in\{M_{l+1},\ldots,
M_n,A_1,B_1,\ldots, A_{q-1},B_{q-1}\}\nonumber\\
&A_q\mapsto G_{f,\lambda}^{-t} \cdot A_q\cdot
G_{f,\lambda}^{t}\nonumber\\
&B_q\mapsto  B_q\cdot G_{f,\lambda}^{t}\nonumber,
\end{align}
where we omit the argument $(M_1,\ldots,B_g)$ of the functions
$G_{f,\eta}^t$, list only those generators which do not transform
trivially and set
\begin{align}
\label{lambdaconj}
&\lambda_{0}=\lambda\\
&\lambda_1=m_i\circ\lambda\circ m_i^\inv=m_i\circ a_q\circ
m_l\circ m_j^\inv\circ m_k\circ m_j\circ m_l^\inv\nonumber\\
&\lambda_2=(m_l^\inv m_i)\circ\lambda\circ (m_l^\inv
m_i)^\inv=m_l^\inv \circ m_i\circ a_q\circ
m_l\circ m_j^\inv\circ m_k\circ m_j\nonumber\\
&\lambda_3=(a_q m_l)^\inv\circ\lambda\circ (a_q m_l)=
m_j^\inv\circ m_k\circ
m_j\circ m_l^\inv\circ a_q\circ m_l\nonumber\\
 &\lambda_4=(a_q m_l m_j^\inv)^\inv\circ\lambda\circ
(a_q m_l m_j^\inv)= m_k\circ
m_j\circ m_l^\inv\circ a_q\circ m_l\circ m_j^\inv\nonumber\\
\intertext{} &\lambda_5=(m_j m_l^\inv m_i)\circ\lambda\circ (m_j
m_l^\inv m_i)^\inv= m_j\circ m_l^\inv\circ m_i\circ a_q \circ
m_l\circ
m_j^\inv\circ m_k\nonumber\\
&\lambda_6=a_q^\inv\circ\lambda\circ a_q=m_l\circ m_j^\inv\circ
m_k\circ m_j\circ m_l^\inv\circ a_q.\nonumber
\end{align}
The ordering of the horizontal segments in Fig.~\ref{csex} then
implies that the transformation of the holonomies along the cyclic
permutations $\lambda_i$ is given by
\begin{align}
&\rho_{\lambda}\circ T^t_{f,\lambda}=\rho_{\lambda}\\
&\rho_{\lambda_1}\circ T^t_{f,\lambda}=G_{f,\lambda}^{-t}\cdot \rho_{\lambda_1}\cdot G_{f,\lambda}^{-t}\nonumber\\
&\rho_{\lambda_2}\circ T^t_{f,\lambda}=G_{f,\lambda}^{-t}G_{f,\lambda_1}^{t}\cdot \rho_{\lambda_2}\cdot G_{f,\lambda_1}^{t}G_{f,\lambda}^{-t}\nonumber\\
&\rho_{\lambda_3}\circ T^t_{f,\lambda}=G_{f,\lambda}^{-t}G_{f,\lambda_1}^{t}G_{f,\lambda_2}^{t}\cdot \rho_{\lambda_3}\cdot G_{f,\lambda_2}^{-t} G_{f,\lambda_1}^{t}G_{f,\lambda}^{-t}\nonumber\\
&\rho_{\lambda_4}\circ T^t_{f,\lambda}=G_{f,\lambda}^{-t}G_{f,\lambda_1}^{t}G_{f,\lambda_2}^{t}G_{f,\lambda_3}^{-t}\cdot \rho_{\lambda_4}\cdot G_{f,\lambda_3}^{t}G_{f,\lambda_2}^{-t} G_{f,\lambda_1}^{t}G_{f,\lambda}^{-t}\nonumber\\
&\rho_{\lambda_5}\circ T^t_{f,\lambda}=G_{f,\lambda}^{-t}G_{f,\lambda_1}^{t}G_{f,\lambda_2}^{t}G_{f,\lambda_3}^{-t}G_{f,\lambda_4}^{t}\cdot \rho_{\lambda_5}\cdot G_{f,\lambda_4}^{-t}G_{f,\lambda_3}^{t}G_{f,\lambda_2}^{-t} G_{f,\lambda_1}^{t}G_{f,\lambda}^{-t}\nonumber\\
&\rho_{\lambda_6}\circ T^t_{f,\lambda}=G_{f,\lambda}^{-t}\cdot
\rho_{\lambda_6}\cdot G_{f,\lambda}^{-t}\nonumber.
\end{align}
A straightforward but lengthy calculation shows that this agrees
with the transformation obtained from the expressions
\eqref{lambdaconj} and the transformation \eqref{exholtr} for the
holonomies along the generators.


\section{Dual generators and the mapping class group}
\label{mapsect}

In this section, we apply the results of Sect.~\ref{wloopsect} to
a specific set of generalised Wilson loop observables which are
constructed from the $\Ad$-invariant symmetric bilinear form
$\langle\,,\,\rangle$ in the Chern-Simons action. We show that the
associated one-parameter groups of diffeomorphisms are related to
the Dehn twists around embedded curves on the surface
$\surf\mindee$. This allows us to determine the transformation of
Fock and Rosly's Poisson structure under the action of the mapping
class group $\mapcld$ and under general automorphisms of
$\pi_1(\surf\mindee)$ which arise from homeomorphisms of
$\surf\mindee$.

The mapping class group $\mapcld$ is the group of equivalence
classes of orientation preserving homeomorphisms of $\surf\mindee$
which fix the punctures as a set and fix the boundary of the disc
$D$ pointwise. Two homeomorphisms are identified  if they differ
by one which is isotopic to the identity. The pure mapping class
group $\pmapcld$ is
 the subgroup of $\mapcld$ which contains the equivalence classes of
homeomorphisms that fix each of the punctures and is related to
the mapping class group $\mapcld$ by the short exact sequence
\begin{align} \label{sequen} 1\rightarrow\pmapcld\xrightarrow{i}
\mapcld \xrightarrow{\pi} S_n \rightarrow 1, \end{align} where $i$
is the canonical embedding and $\pi:\mapcld\rightarrow S_n $
 the projection onto the
symmetric group that assigns to each element of the mapping class
group the associated permutation of the punctures. As explained in
\cite{birmorange, birmgreen}, the pure mapping class group
$\pmapcld$ is generated by Dehn twists around a set of embedded
curves, and a set of generators of the full mapping class group
$\mapcl$ is obtained by supplementing this set with
 $n-1$ elements
  which get mapped to the elementary
transpositions via $\pi$. A set of generators of the pure and full
mapping class group and their action on the fundamental group is
given in the appendix. Each element of the mapping class group
induces an unique automorphism of the fundamental group
$\pi_1(\surf\mindee)$ which maps the loop $m_D$ around the disc to
itself and acts on the loops around the punctures by conjugation
and permutations. Via the identification of the generators
$m_i,a_j,b_j$ with the different copies of $H$ in the product
$H^\ntg$ each element $\varphi\in\mapcld$ then induces a
diffeomorphism $\Phi_\varphi:H^\ntg\rightarrow H^\ntg$.

We consider the generalised Wilson loop observables associated to
an element $\lambda\in\pi_1(\surf\mindee)$ with an embedded
representative and to the bilinear form $\langle\,,\,\rangle$ in
the Chern-Simons action. Using the parametrisation via the
exponential map and composing, we obtain a conjugation invariant
function $\tilde t: H\rightarrow \RR$ by setting
\begin{align}
\label{casfunct} \tilde t (e^{p^a J_a}):=\tfrac{1}{2}\langle
p^aJ_a, p^aJ_a\rangle.
\end{align}
As we require the exponential map to be locally but not globally
bijective, the parametrisation by elements of the Lie algebra
$\gothh$ is in general not unique. To obtain a unique
parametrisation, one has to restrict the Lie algebra elements
$p^aJ_a$ appropriately, which implies that the function $\tilde t$
defined in \eqref{casfunct} is locally but not globally $\cif$.
However, as we are only interested in the local properties of this
map, we will not address this issue further. To determine the
flows generated by the associated Wilson loop observables $\tilde
t_\lambda$, we need to derive the maps $g_{\tilde t}: H\rightarrow
\gothh$, $G_{\tilde t}: H\rightarrow H$ defined in \eqref{gfdef}.
We use a result by Goldman \cite{goldman} summarised in the
following lemma.
\begin{lemma} (Goldman \cite{goldman})
\label{hlem2}

Let $s\in\gothh^*\otimes\gothh^*$ be an $Ad$-invariant, symmetric
bilinear form on $\gothh$ and consider the associated function
$\tilde s: H\rightarrow \RR$, $\tilde s (e^{p^a J_a})=s(p^aJ_a,p^a
J_a)$. Then, the action of the left-and right-invariant vector
fields on $\tilde s$ is given by
\begin{align}
\label{sid} R_a \tilde s (e^{p^bJ_b})=-L_a \tilde
s(e^{p^bJ_b})=2\;s(J_a,J_b) p^b.
\end{align}
\end{lemma}
By applying formula \eqref{sid} to the function $\tilde t:
H\rightarrow \RR$, we find that the maps $g_{\tilde t}:
H\rightarrow \gothh$, $G_{\tilde t}: H\rightarrow H$  in
\eqref{gfdef} take the form
\begin{align}
g_{\tilde t}(e^{p^a J_a})=p^a\qquad G_{\tilde
t}^t(e^{p^aJ_a})=e^{tp^aJ_a}.
\end{align}
For flow parameter $t=1$, the diffeomorphism $G^t_{\tilde
t}:H\rightarrow H$ is the identity map on $H^\ntg$. This implies
that the associated flows $T^1_{\tilde t,\lambda}$ defined by
\eqref{transfdef}  act on the group elements associated to the
generators of the fundamental group by left- and
right-multiplication with the holonomies of certain elements in
the conjugacy class of $\lambda$ and correspond to an automorphism
of the fundamental group $\pi_1(\surf\mindee)$.
 Together with the dependence of
this transformation on intersection points, this indicates that
the transformation $T^1_{\tilde t,\lambda}$ should be related to
the diffeomorphism of $H^\ntg$ induced by the Dehn twist around
$\lambda$.
\begin{theorem}$\quad$
\label{dttheor}
\begin{enumerate}
\item For any embedded $\lambda\in\pi_1(\surf\mindee)$, the
associated one-parameter group $T^t_{\tilde
t,\lambda}:H^\ntg\rightarrow H^\ntg$ of diffeomorphisms of
represents an infinitesimal Dehn twist around $\lambda$
\begin{align}
\label{tid} T^1_{\tilde t,\lambda}=D_\lambda,
\end{align}
where $D_\lambda: H^\ntg\rightarrow H^\ntg$ is the diffeomorphism
of $H^\ntg$ induced by the action
$d_\lambda\in\text{Aut}(\pi_1(\surf\mindee))$ of the Dehn twist
around $\lambda$ on the fundamental group $\pi_1(\surf\mindee)$.

\item The mapping class group $\mapcld$ acts by Poisson
isomorphisms
\begin{align}
\label{mappoiss} \{f\circ \Phi_\varphi, g\circ
\Phi_\varphi\}=\{f,g\}\circ \Phi_\varphi\qquad\forall
f,g\in\cif(H^\ntg), \varphi\in\mapcld.
\end{align}
\end{enumerate}
\end{theorem}

{\bf Proof:} The general reasoning follows the proof in
\cite{we3}, where an analogous statement is proved for
Chern-Simons theory with gauge groups of the form
$H=G\ltimes\gothg^*$.

1. We start by proving identity \eqref{tid} for the set of
generating Dehn twists given in the appendix. For this, we have to
determine the Poisson brackets of the associated Wilson loop
observables, which can be done  either by expressing the curves in
terms of the dual generators $\dcm_i,\dca_j,\dcb_j$ and using
\eqref{caspb} or via the graphical procedure in
Sect.~\ref{factassign} and formula \eqref{intpb0}. Expressed in
both, the generators $m_i,a_j,b_j$ and their duals, the curves
\eqref{dtgen} for the generating Dehn twists in the appendix are
given by
\begin{align}
\label{dtgendual}
&a_i=(\dcm_1^\inv\cdots\dch_{i-1}^\inv)\circ\dcb_i\circ(\dch_{i-1}\cdots\dcm_1)\qquad i=1,\ldots,g\\
&\delta_i=a_i^{-1}\circ b_i^{-1}\circ a_i=(\dcm_1^\inv\cdots\dch_{i-1}^\inv)\circ\dca_i^\inv\circ(\dch_{i-1}\cdots\dcm_1)\qquad i=1,\ldots,g\nonumber\\
&\alpha_i=a_i^{-1}\circ b_i^{-1}\circ a_i\circ b_{i-1}=(\dcm_1^\inv\cdots\dch_{i-1}^\inv)\circ\dca_i^\inv\circ\dca_{i-1}\circ\dch_{i-1}^\inv\circ(\dch_{i-1}\cdots\dcm_1)\nonumber\\
&\epsilon_i=a_i^{-1}\circ b_i^{-1}\circ a_i  \circ h_{i-1} \cdots
h_1=(\dcm_1^\inv\cdots\dcm_n)\circ\dch_1^\inv\cdots\dch_{i-1}^\inv\circ\dca_i^\inv\circ(\dcm_n\cdots\dcm_1)\qquad\quad i=2,\ldots,g\nonumber\\
&\kappa_{\nu,\mu}=m_\mu \circ
m_\nu=(\dcm_1^\inv\cdots\dcm_{\nu-1}^\inv)\circ
(\dcm_\nu^\inv\cdots\dcm_{\mu}^\inv\circ
\dcm_{\mu-1}\cdots\dcm_{\nu+1})\circ(\dcm_{\nu-1}\cdots\dcm_{1})\;\; 1\leq \nu<\mu\leq n\nonumber\\
&\kappa_{\nu,
n+2i-1}=a_i^{-1}\!\!\circ\! b_i^{-1}\!\!\circ \!a_i\!\circ m_\nu\!\!=\!(\dcm_1^\inv\cdots\dcm_{\nu-1}^\inv)\circ(\dcm_{\nu}^\inv\cdots\dch_{i-1}^\inv\circ\dca_i^\inv\circ\dch_{i-1}\cdots\dcm_{\nu+1})\circ(\dcm_{\nu-1}\cdots\dcm_{1})\nonumber\\
&\kappa_{\nu, n+2i}=b_i \circ
m_\nu=(\dcm_1^\inv\cdots\dcm_{\nu-1}^\inv)\circ(\dcm_{\nu}^\inv\cdots\dch_{i}^\inv\circ\dca_i\circ\dch_{i-1}\cdots\dcm_{\nu+1})\circ(\dcm_{\nu-1}\cdots\dcm_{1})\nonumber,
\end{align}
and a lengthy but direct calculation yields the Poisson brackets
of the associated Wilson loop observables with a general function
$f\in\cif(H^\ntg)$
\begin{align}
\label{dtgenpbs} &\{f, \tilde t_{a_i}\}=p_{a_i}^a
R_a^{\bi}f\\
 &\{f, \tilde t_{\delta_i}\}=p_{b_i}^a
L_a^{\ai}f\nonumber\\
 &\{f, \tilde t_{\alpha_i}\}=p_{\alpha_i}^a(
 R^\ai_a+L^{A_{i-1}}_a+R^{B_{i-1}}_a+L^{B_{i-1}}_a)f\nonumber\\
 &\{f, \tilde t_{\epsilon_i}\}=p_{\epsilon_i}^a(R^\ai_a+\sum_{j=1}^{i-1}R^{\aj}_a+L^\aj_a+R^\bjj_a+L^\bjj_a)f\nonumber\\
 &\{f,
 \tilde t_{\kappa_{\nu,\mu}}\}=p_{\kappa_{\nu,\mu}}^a(R^{M_\nu}_a\!+\!L^{M_\nu}_a\!+\!R^{M_\mu}_a\!+\!L^{M_\mu}_a)f \!+\!  (p_{\kappa_{\nu,\mu}}^a- p_{m_\nu\circ\kappa_{\nu,\mu}\circ m_\nu^\inv}^a)\sum_{j=\nu+1}^{\mu-1} (R^{\mj}_a\!+\!L^\mj_a)f\nonumber\\
 &\{f,
 \tilde t_{\kappa_{\nu,n+2i-1}}\}=p_{\kappa_{\nu, n+2i-1}}^a(R^{M_\nu}_a+L^{M_\nu}_a+R^{\ai}_a)f\nonumber\\
& \qquad\qquad+ (p_{\kappa_{\nu,n+2i-1}}^a- p_{m_\nu\circ\kappa_{\nu,n+2i-1}\circ m_\nu^\inv}^a)\left(\sum_{j=\nu+1}^{n} R^{\mj}_a+L^\mj_a+\sum_{j=1}^{i-1}R^{\aj}_a+L^\aj_a+R^\bjj_a+L^\bjj_a\right)f\nonumber\\
\intertext{}
 &\{f, \tilde t_{\kappa_{\nu,n+2i}}\}=p_{\kappa_{\nu, n+2i}}^a(R^{M_\nu}_a+L^{M_\nu}_a+L^{\ai}_a+R^\bi_a+L^\bi_a)f+( p_{\kappa_{\nu,n+2i}}^a-p_{m_\nu\circ\kappa_{\nu,n+2i}\circ m_\nu^\inv}^a) R^\ai_af\nonumber\\
& \qquad\qquad\qquad+(
p_{\kappa_{\nu,n+2i}}^a-p_{m_\nu\circ\kappa_{\nu,n+2i}\circ
m_\nu^\inv}^a)\left(\sum_{j=\nu+1}^{n}
R^{\mj}_a+L^\mj_a+\sum_{j=1}^{i-1}R^{\aj}_a+L^\aj_a+R^\bjj_a+L^\bjj_a\right)f\nonumber,
\end{align}
where we denote by $p^a: H\rightarrow \RR$,
$a=1,\ldots,\text{dim}\;\gothh$, the maps $p^a:
u=e^{k^bJ_b}\mapsto k^a$   and set
$p^a_{\lambda}=p^a\circ\rho_\lambda$.

By using the graphical representation defined in
Sect.~\ref{factassign}, we can determine the intersection points
of these curves with representatives of the generators
$m_i,a_j,b_j$  and assign them between the different factors in
the expression \eqref{dtgendual}. Using this assignment of
intersection points, we can derive the one parameter groups of
transformations associated to the brackets \eqref{dtgenpbs}
\begin{align}
\label{adinf}
T_{\tilde t, a_i}^t: &B_i \mapsto \bi \ai^t\\
\nonumber\\
\label{ddinf}
T_{\tilde t,\delta_i}: &A_i \mapsto A_iH_{\delta_i}^t=B_i^{-1}A_i\\
\nonumber\\
\label{alphainf}
T_{\tilde t, \alpha_i}^t: &A_i\mapsto A_iH_{\alpha_i}^t\\
&B_{i-1}\mapsto
H_{\alpha_i}^{-t}B_{i-1}H_{\alpha_i}^t\nonumber\\
&A_{i-1}\mapsto
H_{\alpha_i}^{-t}A_{i-1}\nonumber\\
\nonumber\\ \label{epsinf}
T_{\tilde t,\epsilon_i}^t: &A_i\mapsto A_iH_{\epsilon_i}^t\\
&A_k\mapsto H_{\epsilon_i}^{-t} A_k
H_{\epsilon_i}^t\nonumber\\
&B_k\mapsto H_{\epsilon_i}^{-t}B_kH_{\epsilon_i}^t\qquad\qquad\forall 1\leq k <i\nonumber\\
\nonumber\\\label{etappinf} T^t_{\tilde t,\kappa_{\nu,\mu}}:
&M_\nu\mapsto H_{\kappa_{\nu,\mu}}^{-t}M_\nu H_{\kappa_{\nu,\mu}}^t\\
&M_\mu\mapsto H_{\kappa_{\nu,\mu}}^{-t}M_\mu H_{\kappa_{\nu,\mu}}^t\nonumber\\
&M_j\mapsto[H_{\kappa_{\nu,\mu}}^{-t},M_\nu]M_j[M_\nu,H_{\kappa_{\nu,\mu}}^{-t}]\qquad\qquad\forall
\nu<j<\mu    \nonumber
\\  \nonumber
\\  \label{etapdeltainf}
T_{\tilde t,\kappa_{\nu, n+2i-1}}^t: &M_\nu\mapsto
H_{\kappa_{\nu,n+2i-1}}^{-t} M_\nu H_{\kappa_{\nu, n+2i-1}}^t\\
&A_i\mapsto A_i H_{\kappa_{\nu,n+2i-1}}^t\nonumber\\
& X\mapsto [H_{\kappa_{\nu,n+2i-1}}^{-t}, M_\nu] X
[M_\nu,H_{\kappa_{\nu,n+2i-1}}^{-t}] \qquad\qquad X\in\{M_{\nu+1},\ldots,M_n,A_1,\ldots,B_{i-1}\}\nonumber\\
\nonumber\\
 \label{etapbinf} T^t_{\tilde t,\kappa_{\nu,n+2i}}:
&M_\nu\mapsto H_{\kappa_{\nu,n+2i}}^{-t}M_\nu H_{\kappa_{\nu,n+2i}}^t\\
 &B_i\mapsto H_{\kappa_{\nu,n+2i}}^{-t}B_i H_{\kappa_{\nu,n+2i}}^t\nonumber\\
&A_i\mapsto H_{\kappa_{\nu,n+2i}}^{-t}A_i[M_\nu, H_{\kappa_{\nu,n+2i}}^{-t}]\nonumber\\
&X\mapsto [H_{\kappa_{\nu,n+2i}}^{-t}, M_\nu] X [M_\nu,
H_{\kappa_{\nu,n+2i}}^{-t}]\qquad\qquad\qquad
X\in\{M_{\nu+1},\ldots,M_n, A_1,\ldots B_{i-1}\} \nonumber,
\end{align}
where we write $X^t:=e^{tp^aJ_a}$ for $X=e^{p^aJ_a}\in H$  and set
$H_\lambda=\rho_\lambda(M_1,\ldots,B_g)$. By comparing expressions
\eqref{adinf} to \eqref{etapbinf} with the formulas \eqref{ad} to
\eqref{etapb} for the action of the Dehn twists  on the generators
of the fundamental group $\pi_1(\surf\mindee)$, we see that these
expressions agree if we set $t=1$ and replace $m,a,b$ in
\eqref{ad} to \eqref{etapb} with the corresponding capital
letters. Hence, the identity \eqref{tid} holds for the generating
Dehn twists \eqref{ad} to \eqref{etapb}.

2. As Corollary \ref{properties} implies that the transformations
$T^t_{f,\lambda}: H^\ntg\rightarrow H^\ntg$ act by Poisson
isomorphisms, the same holds for the generating Dehn twists the
transformations \eqref{adinf} to \eqref{etapbinf} and hence for
all elements of the pure mapping class group $\pmapcld$. To prove
the invariance of the Poisson structure under the full mapping
class group $\mapcld$, it is therefore sufficient to demonstrate
that \eqref{mappoiss} for the
 transformations on $H^\ntg$ induced by the transformations
 \eqref{braid} which permute the punctures. This can be verified by direct computation using
 the
expressions \eqref{holfunctpb}, \eqref{ord}, \eqref{finholf} for
the Poisson bracket. For a proof the analogous statement for the
(2+1)-dimensional Poincar\'e group and gauge groups of the form
$G\ltimes \mathfrak{g}^*$, see also \cite{we1}, \cite{we3}.

3. To prove that identity \eqref{tid} holds for any embedded
curve, we make use of the fact that the mapping class group
$\mapcld$ acts by Poisson isomorphisms. We consider an embedded
curve $\lambda\in\pi_1(\surf\mindee)$ and an embedded curve
$\eta\in\pi_1(\surf\mindee)$ obtained from $\lambda$ via the
action of an element $\varphi\in\mapcld$ on $\pi_1(\surf\mindee)$
\begin{align}
\eta=\varphi(\lambda)\qquad\varphi\in\mapcld .
\end{align}
 From the geometrical definition of the Dehn
twists it follows that the action $D_\eta: H^\ntg\rightarrow
H^\ntg$ of the Dehn twist along $\eta$ is given by
\begin{align}
\label{dtrafo} D_\eta=\Phi_\varphi^\inv\circ
D_\lambda\circ\Phi_\varphi,
\end{align}
where $\Phi_\varphi: H^\ntg\rightarrow H^\ntg$ is the
diffeomorphism induced by $\varphi$. On the other hand the
definition \eqref{rhoxdef} of the maps
$\rho_\lambda:H^\ntg\rightarrow H$ implies
\begin{align}
\label{rhotrafo}
\rho_{\varphi(\lambda)}=\rho_\lambda\circ\Phi_\varphi.
\end{align}
Using identity \eqref{poissid} and the fact that the
diffeomorphism $\Phi_\varphi$ is a Poisson isomorphism, we obtain
\begin{align}
\frac{d}{dt}|_{t=0} f\circ\Phi_\varphi\circ T^t_{\tilde
t,\eta}=\{f\circ\Phi_\varphi,\tilde
t_\eta\}=\{f\circ\Phi_\varphi,\tilde
t_\lambda\circ\Phi_\varphi\}=\{f,\tilde
t_\lambda\}\circ\Phi_\varphi= \frac{d}{dt}|_{t=0} f\circ
T^t_{\tilde t,\lambda}\circ\Phi_\varphi
\end{align}
and therefore
\begin{align}
\label{ttrafo} T^t_{\tilde t,\eta}=\Phi_\varphi^\inv\circ
T^t_{\tilde t, \lambda}\circ\Phi_\varphi.
\end{align}
By setting $t=1$ and combining this identity  with the
corresponding identity \eqref{dtrafo} for the Dehn twist around
$\eta$, we find that identity \eqref{tid} holds for $\eta$ if and
only if it holds for $\lambda$.

\begin{figure}
 \vskip .3in \protect\input epsf
\protect\epsfxsize=12truecm
\protect\centerline{\epsfbox{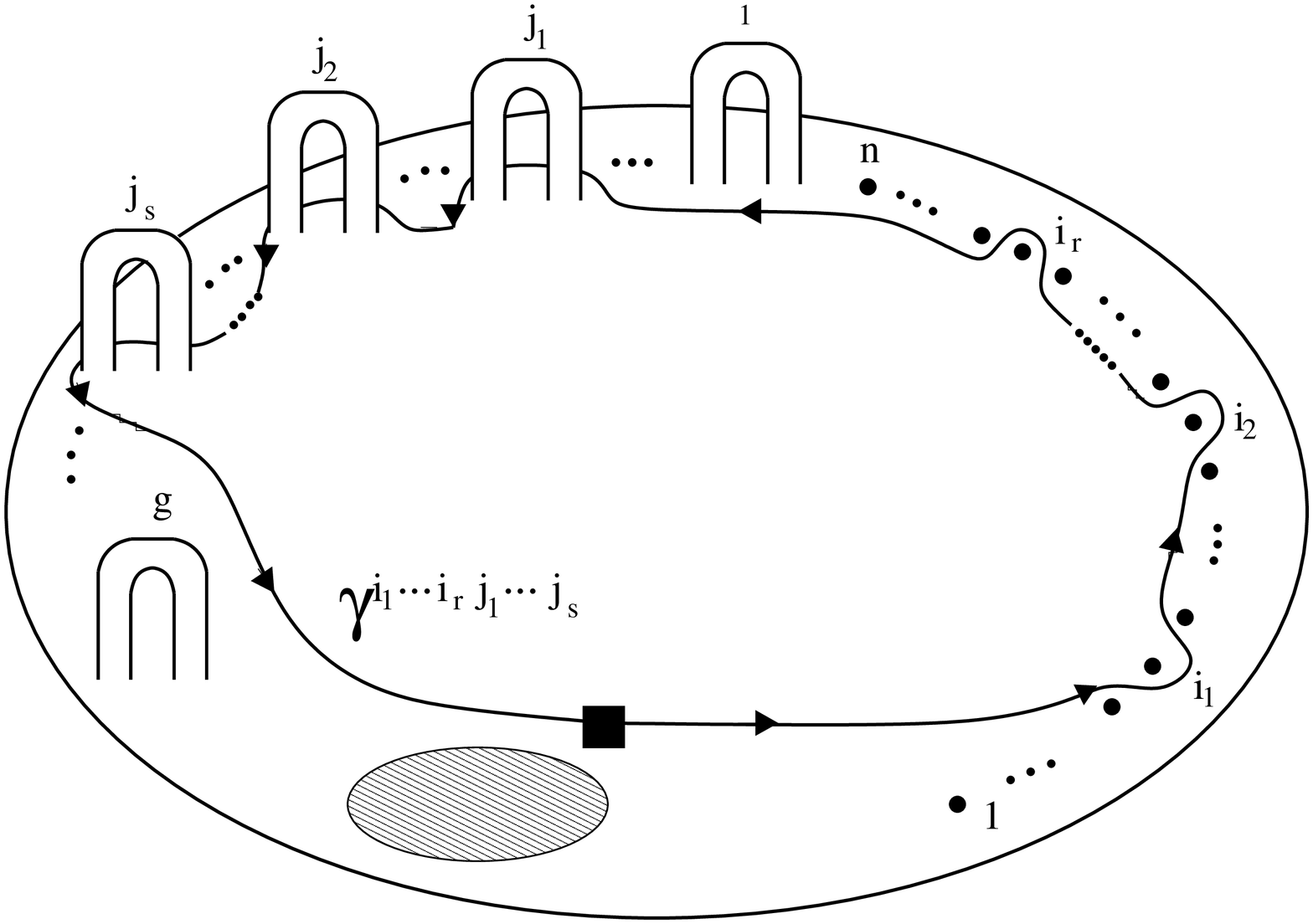}} \caption{Separating curves
on the surface $\surf\mindee$}\label{sepc}
\end{figure}

4. Hence, it is sufficient to prove \eqref{tid} for one element in
each orbit of the action of $\mapcld$ on $\pi_1(\surf\mindee)$. A
set of curves $\lambda\in\pi_1(\surf\mindee)$ spanning the orbits
of the action of $\mapcld$ can be obtained using results from
geometric topology. It has been shown, see for instance Lemma
  2.3.A. in \cite{geomtop}, that the equivalence classes of all
  non-separating curves $\gamma$ on the surface $S_{g,n}\mindee$, are in the same orbit
  under the action of the mapping class group. We can apply the same
  argument to separating curves if we keep in mind that the punctures of the surface $S_{g,n}\mindee$
  can be distinguished via the conjugacy classes assigned to them.
Hence, any two separating curves $\gamma, \gamma'$ on
  $S_{g,n}\mindee $ such that the
two components of $(S_{g,n}\mindee)\setminus\gamma$ and
  $(S_{g,n}\mindee)
\setminus\gamma'$
 contain the same number of handles and the same sets of punctures
are in the same orbit under the action of the mapping class
  group. We therefore have to prove \eqref{tid} for a single
  non-separating curve, such as the generators $a_i$, $\delta_i$, $\alpha_i$,
  $\epsilon_i$, $\kappa_{\nu,n+2i-1}$ or $\kappa_{\nu,n+2i}$ in
  \eqref{dtgen} and the separating curves  $\gamma^{i_1\ldots i_r j_1\ldots
j_s}$  show in Fig.~\ref{sepc}.
\begin{align}
\label{sepcurves} \gamma^{i_1\ldots i_r j_1\ldots
j_s}&=h_{j_s}\circ h_{j_{s-1}}\cdots h_{j_2}\circ h_{j_1}\circ
m_{i_r}\circ
m_{i_{r-1}}\cdots m_{i_2}\circ m_{i_1}\\
&=(\dcm_1^\inv\cdots\dcm_{i_1-1}^\inv)\circ(\dcm_{i_1}^\inv\cdots\dch_{j_s}^\inv)\circ(\dch_{j_s-1}\cdots\dch_{j_{s-1}+1})\circ\ldots\circ
(\dch_{j_2-1}\cdots\dch_{j_1+1})\nonumber\\
&\circ(\dcm_{i_r-1}\cdots\dcm_{i_{r-1}+1})\circ\ldots\circ(\dcm_{i_2-1}\cdots\dcm_{i_1+1})\circ(\dcm_{i_1-1}\cdots\dcm_1)\nonumber\\
\nonumber\\ &\qquad 1\leq j_1<j_2<\ldots<j_s\leq j_{s+1}:=g,\;
1\leq i_1<i_2<\ldots<i_r\leq i_{r+1}:=n. \nonumber
\end{align}
Using formula \eqref{caspb} or \eqref{intpb0}, we can determine
the Poisson bracket of a general function $f\in\cif(H^\ntg)$ with
the Wilson loop observable $\tilde t_{\gamma^{i_1\ldots i_r
j_1\ldots j_s}}$ and obtain
\begin{align}
\label{sepdtpb} &\{f, \tilde t_{\gamma^{i_1\ldots i_r j_1\ldots
j_s}}\}=p_{\gamma^{i_1\ldots i_r j_1\ldots
j_s}}^a\left(\sum_{k=1}^r
R^{M_{i_k}}_a+L^{M_{i_k}}_a+\sum_{k=1}^sR^{A_{j_k}}_a+L^{A_{j_k}}_a+R^{B_{j_k}}_a+L^{B_{j_k}}_a
\right)f\\
&+\sum_{k=1}^r(p_{\gamma^{i_1\ldots i_r j_1\ldots j_s}}^a-p_{
(m_{i_k}\cdots m_{i_1})\circ\gamma^{i_1\ldots i_r j_1\ldots
j_s}\circ(m_{i_k}\cdots m_{i_1})^\inv}^a) \!\!\!\!\!\!\sum_{i_k<j<i_{k+1}}\!\!\!\!\!\!(R^\mj_a+L^\mj_a)f\nonumber\\
&+\sum_{k=0}^{s-1}(p_{\gamma^{i_1\ldots i_r j_1\ldots j_s}}^a-p_{
(h_{j_k}\cdots h_{j_1}m_{i_r}\cdots m_{i_1})\circ\gamma^{i_1\ldots
i_r j_1\ldots j_s}\circ(h_{j_k}\cdots h{j_1}m_{i_r}\cdots
m_{i_1})^\inv}^a)\!\!\!\!\!\!\sum_{j_k<j<j_{k+1}}\!\!\!\!\!\!(R^\aj_a\!+\!L^\aj_a\!+\!R^\bjj_a\!+\!L^\bjj_a)f\nonumber
\end{align}
where we set $i_{r+1}=n$, $j_0=0$, $h_{j_0}=1$. Using the
graphical representation of $\gamma^{i_1\ldots i_r j_1\ldots j_s}$
defined in Sect.~\ref{factassign}, we can determine  the
associated one-parameter group of diffeomorphisms
\begin{align} \label{sepcurvedtinf}
T^t_{\tilde t,\gamma_{i_1\ldots i_r j_1\ldots j_s}}:
&M_{i_k}\mapsto H_{\gamma_{i_1\ldots i_r j_1\ldots j_s}}^{-t}
M_{i_k}H_{\gamma_{i_1\ldots i_r j_1\ldots j_s}}^t\qquad k=1,\ldots,r\\
&A_{j_k}\mapsto H_{\gamma_{i_1\ldots i_r j_1\ldots j_s}}^{-t}
A_{j_k}H_{\gamma_{i_1\ldots i_r j_1\ldots j_s}}^t\qquad
k=1,\ldots,s\nonumber\\
&B_{j_k}\mapsto H_{\gamma_{i_1\ldots i_r j_1\ldots j_s}}^{-t}
B_{j_k} H_{\gamma_{i_1\ldots i_r j_1\ldots j_s}}^t\qquad
k=1,\ldots,s\nonumber\\
&M_j\mapsto[H_{\gamma_{i_1\ldots i_r j_1\ldots j_s}}^{-t},
M_{i_l}\cdots M_{i_1}] M_j[M_{i_l}\cdots
M_{i_1},H_{\gamma_{i_1\ldots i_r j_1\ldots j_s}}^{-t} ]\qquad
i_l<j<i_{l+1}\nonumber\\
&M_j\mapsto[H_{\gamma_{i_1\ldots i_r j_1\ldots j_s}}^{-t},
M_{i_r}\cdots M_{i_1}] M_j[M_{i_r}\cdots
M_{i_1},H_{\gamma_{i_1\ldots i_r j_1\ldots j_s}}^{-t} ]\qquad
i_r<j\leq n\nonumber\\
&A_j\mapsto[H_{\gamma_{i_1\ldots i_r j_1\ldots j_s}}^{-t},
M_{i_r}\cdots M_{i_1}] A_j[M_{i_r}\cdots
M_{i_1},H_{\gamma_{i_1\ldots i_r j_1\ldots j_s}}^{-t} ]\qquad
1\leq j<j_1\nonumber\\
&B_j\mapsto[H_{\gamma_{i_1\ldots i_r j_1\ldots j_s}}^{-t},
M_{i_r}\cdots M_{i_1}] B_j[M_{i_r}\cdots
M_{i_1},H_{\gamma_{i_1\ldots i_r j_1\ldots j_s}}^{-t} ]\qquad
1\leq <j<j_1\nonumber\\
&A_j\mapsto[H_{\gamma_{i_1\ldots i_r j_1\ldots j_s}}^{-t},
H_{j_l}\cdots M_{i_1}] A_j[H_{j_l}\cdots M_{i_1},
H_{\gamma_{i_1\ldots i_r j_1\ldots j_s}}^{-t} ]\qquad
j_l<j<j_{l+1}\nonumber\\
&B_j\mapsto[H_{\gamma_{i_1\ldots i_r j_1\ldots j_s}}^{-t},
H_{j_l}\cdots M_{i_1}] B_j[H_{j_l}\cdots M_{i_1},
H_{\gamma_{i_1\ldots i_r j_1\ldots j_s}}^{-t} ]\qquad
j_l<j<j_{l+1}\nonumber
\end{align}
where we write $H_\lambda=\rho_\lambda(M_1,\ldots,B_g)$ and for
$X=e^{p^aJ_a}\in H$, $X^{t}:=e^{tp^a_\lambda J^a}$. The action of
the  Dehn twists around $\gamma_{i_1\ldots i_r j_1\ldots j_s}$ on
 the generators of the fundamental group $\pi_1(\surf\mindee)$ is given by
\begin{align}\label{sepcurvedt} \text{d}_{\gamma_{i_1\ldots i_r
j_1\ldots j_s}}: &m_{i_k}\mapsto \gamma_{i_1\ldots i_r j_1\ldots
j_s}^\inv
m_{i_k}\gamma_{i_1\ldots i_r j_1\ldots j_s}\qquad k=1,\ldots,r\\
&a_{j_k}\mapsto\gamma_{i_1\ldots i_r j_1\ldots j_s}^\inv
a_{j_k}\gamma_{i_1\ldots i_r j_1\ldots j_s}\qquad
k=1,\ldots,s\nonumber\\
&b_{j_k}\mapsto\gamma_{i_1\ldots i_r j_1\ldots j_s}^\inv
b_{j_k}\gamma_{i_1\ldots i_r j_1\ldots j_s}\qquad
k=1,\ldots,s\nonumber\\
&m_j\mapsto[\gamma_{i_1\ldots i_r j_1\ldots j_s}^\inv,
m_{i_l}\cdots m_{i_1}] m_j[\gamma_{i_1\ldots i_r j_1\ldots
j_s}^\inv, m_{i_l}\cdots m_{i_1}]^\inv\qquad
i_l<j<i_{l+1}\nonumber\\
&m_j\mapsto[\gamma_{i_1\ldots i_r j_1\ldots j_s}^\inv,
m_{i_r}\cdots m_{i_1}] m_j[\gamma_{i_1\ldots i_r j_1\ldots
j_s}^\inv, m_{i_r}\cdots m_{i_1}]^\inv\qquad
i_r<j\leq n\nonumber\\
&a_j\mapsto[\gamma_{i_1\ldots i_r j_1\ldots j_s}^\inv,
m_{i_r}\cdots m_{i_1}] a_j[\gamma_{i_1\ldots i_r j_1\ldots
j_s}^\inv, m_{i_r}\cdots m_{i_1}]^\inv\qquad
1\leq j<j_1\nonumber\\
&b_j\mapsto[\gamma_{i_1\ldots i_r j_1\ldots j_s}^\inv,
m_{i_r}\cdots m_{i_1}] b_j[\gamma_{i_1\ldots i_r j_1\ldots
j_s}^\inv, m_{i_r}\cdots m_{i_1}]^\inv\qquad
1\leq j<j_1\nonumber\\
&a_j\mapsto[\gamma_{i_1\ldots i_r j_1\ldots j_s}^\inv,
h_{j_l}\cdots m_{i_1}] a_j[\gamma_{i_1\ldots i_r j_1\ldots
j_s}^\inv, h_{j_l}\cdots m_{i_1}]^\inv\qquad
j_l<j<j_{l+1}\nonumber\\
&b_j\mapsto[\gamma_{i_1\ldots i_r j_1\ldots j_s}^\inv,
h_{j_l}\cdots m_{i_1}] b_j[\gamma_{i_1\ldots i_r j_1\ldots
j_s}^\inv, h_{j_l}\cdots m_{i_1}]^\inv\qquad
j_l<j<j_{l+1},\nonumber
\end{align}
and comparing with expression \eqref{sepcurvedtinf}, we find that
the diffeomorphism of $H^\ntg$ induced by the Dehn twist
\eqref{sepcurvedt} agrees with the transformation
\eqref{sepcurvedtinf} for $t=1$. Hence, equation \eqref{tid} holds
for the separating curves $\gamma^{i_1\ldots i_r j_1\ldots j_s}$
and therefore for all embedded curves
$\lambda\in\pi_1(\surf\mindee)$.\hfill$\Box$


The Wilson loop observables $\tilde t_{\lambda}$ associated to the
bilinear form $\langle\,,\,\rangle$ in the Chern-Simons action
therefore act as the Hamiltonians for infinitesimal Dehn twists
along embedded curves $\lambda\in\pi_1(\surf\mindee)$.  While the
number and form of other $\Ad$-invariant functions $f\in\cif(H)$
is a property of the gauge group $H$, the observables $\tilde
t_\lambda$ are present generically in Chern-Simons theory
 and the associated transformations have a geometrical
 interpretation.
The fact that Dehn twists are infinitesimally generated via the
Poisson bracket allows us to determine the transformation
behaviour of the Poisson bracket and the transformations
$T^t_{f,\lambda}$ under a general automorphism of the fundamental
group $\pi_1(\surf\mindee)$ which arises from a homeomorphism of
$\surf\mindee$. We have the following theorem.

\begin{theorem}
Let $\varphi\in\text{Aut}(\pi_1(\surf\mindee))$ be an automorphism
of the fundamental group which satisfies the requirement
\eqref{invprops2} with $w=1$, $\varphi(m_D)=m_D^{\pm 1}$, and
denote by $\Phi_{\varphi}: H^\ntg\rightarrow H^\ntg$ the induced
diffeomorphism of $H^\ntg$.

\begin{enumerate}
\item Then, the transformation of the Fock-Rosly Poisson bracket
$\{\,,\,\}_r$ associated to the classical $r$-matrix $r$ under
$\Phi_\varphi$ is given by
\begin{align}
\label{poisstransf} \{f\circ\Phi_\varphi,
f\circ\Phi_\varphi\}_r=\begin{cases} \{f,g\}_r\circ \Phi_{\varphi}
& \text{if}\;\;\varphi(m_D)=m_D\\\{f,g\}_{-\sigma(r)}\circ
\Phi_{\varphi} &
\text{if}\;\;\varphi(m_D)=m_D^\inv\end{cases}\qquad\forall
f,g\in\cif(H^\ntg).
\end{align}

\item For any conjugation invariant $f\in\cif(H)$ and any
$\lambda\in\pi_1(\surf\mindee)$ with an embedded representative,
the flow $T^t_{f,\lambda}$ generated by the observable $f_\lambda$
satisfies
\begin{align}
\label{ttransf} T^t_{f,\varphi(\lambda)}=\begin{cases}
\Phi_{\varphi}^\inv\circ T^{t}_{f,\lambda}\circ\Phi_{\varphi} &
\text{if}\;\;
\varphi(m_D)=m_D\\
\Phi_{\varphi}^\inv\circ T^{-t}_{f,\lambda}\circ\Phi_{\varphi} &
\text{if}\;\; \varphi(m_D)=m_D^\inv.
\end{cases}
\end{align}
\end{enumerate}
\end{theorem}

{\bf Proof:} To prove identities \eqref{poisstransf},
\eqref{ttransf} we note that automorphisms
$\varphi\in\text{Aut}(\pi_1(\surf\mindee))$ satisfying the
conditions \eqref{invprops2} with $\varphi(m_D)=m_D$ correspond to
elements of the mapping class group $\mapcld$, while automorphisms
with $\varphi(m_D)=m_D^\inv$ are obtained by composing elements of
the mapping class group $\mapcld$ with the involution I
\begin{align}
\label{iphi} \varphi(m_D)=m_D^\inv\quad\Rightarrow\quad \exists
\psi\in\mapcld:\; \varphi=\psi\circ I.
\end{align}
In the first case, identities \eqref{poisstransf}, \eqref{ttransf}
 follow from the proof of  Theorem \ref{dttheor}. For automorphisms of the
form \eqref{iphi}, we note that the diffeomorphisms $\Phi_\eta,
\Phi_\tau:H^\ntg\rightarrow H^\ntg$ associated to arbitrary
$\eta,\tau\in\text{Aut}(\pi_1(\surf\mindee))$ satisfy
$\Phi_{\tau\circ\eta}=\Phi_\eta\circ\Phi_{\tau}$. Recalling
identity \eqref{invid} from Theorem \ref{frdualth} which specifies
the transformation of the Poisson bracket under the diffeomorphism
$\Phi_I$ associated to the involution, we then find
\begin{align}
\label{finhelp} \{f\circ\Phi_{\psi\circ I}, g\circ\Phi_{\psi\circ
I}\}_r=\{f\circ \Phi_I,
g\circ\Phi_I\}_r\circ\Phi_\psi=\{f,g\}_{-\sigma(r)}\circ
\Phi_I\circ\Phi_\psi=\{f,g\}_{-\sigma(r)}\circ \Phi_{\psi\circ I}.
\end{align}
To prove identity \eqref{ttransf}, we apply \eqref{poisstransf} to
the bracket of the observable $f_{I(\lambda)}$ with a general
function $g\in\cif(H^\ntg)$
\begin{align}
&\frac{d}{dt}|_{t=0} g\circ T^t_{f, I(\lambda)}=\{ f_{I(\lambda)},
g\}=\{f_\lambda\circ \Phi_I, g\}=-\{f_\lambda, g\circ
\Phi_I\}\circ \Phi_I=-\frac{d}{dt}|_{t=0} g\circ \Phi_I\circ
T_{f,\lambda}^t\circ \Phi_I\nonumber\\
\label{invtrafo} \Rightarrow \quad &T^t_{f,
I(\lambda)}=\Phi_I\circ T^{-t}_{f,\lambda}\circ
\Phi_I\qquad\forall f\in\cif(H).
\end{align}
Using the identity $\Phi_{\psi\circ I}=\Phi_I\circ \Phi_\psi$
together with \eqref{finhelp} then proves the claim.\hfill$\Box$


\section{Outlook and Conclusions}

\label{outlook}

In this paper we defined the dual of a set of generators of the
fundamental groups $\pi_1(\surf)$, $\pi_1(\surf\mindee)$ and
applied it to the moduli space of flat connections on $\surf$.
This dual is given by an involution of the fundamental group
$\pi_1(\surf\mindee)$ and can be viewed as a dual graph for the
set of curves representing these generators. In particular, the
expression of the homotopy equivalence class of a general embedded
curve on $\surf\mindee$ in terms of the dual generators determines
both the number of its intersection points with the
representatives of the generators and the order in which the
intersection points occur on these representatives.

By applying this involution to Fock and Rosly's description
\cite{FR} of the moduli space of flat $H$-connections, we showed
that the Poisson structure takes a particularly simple form when
expressed in terms of both the holonomies along the original
generators and those of their duals. This allowed us to give
explicit expressions for the Poisson brackets of the Wilson loop
observables associated to general conjugation invariant functions
on $H$ and  general embedded curves on $\surf\mindee$ and to
derive the associated flows on phase space. For the generic
observables constructed from the non-degenerate $Ad$-invariant
bilinear form in the Chern-Simons action, we showed that the
associated flows represent infinitesimal Dehn twists. Using the
fact that these Dehn twists are generated via the Poisson bracket,
we determined  the transformation of the Poisson structure under
general automorphism of $\pi_1(\surf\mindee)$ induced by a
homeomorphisms of $\surf\mindee$.

The results in this paper generalise Goldman's classic results on
twist flows \cite{goldman} to surfaces with punctures. However, in
contrast to the formulation in \cite{goldman}, our description of
these flows is purely algebraic and gives concrete expressions for
their action on the holonomies along a set of generators of the
fundamental group $\pi_1(\surf\mindee)$. As the holonomies along
these generators appear as the basic variables in many approaches
to the quantisation of Chern-Simons theory
\cite{AGSI,AGSII,AS,BNR,we2}, this could prove useful in applying
the results to the associated quantum theories.

In the case of infinitesimal Dehn twists the associated quantum
transformations have been determined for some gauge groups. For
Chern-Simons theory with compact semisimple gauge groups, the
quantum action of the Dehn twists has been derived by Alekseev,
Grosse and Schomerus \cite{AGSI, AGSII,AS} who found that it is
implemented via the ribbon element of the quantum group
 arising in the combinatorial quantisation procedure.
The case of semidirect product gauge groups $G\ltimes\gothg^*$ is
investigated in \cite{we3}, where it is shown that the mapping
class group acts on the representation spaces of the quantum
algebra and that the action of Dehn twists is given by the ribbon
element of the quantum double $D(G)$. This raises the question if
 the implementation of Dehn twists in the quantum theory via the ribbon element of an
 associated quantum group is also present for other gauge groups such
 as the group $SL(2,\CC)$ investigated in \cite{BNR}.
It would also be interesting to use the results of this paper to
determine the quantum action of the flows generated by other
Wilson loop observables and to see how these flows reflect the
properties of the quantum group arising in the quantisation of the
theory.

Finally, Wilson loop observables and the transformations these
observables generate via the Poisson bracket play an important
role in the Chern-Simons formulation of (2+1)-dimensional gravity.
For (2+1)-dimensional gravity with vanishing cosmological
constant, it is shown in \cite{ich} that the flows generated by a
particular set of Wilson loop observables correspond to the
construction of evolving (2+1)-spacetimes via grafting. The
results of this paper allow one to generalise these results to
other values of the cosmological constant, for which the Poisson
structure is more involved, and to establish a common framework
which treats the cosmological constant as a deformation parameter
\cite{ich3}.

\subsection*{Acknowledgements}

I thank Bernd Schroers for answering  questions, pointing out
references and for comments on the draft of this paper. Research
at Perimeter Institute is supported in part by the Government of
Canada through NSERC and by the Province of Ontario through MEDT.


\begin{appendix}

\section{The generators of the   mapping class group}

In this appendix we give a set of generators of the mapping class
groups $\mapcld$ and $\mapcl$ and provide explicit expressions for
their action on a set of generators of the fundamental groups
$\pi_1(\surf\mindee)$, $\pi_1(\surf)$. A set of generators and
relations of the  mapping class group on oriented two-surfaces was
first derived by Birman \cite{birmgreen, birmorange}. In this
paper we work with the set used in \cite{AS}, which was first
presented in \cite{Wajnryb}, see also \cite{MatPol}.

\begin{figure}
 \vskip .3in \protect\input epsf
\protect\epsfxsize=12truecm
\protect\centerline{\epsfbox{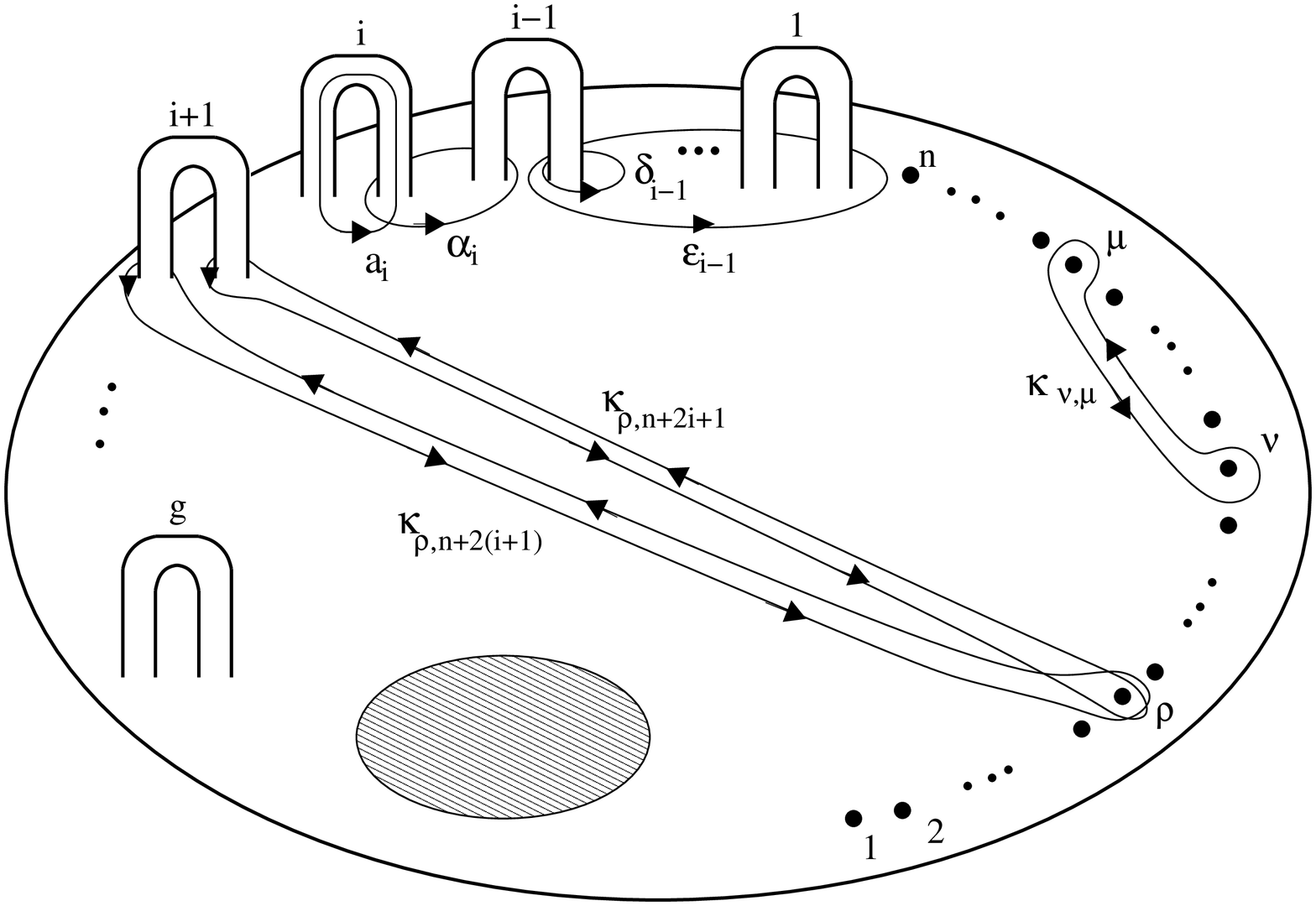}} \caption{The curves
associated to the generators of the mapping class group
$\pmapcld$}\label{dtgenfig}
\end{figure}

Both the pure mapping class group $\pmapcld$ and $\pmapcl$ are
generated by a finite set of Dehn twists around certain curves on
the surfaces $\surf\mindee$, $\surf$, but the latter with
additional relations. These curves are depicted in
Fig.~\ref{dtgenfig}, and in terms of the generators $m_i,a_j,b_j$
of $\pi_1(\surf\mindee)$, $\pi_1(\surf)$ their homotopy
equivalence classes are given by
\begin{align}
\label{dtgen}
&a_i\qquad i=1,\ldots,g\\
&\delta_i=a_i^{-1}\circ b_i^{-1}\circ a_i\qquad i=1,\ldots,g\nonumber\\
&\alpha_i=a_i^{-1}\circ b_i^{-1}\circ a_i\circ b_{i-1}\nonumber\\
&\epsilon_i=a_i^{-1}\circ b_i^{-1}\circ a_i  \circ h_{i-1} \cdots
h_1=\qquad i=2,\ldots,g\nonumber\\
&\kappa_{\nu,\mu}=m_\mu \circ
m_\nu\qquad\;\; 1\leq \nu<\mu\leq n\nonumber\\
&\kappa_{\nu,
n+2i-1}=a_i^{-1}\circ b_i^{-1}\circ a_i\circ m_\nu\nonumber\\
&\kappa_{\nu, n+2i}=b_i \circ m_\nu\nonumber.
\end{align}
Dehn twists around embedded curves representing the elements
\eqref{dtgen} induce automorphisms of the fundamental groups
$\pi_1(\surf\mindee)$, $\pi_1(\surf)$ which are given by their
action on the generators $m_i,a_j,b_j$. in the following we list
only those generators which do not transform trivially.
\begin{align}
\label{ad}
\text{d}_{a_i}: &b_i \mapsto b_ia_i\\
\nonumber\\ \label{dd}
\text{d}_{\delta_i}: &a_i \mapsto a_i\delta_i=b_i^{-1}a_i\\
\nonumber\\\label{alpha}
\text{d}_{\alpha_i}: &a_i\mapsto b_i^{-1}a_ib_{i-1}=a_i\alpha_i\\
&b_{i-1}\mapsto
\alpha_i^{-1}b_{i-1}\alpha_i\nonumber\\
&a_{i-1}\mapsto \alpha_i^{-1}a_{i-1}\nonumber\\ \nonumber\\
\label{eps}
\text{d}_{\epsilon_i}: &a_i\mapsto b_i^{-1}a_ik_{i-1}\ldots k_1=a_i\epsilon_i\\
&a_k\mapsto \epsilon_i^{-1} a_k\epsilon_i\nonumber
\\
&b_k\mapsto \epsilon_i^{-1}b_k\epsilon_i\qquad\qquad\forall 1\leq k <i\nonumber\\
\nonumber\\ \label{etapp} \text{d}_{\kappa_{\nu,\mu}}:
&m_\nu\mapsto\kappa_{\nu,\mu}^{-1}m_\nu\kappa_{\nu,\mu}\\
&m_\mu\mapsto\kappa_{\nu,\mu}^{-1}m_\mu\kappa_{\nu,\mu}\nonumber\\
&m_j\mapsto  [\kappa_{\nu,\mu}^{-1},
m_\nu]m_j[m_\nu,\kappa_{\nu,\mu}^{-1} ]
\quad\qquad\qquad\qquad\qquad\qquad\quad\quad\forall \nu<j<\mu
\nonumber\\ \nonumber\\\label{etapdelta} \text{d}_{\kappa_{\nu,
n+2i-1}}: &m_\nu\mapsto
\kappa_{\nu,n+2i-1}^{-1}m_\nu\kappa_{\nu, n+2i-1}\\
&a_i\mapsto a_i\kappa_{\nu,n+2i-1}\nonumber\\
& x\mapsto [\kappa_{\nu,n+2i-1}^{-1},m_\nu]x [m_\nu,
\kappa_{\nu,n+2i-1}^{-1}] \qquad\qquad
x\in\{m_{\nu+1},\ldots,m_n,a_1,\ldots,b_{i-1}\}\nonumber\\\nonumber\\\label{etapb}
\text{d}_{\kappa_{\nu,n+2i}}:
&m_\nu\mapsto \kappa_{\nu,n+2i}^{-1}m_\nu\kappa_{\nu,n+2i}\\
 &b_i\mapsto\kappa_{\nu,n+2i}^{-1}b_i\kappa_{\nu,n+2i}\nonumber\\
&a_i\mapsto \kappa_{\nu,n+2i}^{-1}a_i [m_\nu, \kappa_{\nu,n+2i}^\inv]\nonumber\\
&x\mapsto [\kappa_{\nu,n+2i}^{-1},m_\nu] x
[m_\nu,\kappa_{\nu,n+2i}^\inv]\qquad\qquad\qquad
x\in\{m_{\nu+1},\ldots,m_n, a_1,\ldots b_{i-1}\} \nonumber
\end{align}

\begin{figure}
 \vskip .3in \protect\input epsf
\protect\epsfxsize=12truecm \protect\centerline{\epsfbox{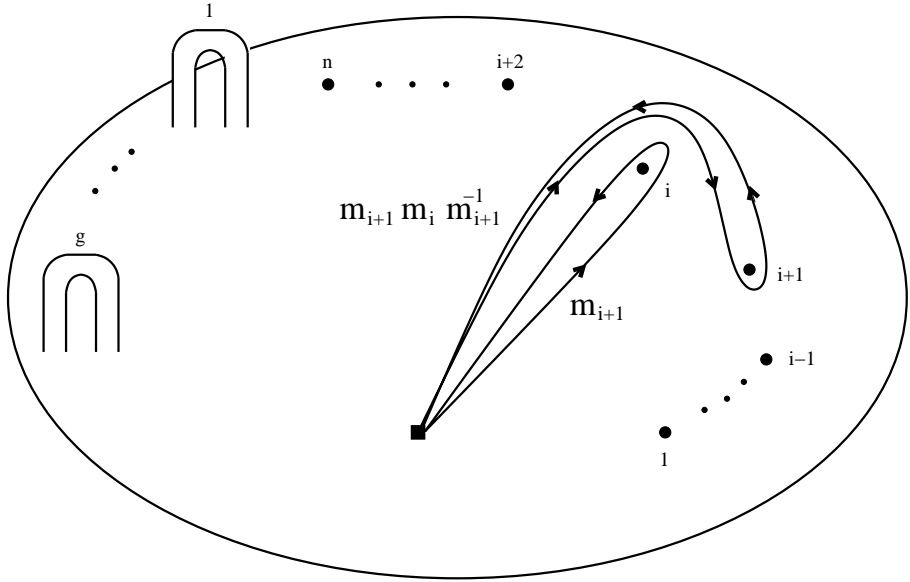}}
\caption{The generators of the braid group on the
  surface
$S_{g,n}\mindee$}\label{br1}
\end{figure}

A set of generators of the full mapping class group of the surface
 $S_{g,n}\mindee$ is obtained by supplementing this set of generators with
 the  generators $\sigma^i$, $i=1,\ldots,n$ of the braid group.
The action of these generators on the loops $m_i$ around the
punctures is shown in Fig.~\ref{br1}. They leave invariant all
generators of the fundamental group except $m_i$ and $m_{i+1}$, on
which they act according to
\begin{align}
\label{braid} \sigma^i:
&m_i\mapsto m_{i+1}\\
&m_{i+1}\mapsto m_{i+1}m_im_{i+1}^{-1}\;.\nonumber
\end{align}
\end{appendix}


\begin{thebibliography}{99}



\bibitem{AT} Achucarro, A., Townsend, P.: {A Chern--Simons
 action for three-dimensional anti-de Sitter supergravity
 theories}.  { Phys.~Lett.} B { 180},  85--100 (1986)

\bibitem{Witten1} Witten, E.: {2+1 dimensional gravity as an exactly
 soluble system}. { Nucl.~Phys.} B {311},  46--78 (1988),
 {\em Nucl.~Phys.} B {339},  516--32 (1988)

\bibitem{FR} Fock, V.~V.,  Rosly, A.~A.: {Poisson structures on
moduli of flat connections on Riemann surfaces and $r$-matrices}.
{\em  ITEP preprint} (1992) { 72-92} (see also {math.QA/9802054}).




\bibitem{we1} Meusburger, C., Schroers, B.~J.:
Poisson structure and symmetry in the Chern-Simons formulation of
(2+1)-dimensional gravity. Class.~Quant.~Grav.{\bf 20}, 2193--2234
(2003)

\bibitem{ich}Meusburger, C.: Grafting and Poisson structure in
(2+1)-gravity with vanishing cosmological constant. to appear in
Commun. Math. Phys., gr-qc/0508004 (2005)

\bibitem{AGSI} Alekseev, A.~Y., Grosse, H., Schomerus, V.: {Combinatorial
quantization of the Hamiltonian Chern-Simons Theory}. {
Commun.~Math.~Phys.} { 172}, 317--58 (1995)

\bibitem{AGSII}
Alekseev, A.~Y., Grosse, H., Schomerus, V.: {Combinatorial
quantization of the Hamiltonian Chern-Simons Theory II}.
 { Commun.~Math.~Phys.} { 174}, 561--604 (1995)
\bibitem{AS} A.~Yu.~Alekseev and V.~Schomerus, {\em Representation
theory of Chern-Simons observables,} Duke Math.~Journal {\bf 85}
 447--510 (1996)

\bibitem{BNR} Buffenoir, E., Noui, K., Roche, P.: Hamiltonian
Quantization of Chern-Simons theory with $SL(2,\CC)$ Group. {
Class.~Quant.~Grav.} { 19},  4953-5016 (2002)

\bibitem{we2} Meusburger, C., Schroers, B.~J.: The quantisation of Poisson
    structures arising in Chern-Simons theory with gauge group
    $G\ltimes \mathfrak{g}^*$. Adv.~Theor.~Math.~Phys.~7,
    1003--1043 (2004)

\bibitem{goldman} Goldman, W.~M.: Invariant functions on Lie groups
and Hamiltonian flows of surface group representations.
Invent.~math. 85,  263--302 (1986)

\bibitem{ich3} Meusburger, C.: Geometrical (2+1)-gravity and the
Chern-Simons formulation: Grafting, Dehn twists, Wilson loop
observables and the cosmological constant. In Preparation


\bibitem{combgr} Collins, D.~J., Grigorchuk, R.~J.,  Kurchanov,
P.~F., Zieschang,H.: Combinatorial Group Theory and Applications
to Geometry. Berlin Heidelberg: Springer Verlag,  1998


\bibitem{AMII}   Alekseev, A.~Y., Malkin, A.~Z.: { Symplectic
structure of the moduli space of flat connections on a Riemann
surface}. Commun.~Math.~Phys. { 169}, 99--119 (1995)

\bibitem{we3} Meusburger, C.,  Schroers, B.~J.:
Mapping class group actions in Chern-Simons theory with gauge
group $G\ltimes \mathfrak{g}^*$. Nucl.~Phys.~B 706, 569-597 (2005)

\bibitem{birmorange} Birman, J.~S.: {Braids, links and mapping
    class groups},  {Ann.~of Math.~Studies} { 82} . Princeton: (Princeton
    Univ. Press,  1975



\bibitem{birmgreen} Birman, J.~S.:  {Mapping class groups and their
    relationship to braid groups}. {Comm.~Pure Appl.~Math.} { 22},
    213--38 (1969)















\bibitem{BaRa} A.~O.~Barut, R.~Raczka, {Theory of group
representations and applications}, World Scientific, Singapore,
1986.



\bibitem{geomtop}  N.~V.~Ivanov, Mapping Class Groups, in:
  R.~J.~Daverman  and  R.~B.~Sher  (Eds.), Handbook of Geometric Topology
  (Elsevier Science, Amsterdam, 2002).

\bibitem{Wajnryb} B.~Wajnryb, A simple presentation of the mapping
class group of an orientable surface, Israel Journal of
Mathematics {45}  (1983) 157--174.

\bibitem{MatPol} S.~Mateev  and  M.~Polyakov, A geometrical presentation
of the surface mapping class group and surgery,
Commun.~Math.~Phys. { 160} (1994)  537--550.



























\end{thebibliography}
\end{document}